\documentclass[twocolumn]{aastex631}

\def\lesssim{\mathrel{\hbox{\rlap{\hbox{\lower5pt\hbox{$\sim$}}}\hbox{$<$}}}}
\def\gtrsim{\mathrel{\hbox{\rlap{\hbox{\lower5pt\hbox{$\sim$}}}\hbox{$>$}}}}

\def\til{\raise.17ex\hbox{$\scriptstyle\mathtt{\sim}$}}
\newcommand{\um}{\,$\,\mu$m}
\newcommand{\sqdeg}{\,\,deg$^2$}
\newcommand{\m} {\,\,m}

\usepackage{color,soul}
\usepackage{multirow}
\usepackage{amsmath}

\usepackage{soul}

\begin{document}
\title{A spectral library and census of near-infrared stellar large-amplitude variables from Palomar Gattini-IR}

\author[0000-0001-6627-9903]{Nicholas Earley}
\affiliation{Cahill Center for Astrophysics, California Institute of Technology, Pasadena, CA 91125, USA}
\correspondingauthor{Nicholas Earley}
\email{nearley@caltech.edu}

\author[0000-0003-2758-159X]{Viraj Karambelkar} 
\altaffiliation{NASA Hubble Fellow} 
\affiliation{Columbia Astrophysics Laboratory, Columbia University, 538 West 120th Street 704, MC 5255, New York, NY 10027} 

\author[0000-0002-5619-4938]{Mansi Kasliwal}
\affiliation{Cahill Center for Astrophysics, California Institute of Technology, Pasadena, CA 91125, USA}

\author[0000-0002-8989-0542]{Kishalay De}
\affil{Department of Astronomy and Columbia Astrophysics Laboratory, Columbia University, 550 W 120th St. MC 5246, New York, NY 10027, USA}
\affil{Center for Computational Astrophysics, Flatiron Institute, 162 5th Ave., New York, NY 10010, USA}

\author{Lynne Hillenbrand}
\affiliation{Cahill Center for Astrophysics, California Institute of Technology, Pasadena, CA 91125, USA}

\author[0000-0002-4622-796X]{Roberto Soria}
\affil{College of Astronomy and Space Sciences, University of the Chinese Academy of Sciences, Beijing 100049, China}
\affil{Sydney Institute for Astronomy, School of Physics A28, The University of Sydney, Sydney, NSW 2006, Australia}

\author[0009-0005-8230-030X]{Aswin Suresh}
\affil{Center for Interdisciplinary Exploration and Research in Astrophysics (CIERA), Northwestern University, Evanston, IL 60201, USA}
\affil{Department of Physics and Astronomy, Northwestern University, Evanston, IL 60201, USA}

\author[0000-0003-1412-2028]{Michael C. B. Ashley}
\affil{School of Physics, University of New South Wales, Sydney NSW 2052, Australia}

\author[0000-0001-9315-8437]{Matthew J. Hankins}
\affil{Arkansas Tech University, Russellville, AR 72801, USA}

\author{Anna M. Moore}
\affil{Research School of Astronomy and Astrophysics, Australian National University, Canberra, ACT 2611, Australia}

\author{Jamie Soon}
\affil{Research School of Astronomy and Astrophysics, Australian National University, Canberra, ACT 2611, Australia}

\author[0000-0001-9304-6718]{Tony Travouillon}
\affil{Research School of Astronomy and Astrophysics, Australian National University, Canberra, ACT 2611, Australia}

\shorttitle{Census of LAVs from PGIR}
\shortauthors{Earley et al.}

\date{\today}
\begin{abstract}
    We present a near-infrared census of stellar large-amplitude variables (LAVs) observed by the Palomar Gattini-IR (PGIR) surveyor from 2019---2021. Over the three-year time period, PGIR performed a brightness-limited survey of the Northern sky ($\sim$\,18{\rm,}000{\sqdeg}) to $J$-band AB magnitudes $\sim$\,13 within and $\sim$\,15 outside the Galactic plane. From $\sim$\,70 million stars detected in PGIR reference images, we provide a spectral and photometric library of the 128 largest amplitude stellar variables detected to median SNR $>$ 10 for more than 50 epochs with more than 5 high-amplitude detections, peak-to-peak magnitudes $\geq$ 2, and von Neumann ratios $\leq$ 0.2. We obtained medium-resolution near-infrared spectra with TripleSpec on the 200-inch Hale Telescope at Palomar Observatory and SpeX at NASA's Infrared Telescope Facility. The spectral census consists of 82 evolved and dust-obscured Asymptotic Giant Branch stars, 16 R Coronae Borealis stars, 13 young-stellar or pre-main-sequence objects, 8 symbiotic binaries, 7 erratic carbon- and oxygen-rich giants, and 2 RV Tauri supergiants. The spectral and photometric dataset serves as an atlas of near-infrared LAVs and a repository of evolved stars, eruptive variables, and binary systems for future deeper infrared surveys.
\end{abstract}

\section{Introduction}
Wide-field, modern time domain surveys have driven the systematic and comprehensive exploration of the multi-band variable sky. With the expansion of wide field-of-view and fast surveyors in the past decade, long-baseline time domain surveys are revealing the photometric signatures of millions of the largest amplitude transient sources and stellar variables, creating an unprecedented volume of targets to be flagged for spectroscopic follow-up, chemical identification, and ontological classification. In optical surveys, populations of variable stars have been characterized using data from the Massive Compact Halo Object (MACHO) project \citep{Alcock:1997ApJ...486..697A, 1998AJ....115.1921A, 2001ApJ...554..298A}, {\it Gaia} \citep{2021A&A...648A..44M}, EROS-2 \citep{2014A&A...566A..43K}, Catalina and Siding Spring Surveys \citep{2009ApJ...696..870D, 2017MNRAS.469.3688D}, the All Sky Automated Survey for Supernovae \citep[ASAS-SN,][]{Jayasinghe:2018MNRAS.477.3145J, Jayasinghe:2021MNRAS.503..200J}, the Optical Gravitational Lensing Experiment \citep[OGLE, e.g.,][and others]{Soszy:2009AcA....59..335S, Pietrukowicz:2017NatAs...1E.166P, Iwanek:2022ApJS..260...46I}, the Asteroid Terrestrial-impact Last Alert System \citep[ATLAS,][]{Heinze:2018AJ....156..241H}, and the Zwicky Transient Facility \citep[ZTF,][]{Chen:2020ApJS..249...18C}. As a result, the landscape of stellar variability in optical wavelengths has been extensively studied and characterized. Of these stellar objects, the broad phenomenological class of large-amplitude variables (LAVs) encompass sources with amplitudes $\gtrsim 1$ magnitude between minima and maxima in timeseries data. Stellar LAVs consist of heterogeneous populations of long-period variables (LPVs; periods $\gtrsim$ hundreds of days), young stellar objects (YSOs), eruptive variables or interacting binary systems. Much of the largest-amplitude LPVs further consist of a diversity of Asymptotic Giant Branch (AGB) stars or massive red supergiants (RSGs), which serve as testbeds for late-stage stellar evolution. 

However, while censuses of these objects have been extensively carried out in optical wavelengths, similar campaigns for stellar variables in infrared wavelengths betwen 1$-$3{\um} have been more limited in terms of sky coverage and cadence. 

Due to the high-cost and limited fields-of-view for early infrared detectors, the first dedicated near-infrared (NIR\footnote{In this paper, we refer to wavelengths between $1-3${\um}, covering the photometric $J$, $H$, and $K$-bandpasses, as the NIR.}) observations for stellar variables focused on fields restricted in area $\lesssim 1${\sqdeg} \citep[e.g., a $24\times24$ arcmin$^2$ $K$-band survey of $\sim 400$ LAVs with the PANIC camera and a 4{\sqdeg} $J$ and $K_s$ survey of $\sim 1000$ variables in the Galactic Bulge with DENIS, detailed respectively in][]{Glass:2001MNRAS.321...77G,Schultheis:2000A&A...362..215S}. 
While the Two Micron All Sky Survey (2MASS) generated an all-sky point-source catalog between 1997$-$2001, stellar variability was only able to be probed in overlapping regions observed multiple times through the course of the survey, limiting sky coverage to $\sim 100${\sqdeg} \citep{Kouzuma:2009AJ....138.1508K}. Other targeted photometric infrared campaigns have since been mounted in the $K$-band with the Wide Field CAMera (WFCAM) on the UK InfraRed Telescope \citep[UKIRT; e.g., in M33,][]{Javadi:2015MNRAS.447.3973J} and the WIRCAM on the Canada-France-Hawaii Telescope \citep[CFHT; e.g., in the M3 globular cluster,][]{Bhardwaj:2020AJ....160..220B}. 

The VISTA Variables in the Via Lactea (VVV) survey has been the most extensive, covering a total of 520{\sqdeg} over a six-year period (2009-2015), discovering more than $10^6$ stellar variables in its survey of the Galactic Bulge and southern disk \citep{Catelan:2011rrls.conf..145C,Guo:2022MNRAS.513.1015G,Nikzat:2022A&A...660A..35N}. In total however, infrared efforts at photometric and spectral atlases of the largest amplitude stellar variables have largely been focused on particular regions, limited below 1000{\sqdeg} of sky coverage. Unveiling the full extent and spectral properties of the dynamic infrared sky is critical for revealing a population of sources hidden due to high line-of-sight extinction values in optical wavelengths \citep{Cardelli:1989ApJ...345..245C}. 

In this paper, we aim to expand near-infrared sky coverage with a three-year, brightness-limited campaign in the $J$-band, employing the Palomar-Gattini IR (PGIR) time domain survey as our photometric discovery engine. PGIR is a 25 deg$^2$ near-infrared $J$-band camera on a 30 cm robotic telescope at the Palomar Observatory in Southern California \citep{2020PASP..132b5001D, 2019NatAs...3..109M}. PGIR survey operations began in July 2019 and the survey is currently scanning the entire Northern sky ($\sim$18{\rm,}000{\sqdeg}) every two nights to a depth of $J_{\text{AB}} \sim 15$, and $J_{\text{AB}} \sim 13$ in the Galactic plane due to confusion noise. As a result of its wide and shallow near-infrared survey, PGIR has targeted specific subclasses of stellar variables such as Galactic novae \citep{De:2021ApJ...912...19D} and R Coronae Borealis (RCB) stars \citep{viraj:2024PASP..136h4201K}, which have constrained rate estimates and implications for these sources' formation channels. We now deliver a dedicated near-infrared spectral library and a census for several classes of large-amplitude stellar variables observed through the course of its survey. While near-infrared spectral libraries for a fraction of these variables form the cornerstone for stellar modeling \citep[e.g.,][]{Rayner:2009ApJS..185..289R}, similar libraries have not been presented for the myriad of large-amplitude variable types. The atlas of LAVs consists of 128 large-amplitude variables identified over a three-year survey of the Northern sky ($\delta> -28.9^{\circ}$), including R Coronae Borealis (RCB) stars, outbursting young stellar objects (YSOs), extremely long-period AGB stars, AGB stars undergoing sudden, intense dust formation episodes, and candidate symbiotic binary systems. The following spectral and photometric data will be useful for comparison with theoretical modeling of large-amplitude variability and stellar evolution in future work. 

In \S\ref{sec:selection}, we detail the methodology and selection criteria for variables included in the census. In \S\ref{sec:data}, we describe the construction of data products, highlighting broad demographic features in public optical and IR color-diagrams. In \S\ref{sec:demographics}, we describe the demographics of the census in detail, highlighting photometric and spectral features of individual objects within the sample selection. In \S\ref{sec:future}, we discuss the discovery prospects for future infrared surveys extending beyond 1{\um} into the $K$-band.

\section{Sample selection}
\label{sec:selection}

\begin{table*}
	\centering
	\caption{PGIR stellar LAV census selection criteria}
	\label{tab:criteria}
	\begin{tabular}{cccc}
		\hline
		 Step & Sample & Selection criteria & Number of unique sources \\
		\hline
		1 & LAVs in PGIR reference images & $ptp > 1$ & 15,320\\
            & & $\eta < 0.5$ & \\
            \hline
            2 & LAVs outside of PGIR LPV catalog  & No source in catalog within 5" & 2,799 \\
            \hline
            & & Median SNR $>$ 10 & \\
            3 & Long-lived, high SNR LAVs & $>$ 50 epochs of detections & 838\\
            & & $>$ 5 detections with (\(\lvert \text{mag} - \text{min(mag)}\rvert\) $>$ 1 mag) & \\ 
            \hline
            \hline
             & & $ptp \geq 2$ &  \\
            4 & Stellar census sources & $\eta \leq 0.2$ & 128 \\
            & & Quality forced aperture photometry\tablenotemark{a} & \\ 
            \hline
            \hline
	\end{tabular}
    \tablecomments{Starting with LAVs in PGIR reference images, the following subsample is selected from the preceding sample, i.e, the criteria are applied successively.}
    \tablenotetext{a}{130 sources satisfy the criteria on match file $ptp$ and $\eta$, however, one source exhibited systematic offsets in photometric measurements and is thus excluded from the sample. One source was cross-matched to a BL-Lac object and was excluded from spectroscopic follow-up.}
\end{table*}

PGIR employs a data reduction pipeline that performs real-time sky subtraction, flat-fielding, astrometry, photometry, image subtraction, and transient candidate identification \citep{2020PASP..132b5001D}. As part of the pipeline, reference images of each field in the survey have been used to construct a master source catalog which is spatially cross-matched to the source catalog for every new observation of the field. These reference images of PGIR fields serve as seeds for source detection---sources identified thereafter in individual epochs through nominal survey operations are then spatially cross-matched to the reference sources, which produces ``match file" lightcurves \citep{2020PASP..132b5001D}. In this census, we employ PGIR match file lightcurves, predicated on source detection in stacked images, which were generated over a three-year period from the beginning of survey operations in 2019 until 2021 July 14 to discover stellar variables. We searched for large-amplitude, erratic variability amongst\,\,$\sim 70$ million stars detected in PGIR reference images. This spectroscopic campaign was mounted prior to the first release of Palomar Gattini-IR J-band Light Curves, which are predicated on PSF-profile fit photometry on source cross-matches with 2MASS point source detections. The complete catalog contains lightcurves for $\sim 286$ million unique sources \citep{2024PASP..136j4501M}. The PGIR reference image source catalog thus represents $\sim 25\%$ of all sources in PGIR's survey, largely limited by confusion noise limiting source detection in the Galactic plane. 

Using the original match files, we employed a lightcurve-based selection criterion, probing for large-amplitude variability through two metrics---the absolute peak-to-peak amplitude for magnitude variations ($ptp$), and the von Neumann ratio ($\eta$). As defined, $ptp$ is the difference between maximum and minimum magnitude in the complete timeseries data and $\eta$ is the the ratio of the mean square successive difference to the distribution variance \citep{vn1, vn2}, which quantifies the degree of correlation between successive datapoints, and as a result, the ``smoothness" of a timeseries. The inverse von Neumann ratio has been previously used as a successful variability index to parse out long-lived variable lightcurves, in which timescales exceed the observing cadence, from those of non-variable sources \citep[e.g.,][and references therein]{Sokolovsky:2017MNRAS.464..274S,viraj:2024PASP..136h4201K, aswin:2024PASP..136h4203S, Bhattacharjee:2025PASP..137b4201B}.  

\begin{figure*}
\centering
\includegraphics[width=0.55\textwidth]{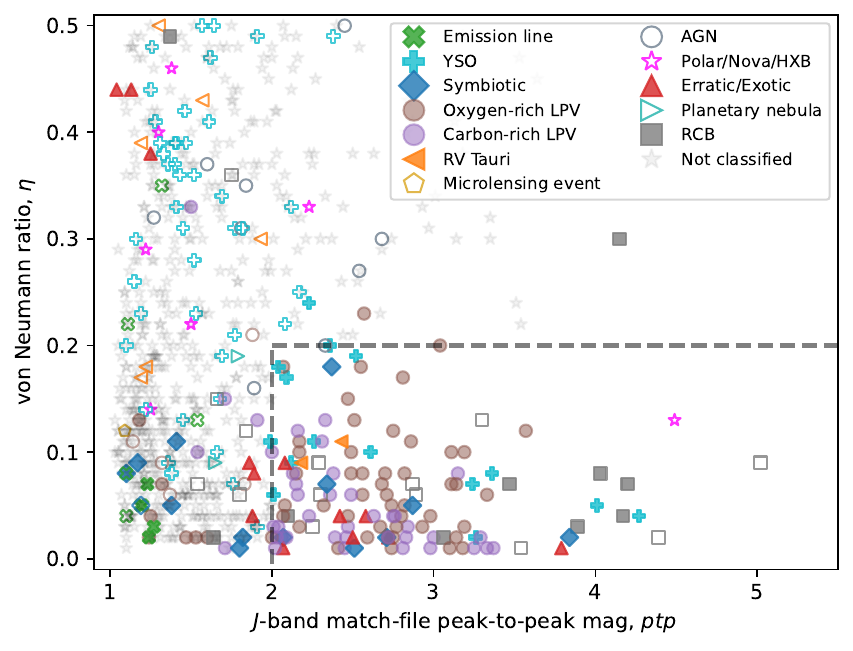}\includegraphics[width=0.45\textwidth]{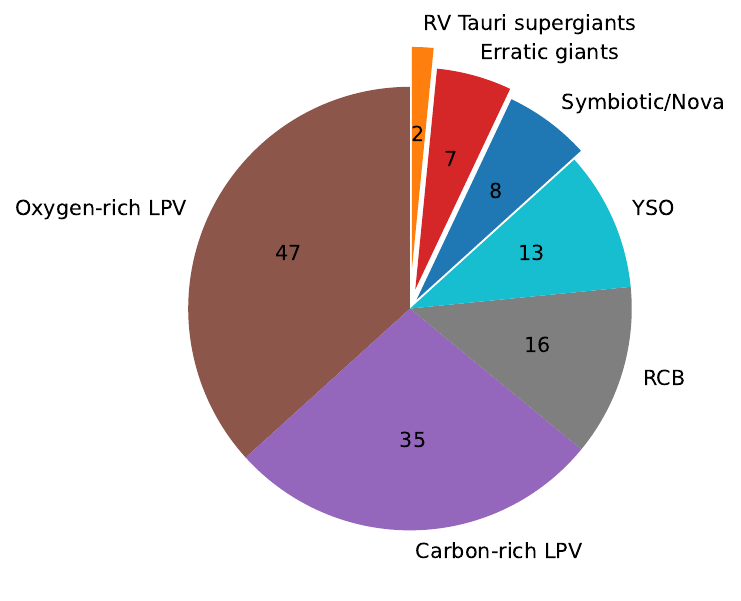}
\caption{Left: 838 long-lived LAVs were identified with peak-to-peak $J$-band magnitudes $ptp\geq1$ and von Neumann ratio $\eta\leq0.5$ from PGIR match file photometry from 2019-2021 (median SNR $>10$). Variables denoted with gray stars are listed as long-period-variables or are not known in Simbad. Unfilled colored markers designate sources we did not obtain spectra for which have defined classifications in the literature. Source classifications from spectroscopy were obtained for all sources in the region bounded by a dashed-line, with $ptp\geq2$ and $\eta\leq0.2$. We did not obtain additional spectra for a sample of previously known R Coronae Borealis stars (RCBs), a BL Lac object, and the classical nova V2891 Cyg. 37 sources outside of this region were also spectroscopically identified. Right: The demographics of the lightcurve-selected stellar near-infrared LAV sample ($ptp\geq 2$, $\eta\leq0.2$).}
\label{fig:census}
\end{figure*}

\begin{figure}
\centering
\includegraphics[width=\columnwidth]{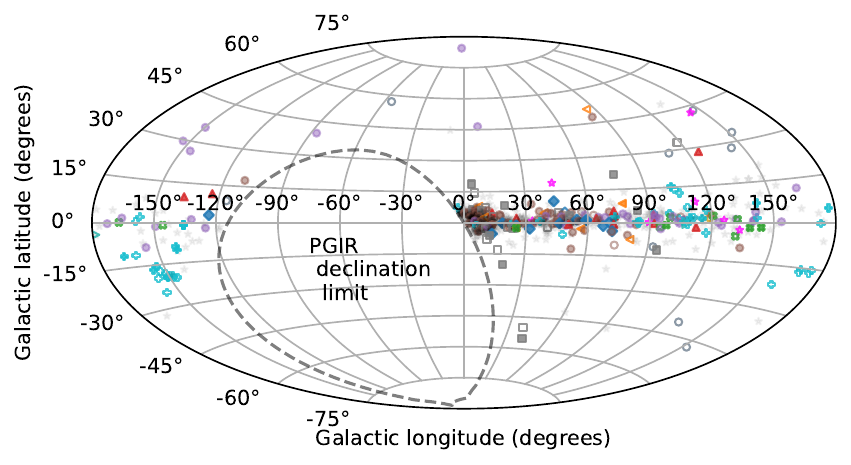}
\includegraphics[width=\columnwidth]{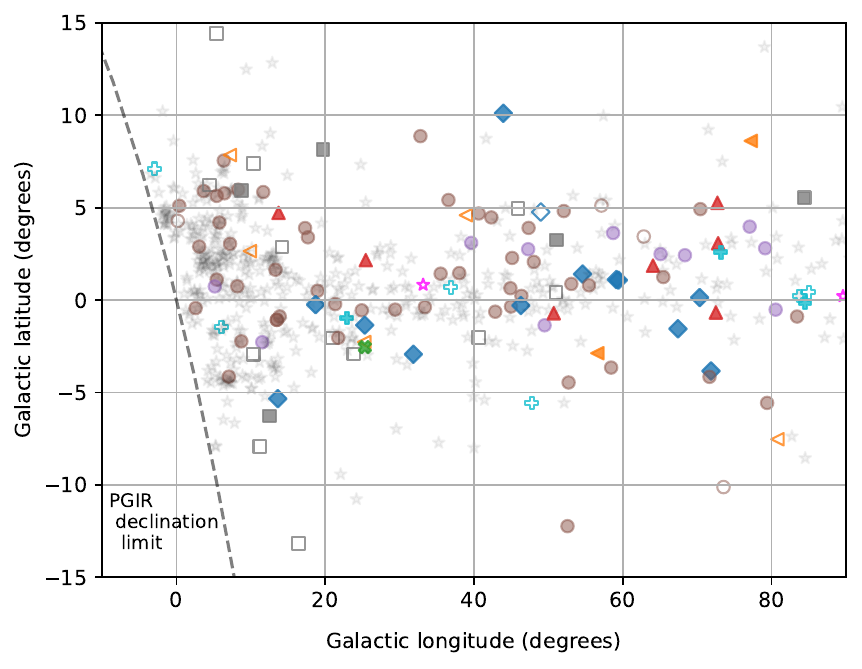}
\caption{Top: The Galactic distribution of all long-lived NIR LAVs identified with PGIR. The region bounded by dashed gray line indicates the coordinates that are inaccessible from Palomar (declinations $< -28.9^{\circ}$) 92\% in the census sample lie within $\pm15^\circ$ of the Galactic plane. Bottom: Zoomed in region of the Galactic Bulge with latitude $|b|<15^\circ$. Source classifications are colored the same as in Figure~\ref{fig:census}.}
\label{fig:distribution}
\end{figure}

We identified 15,320 LAVs from PGIR match files with von Neumann ratios less than 0.5 and peak-to-peak J-band magnitudes exceeding 1 mag. Of these variables, 12,521 were crossmatched within 5" of sources in the PGIR LPV catalog \citep{aswin:2024PASP..136h4203S}, leaving 2,799 LAVs that were considered in this work. In future works, we will seek to further spectroscopically characterize the LPVs that were filtered out by their photometry via machine learning methods (Frostig et al., in prep employs SDSS-V APOGEE $H$-band batch-spectroscopy for sources in the complete catalog; De et al., in prep uses IRTF SpeX observations to characterize 175 LPVs with periods exceeding 1000 days). 

We subsequently selected 838 sources with at least 50 epochs of detections, a minimum of at least 5 detections exceeding the minimum magnitude by 1 mag, and a median SNR of at least 10 in the match file lightcurves. These criteria were found to exclude spurious low-amplitude variability and select for long-lived, high-amplitude behavior.
To create a comprehensive and detailed census of the largest amplitude sources, we define ``large-amplitude" in the census criterion to refer to peak-to-peak amplitudes that exceed 2 magnitudes in PGIR $J$-band match file photometry. Within the sample, 130 sources hosted $ptp\geq2$ and $\eta\leq0.2$. We performed forced aperture photometry within a 3-pixel ($\sim$13") radius centered on the detected source coordinates. One source did not host any quality detections with PGIR aperture photometry and is thus excluded from the census. Large-amplitude variability in the match file lightcurve for this object was attributable to a known issue resulting in spurious measurements when targets are observed on the west side of the telescope meridian axis \citep[e.g., see][]{2024PASP..136j4501M}. One other LAV within the census selection criteria is a confirmed active galactic nucleus (AGN) and extraneous to the scope of the census of stellar variables: 8C 1803+784, a known BL Lac object \citep{Kankkunen:2025A&A...693A.318K}. We did not obtain additional spectroscopy for this object.

In constructing the following lightcurves for the sample of sources in the census, we present forced aperture photometry in addition to PGIR match file photometry. Match file $ptp$ and $\eta$ statistics differ from those calculated on forced photometry lightcurves which limit the completeness of the current sample. In a forthcoming paper, the forced photometry lightcurves for all 838 sources in the long-lived, high SNR LAV sample will be assessed and sources that satisfy the census criteria will be classified.

We present a complete census of the 128 largest amplitude variable stellar objects from the sample of long-lived, high SNR LAVs, hosting $ptp \geq 2$ magnitudes and $\eta \leq 0.2$ in match file lightcurves. All sources are confirmed to have quality forced aperture photometry measurements. The selection criteria are summarized in Table~\ref{tab:criteria}. 

We obtained spectra for 37 additional sources with match file $\eta$ and $ptp$ magnitudes not satisfying the strict selection cuts, with 21 exhibiting non-AGB or LPV-like behavior. For the remaining 671 out of 838 sources, 591 had either ambiguous (e.g., LPV or generic variable designations) or non-existent classifications on Simbad as opposed to 80 with specific well-defined classifications that were independently cross-matched in the literature. In Figure~\ref{fig:census}, we present the selection criteria premised on $\eta$ and $ptp$ in match file photometry and the complete census of stellar variables within the selection. In Figure~\ref{fig:distribution}, we present the the on-sky distribution of the complete long-lived LAV sample.

\section{Ancillary Data and Spectroscopic Follow-up}
\label{sec:data}
In this section, we describe the photometric and spectral data products used in the census of NIR LAVs. Archival photometry was obtained and source cross-matching performed using data from 2MASS, WISE, ZTF, and Gaia. Census sources were classified via NIR spectroscopic follow-up. 

\subsection{Multi-band photometry and color diagrams}
Due to PGIR's coarse 8.7" native pixel scale (4.3"/pixel in drizzled stacked images), we use the closest 2MASS counterpart source coordinates for cross-matching with multiband photometry when possible. 2MASS counterparts in the 2MASS All-Sky Point Source Catalog \citep[PSC,][]{Cutri:2003tmc..book.....C} were found within a 6" cone-search radius centered on PGIR source coordinates. Two objects of the 838 long-lived, high SNR LAVs do not host a 2MASS counterpart---the YSO PTF10nvg and classical nova V2891 Cyg \citep[PGIR19brv,][]{De:2021ApJ...912...19D} which are both included among the 128 census objects. We did not obtain additional spectroscopy for V2891 Cyg.

\begin{figure*}
 \centering
\includegraphics[width=0.5\textwidth]{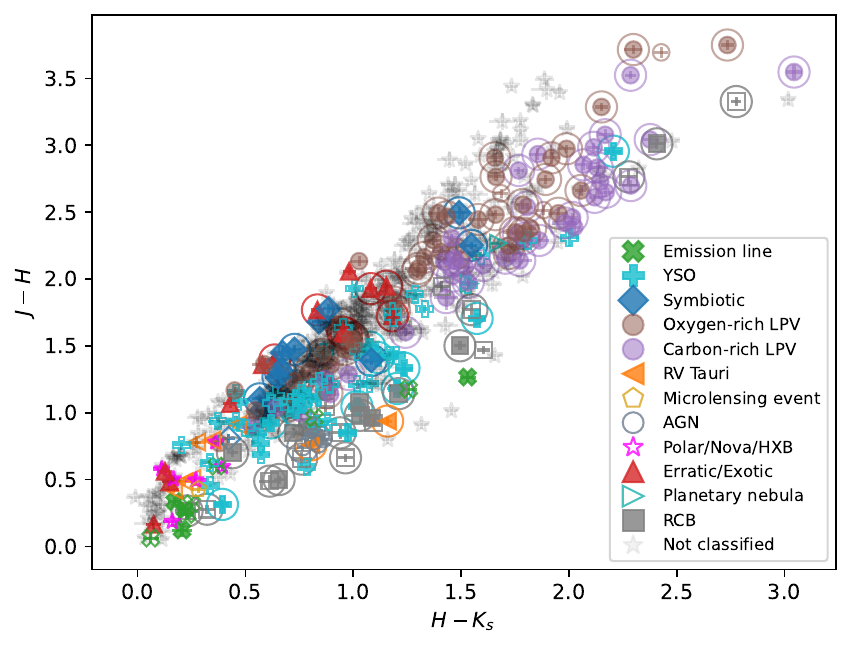}\includegraphics[width=0.5\textwidth]{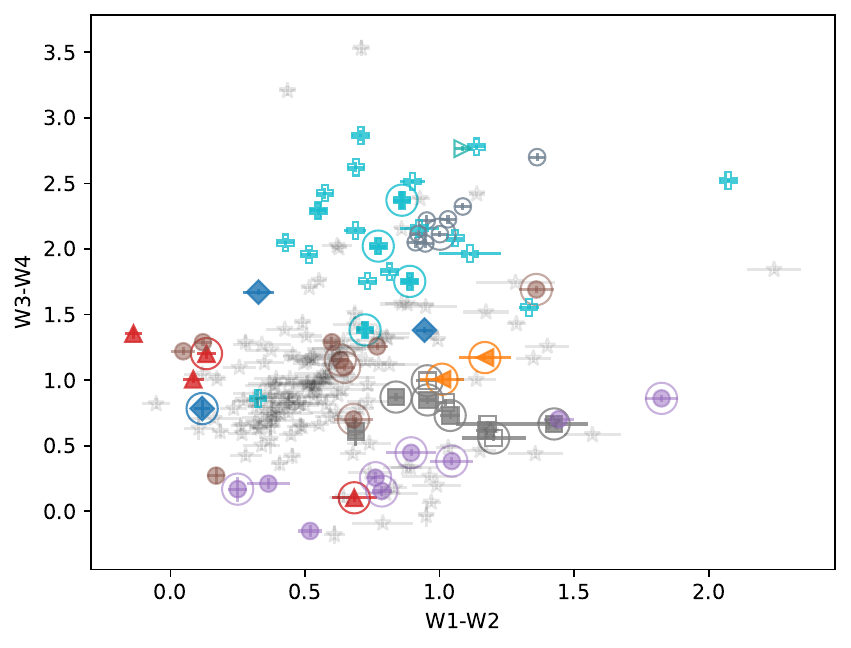}
\includegraphics[width=0.5\textwidth]{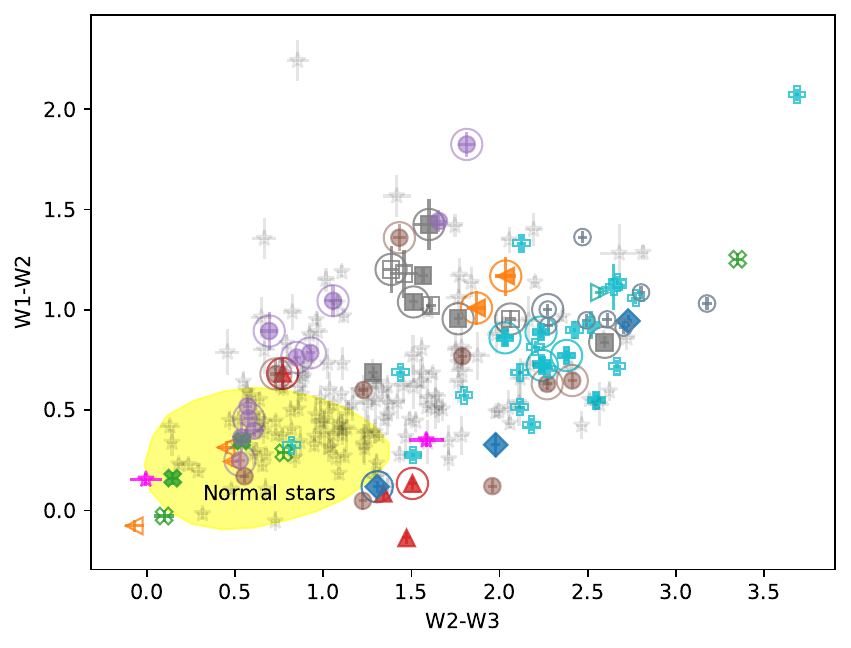}\includegraphics[width=0.565\textwidth]{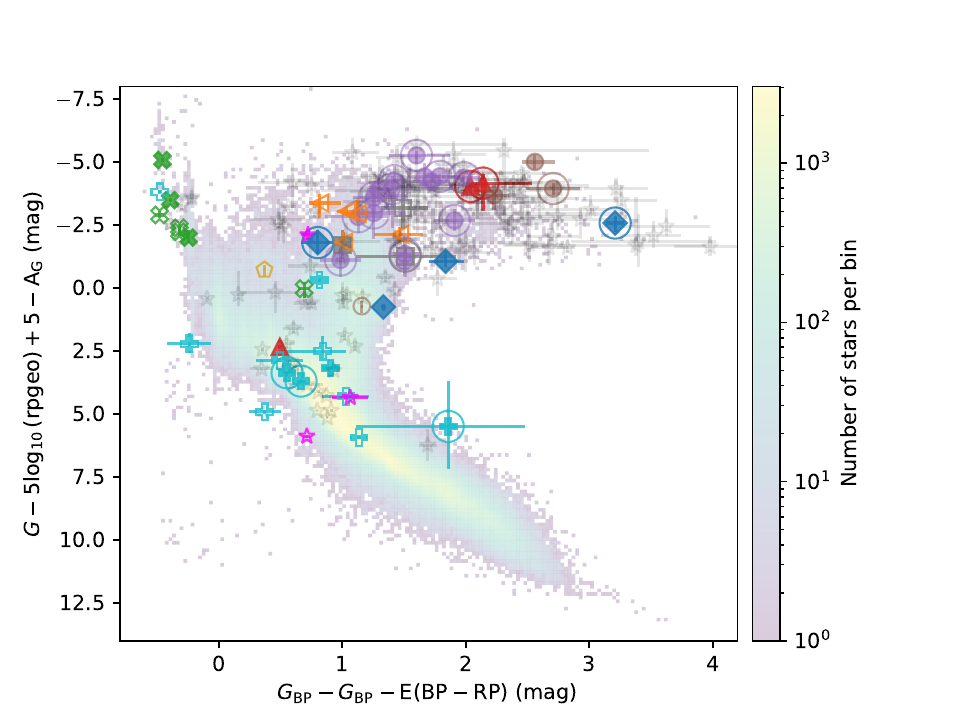}
\caption{Color diagrams for a fraction of long-lived, high SNR LAVs. Only sources with quality photometry in the relevant bands are presented in each color-color diagram (CCD). As in Figure~\ref{fig:census}, unfilled markers denote sources already classified, which we did not acquire spectra for in this work. Markers denoting sources within the census are circled. Top left: Observed 2MASS $J-H$ vs. $H-K_s$ CCD. Top right: Observed ALLWISE $W3-W4$ vs. $W1-W2$ CCD. Bottom left: Observed ALLWISE $W1-W2$ vs. $W2-W3$ CCD. 70\% of sources in the panel lie outside of the locus designating normal stars without large infrared excesses, reproduced from \citet{Wright:2010AJ....140.1868W}. Bottom right: Confirmed and candidate stellar variables with reported $A_G$ and $E(BP-RP)$ in {\it Gaia} DR3 are presented in an extinction-corrected de-reddened color-absolute magnitude diagram. Stellar density for a random sample of sources in Gaia DR3 are overplotted for reference.}
\label{fig:colors}
\end{figure*}

In Figure~\ref{fig:colors}, we present color-color diagrams for the larger subset of all long-lived, high SNR LAVs with quality photometry in the ALLWISE full-sky data release and the 2MASS All-Sky PSC. For each diagram, we apply quality flags to select uncontaminated sources with SNR $\geq10$ (i.e., \texttt{ph\_qual==A} and \texttt{cc\_flags==0} in each of the considered passbands). Out of 838 variables, 214 ($\sim 26\%$) have quality photometry in all WISE passbands, and 243 ($\sim 29\%$) in $W1$, $W2$, and $W3$. The majority of sources (745, $\sim 89\%$) have quality 2MASS photometry. While sources from various class contaminate each region of the CCDs, we observe clear delineations in the the MIR between carbon-rich variables and young-stellar or pre-main-sequence objects. In the $W1-W2$ vs. $W2-W3$ color-color diagram, we mark the region from \citet{Wright:2010AJ....140.1868W} designating normal stars that do not typically exhibit mid-infrared excesses. Approximately $70\%$ of the variables with quality $W1$, $W2$, and $W3$ photometry lie outside of the typical stellar locus, exhibiting large infrared excesses.
Within the de-reddened, extinction-corrected Gaia color-absolute magnitude diagram (CaMD), 178 stars are included using Gaia EDR3 distances from \citet{2021AJ....161..147B} as well as reddening values $E(RP-BP)$ and extinction corrections $A_G$ from DR3. For reference, we overplot the CaMD for a random sample of high-quality sources, demonstrating that the majority of quality sources in the PGIR LAV sample lie on the AGB. The data behind the figures are provided in Appendix~\ref{sec:appendix_color}. Only $\sim5\%$ of the long-lived PGIR LAVs were cross-matched within 2" to sources in the Gaia DR3 alert database\footnote{We acknowledge ESA Gaia, DPAC and the Photometric Science Alerts Team (\url{http://gsaweb.ast.cam.ac.uk/alerts})}. Six of these sources are included within the census: the classical nova Gaia 19ext (PGIR19brv/V2891 Cyg); the confirmed RCB star Gaia 20bbh (FH Sct); the YSOs Gaia 16aft (PTF10nvg), Gaia 19fct (iPTF 15afq), and Gaia 18dvy discussed in \S\ref{sec:yso}; and a previously unclassified red variable, Gaia 21aor, which we discuss in \S\ref{sec:symbiotic}.

\subsection{Multi-band lightcurves}
For the census sources that follow, we construct multi-band lightcurves, supplementing PGIR {\it J}-band forced photometry with photometry from sources within 2" of the 2MASS counterpart in the ZTF DR23 \citep{2019PASP..131a8003M} release covering the {\it g}-, {\it r}-, and {\it i}-bands and 3" in the {\it NEOWISE} Single Exposure-R catalog \citep{NEOWISE:2020ipac.data.I144N} in {\it W1} (3.6{\um}) and {\it W2} (4.5{\um}). Photometry is obtained from the IRSA service\footnote{\url{https://irsa.ipac.caltech.edu/Missions/wise.html}}$^{,}$\footnote{\url{https://irsa.ipac.caltech.edu/Missions/ztf.html}}. NEOWISE lightcurves use profile-fit photometry and are binned and weighted according to their associated photometric errors. The following quality flags are applied:  [(\texttt{qual\_frame>0}) \& (\texttt{qi\_fact>0}) \& (\texttt{cc\_flags==0})]. Four sources in the census do not have any quality NEOWISE photometry (PGIR IDs V1068325830 (YSO), V1243011042 (RCB), V1243308242 and V1103017185 (O-rich LPVs)). Similarly, nineteen sources in the census do not have quality ZTF photometry after ensuring clean event extractions and photometric solutions at each epoch (i.e., requiring \texttt{catflags==0} and non-negative magnitude errors). 

\subsection{NIR spectroscopy}
For the majority of the targets, we employed TripleSpec \citep{2008SPIE.7014E..0XH}, an $R \approx 2700$ near-infrared ($JHK$) spectrograph on the inch telescope at Palomar Observatory. Four spectra were obtained using the SpeX spectrograph ($R \approx 1500$) on the NASA Infrared Telescope Facility \citep[IRTF,][]{2003PASP..115..362R} at Mauna Kea. All spectra were extracted, flux calibrated with standard star observations, and corrected for telluric absorption features using standard practices with the IDL packages {\sc spextool} \citep{2004PASP..116..362C} and {\sc xtellcor} \citep{2003PASP..115..389V}.

In order to present the spectra collectively per class, we scale by an arbitrary flux density and offset each spectrum by an arbitrary value, which preserves the shape of each spectral continuum and allows for comparison between targets of the same class with broad molecular features \citep[e.g., similar to the presentation by][]{2009ApJS..185..289R}. Spectral regions with poor atmospheric transmission are masked or shaded in gray in the resulting figures. To differentiate spectra corresponding to the same target at different time points, lowercase letters are appended to the PGIR identifier in each spectral figure, in which alphabetical order corresponds to chronological order. We do not provide exhaustive names for previously known targets, but rather 2MASS/IRAS/SIMBAD IDs, or other published names, alongside PGIR identifiers in the text and Appendix Table~\ref{tab:lavselection}.

\section{Demographics of lightcurve-selected NIR LAVs}
\label{sec:demographics}
In this section, we describe the demographics of our near-infrared census of LAVs, presenting a library of spectral and multiband photometric data. 

\subsection{Oxygen- and carbon-rich long period variable giants}
\label{sec:lpv}
\begin{figure*}
\includegraphics[width=\textwidth]{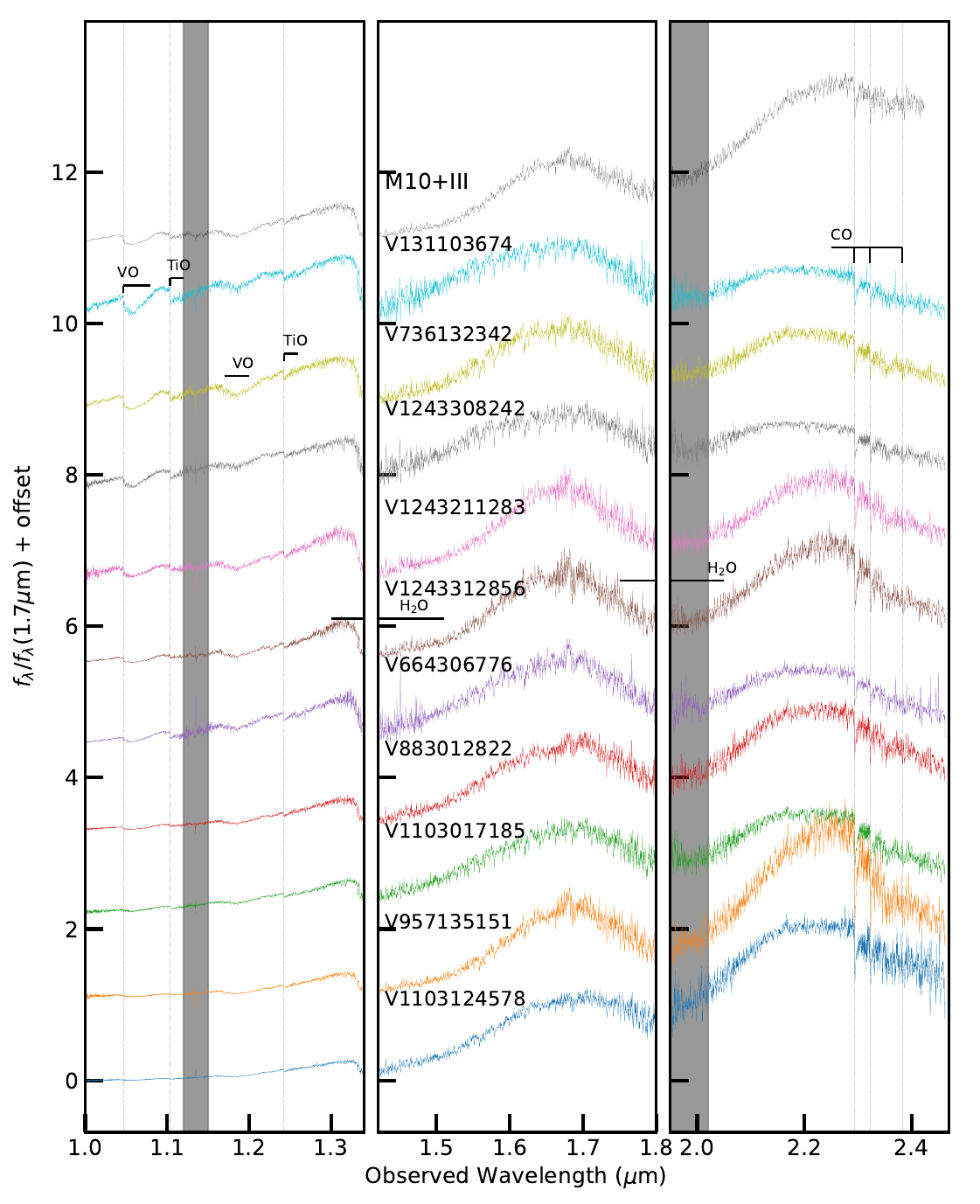}
\caption{Observed O-rich LPV NIR spectra in arbitrary units. Rest frame vacuum wavelengths for relevant atomic and molecular lines are identified by gray vertical lines. A representative observed spectral template from the IRTF library of cool stars \citep{2009ApJS..185..289R} is provided as the topmost spectrum for comparison (IRAS 14086-0703, spectral type M10+ III, in this panel). Strong features near 1.12{\um} and 2{\um} as a result of poor telluric corrections are shaded in gray.}
\label{fig:lpv_spec}
\end{figure*}

\begin{figure*}
\includegraphics[width=\textwidth]{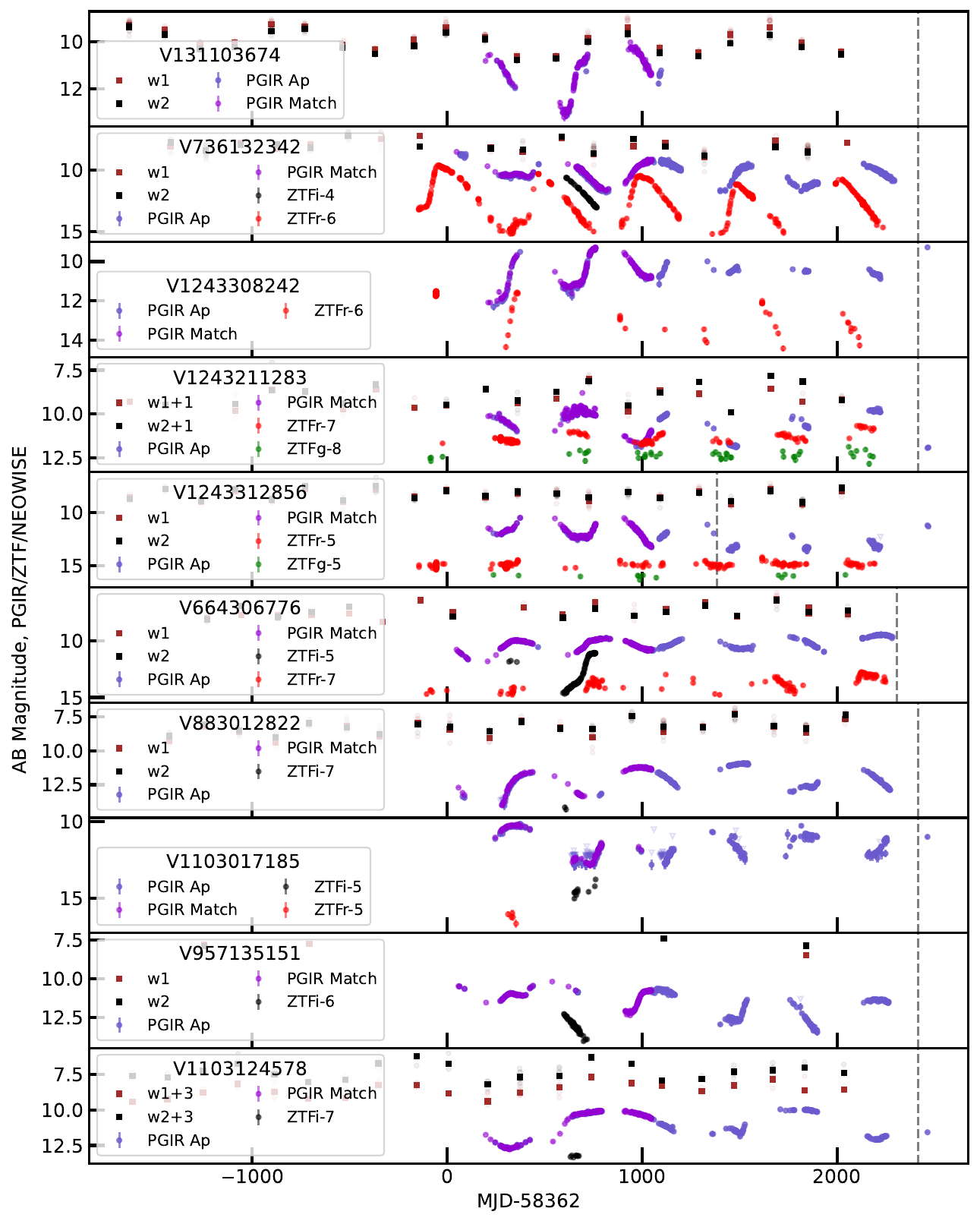}
\caption{O-rich LPV multiband photometry. Archival optical--near-infrared--mid-infrared data spanning roughly twice the length of PGIR's survey duration is provided when available. PGIR $J$-band match file photometry is shown in dark purple with the most recent forced aperture photometry in lighter blue. All magnitudes are AB and some bands are shifted by a constant offset to allow for a visually clearer comparison between filters. Gray, dashed vertical lines designate the time at which the associated spectrum was acquired.}
\label{fig:lpv_fph}
\end{figure*}

\begin{figure*}
\includegraphics[width=\textwidth]{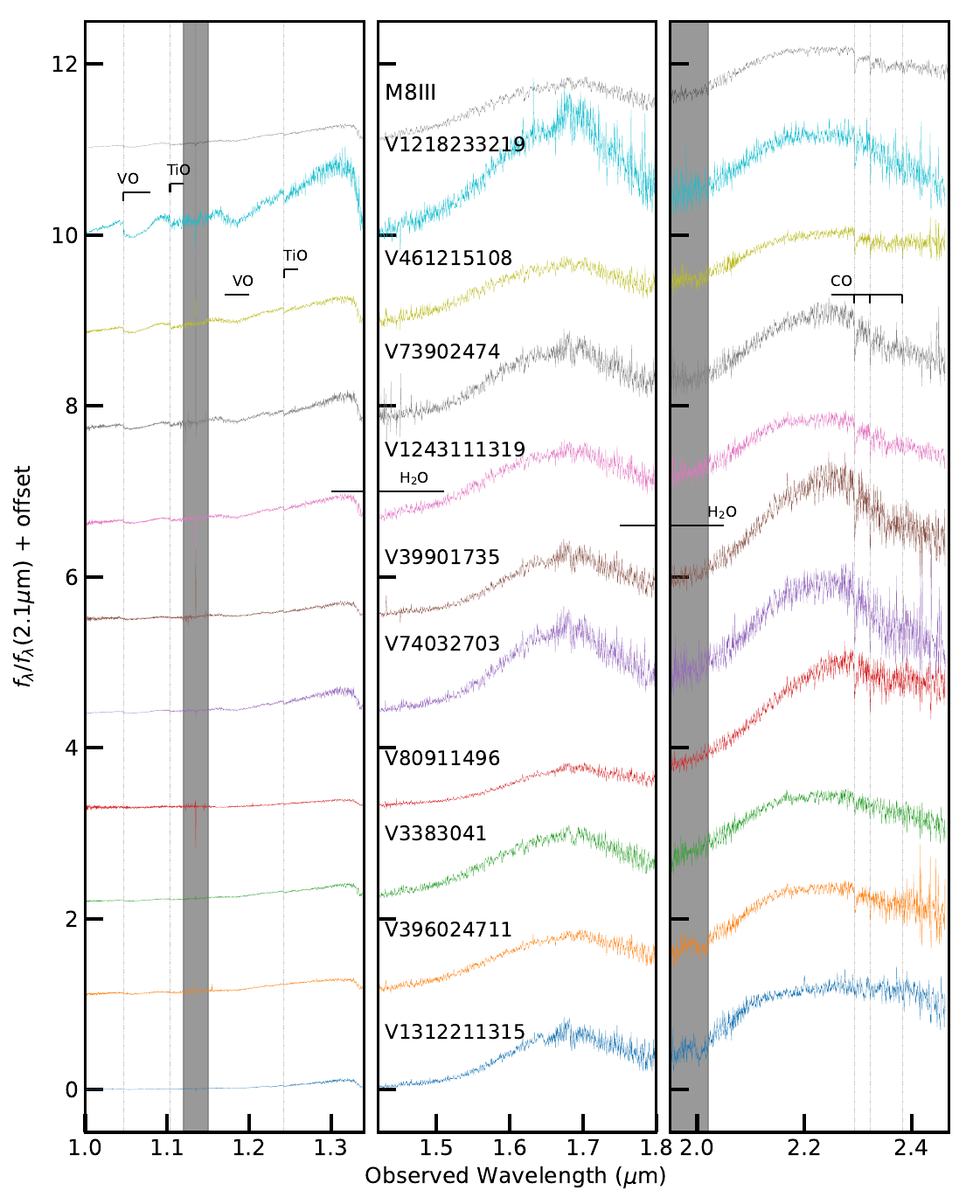}
\caption{O-rich LPV spectra, continued from Figure~\ref{fig:lpv_spec}. All sources in this subsample are cataloged as OH/IR stars with counterpart OH maser emission at 1612 MHz, or SiO and H$_2$O maser emission, except for V39901735, noted as an OH/IR star with suppressed OH maser emission. The reference star at the top is IRAS 01037+1219, spectral type M8 III.}
\label{fig:ohir0_spec}
\end{figure*}

\begin{figure*}
\includegraphics[width=\textwidth]{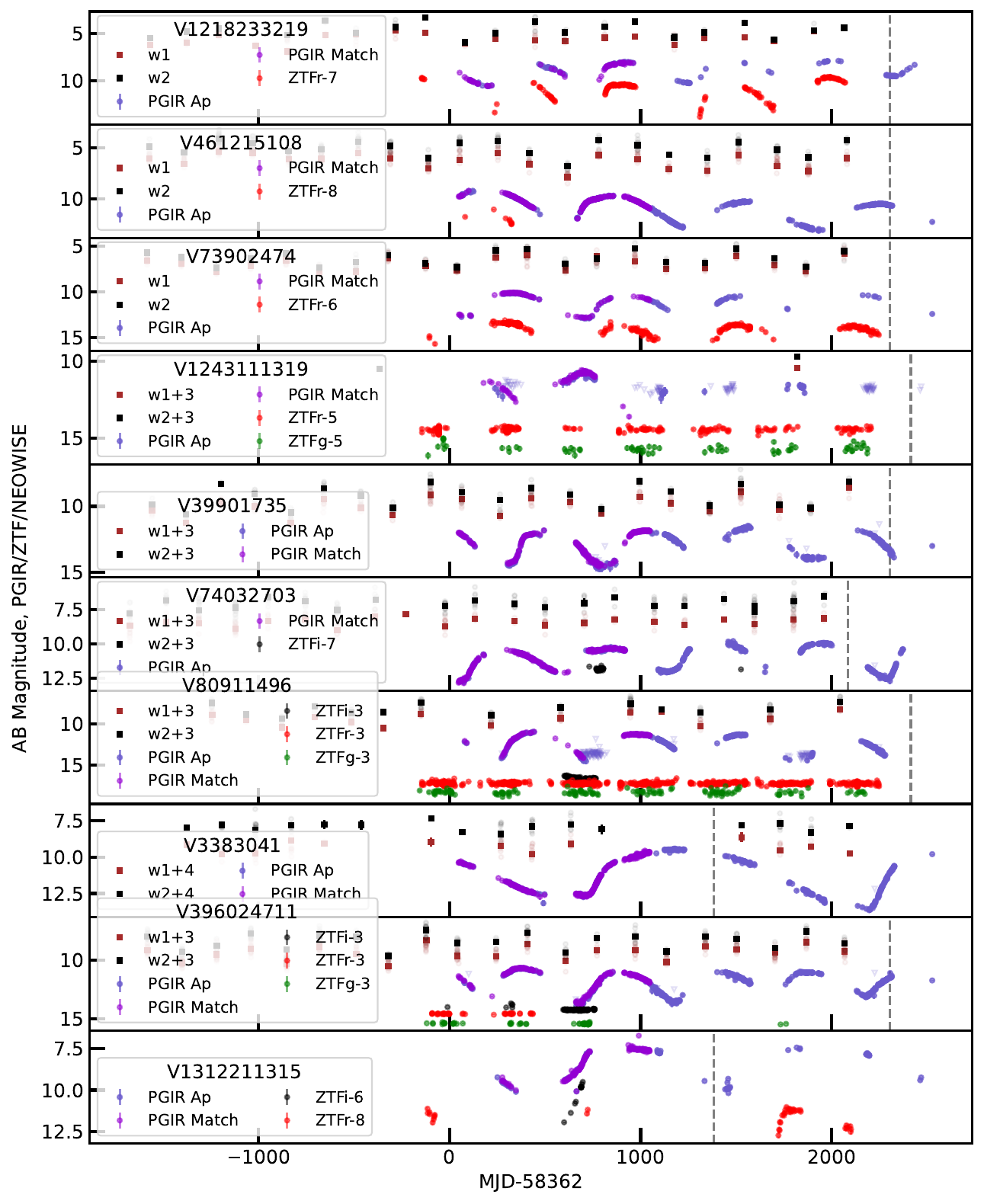}
\caption{O-rich LPV lightcurves corresponding to the sources in Figure~\ref{fig:ohir0_spec}. V3383041 is an extreme OH/IR star with a period exceeding 1000 days.}
\label{fig:ohir0_lc}
\end{figure*}

\begin{figure*}
\includegraphics[width=\textwidth]{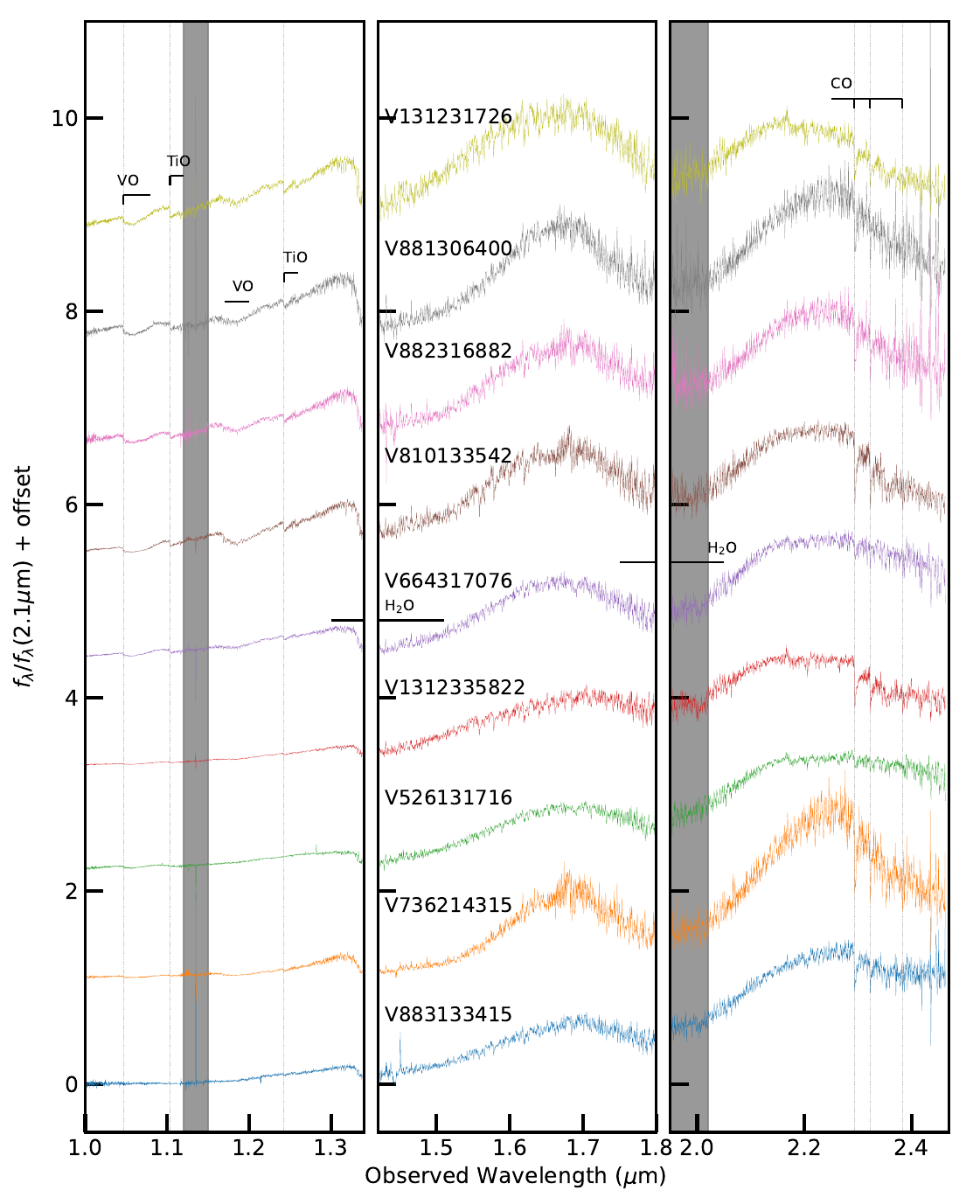}
\caption{O-rich LPV spectra, continued from Figure~\ref{fig:ohir0_spec}. All sources have counterpart maser emission.}
\label{fig:ohir1_spec}
\end{figure*}

\begin{figure*}
\includegraphics[width=\textwidth]{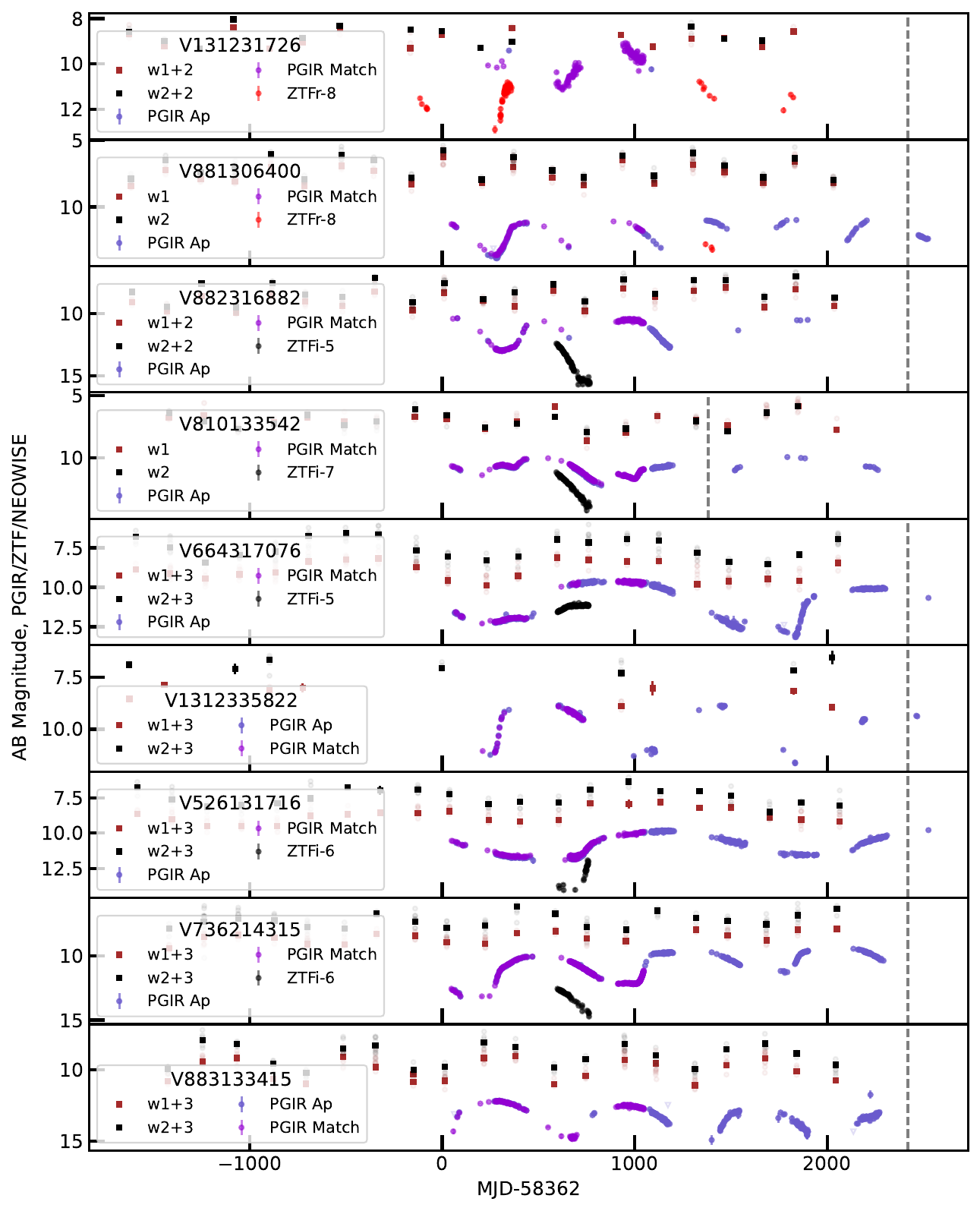}
\caption{O-rich LPV lightcurves corresponding to the sources in Figure~\ref{fig:ohir1_spec}. V664317076 and V526131716 are extreme OH/IR stars, hosting a period $\sim 1500-2000$ days.}
\label{fig:ohir1_lc}
\end{figure*}

\begin{figure*}
\includegraphics[width=\textwidth]{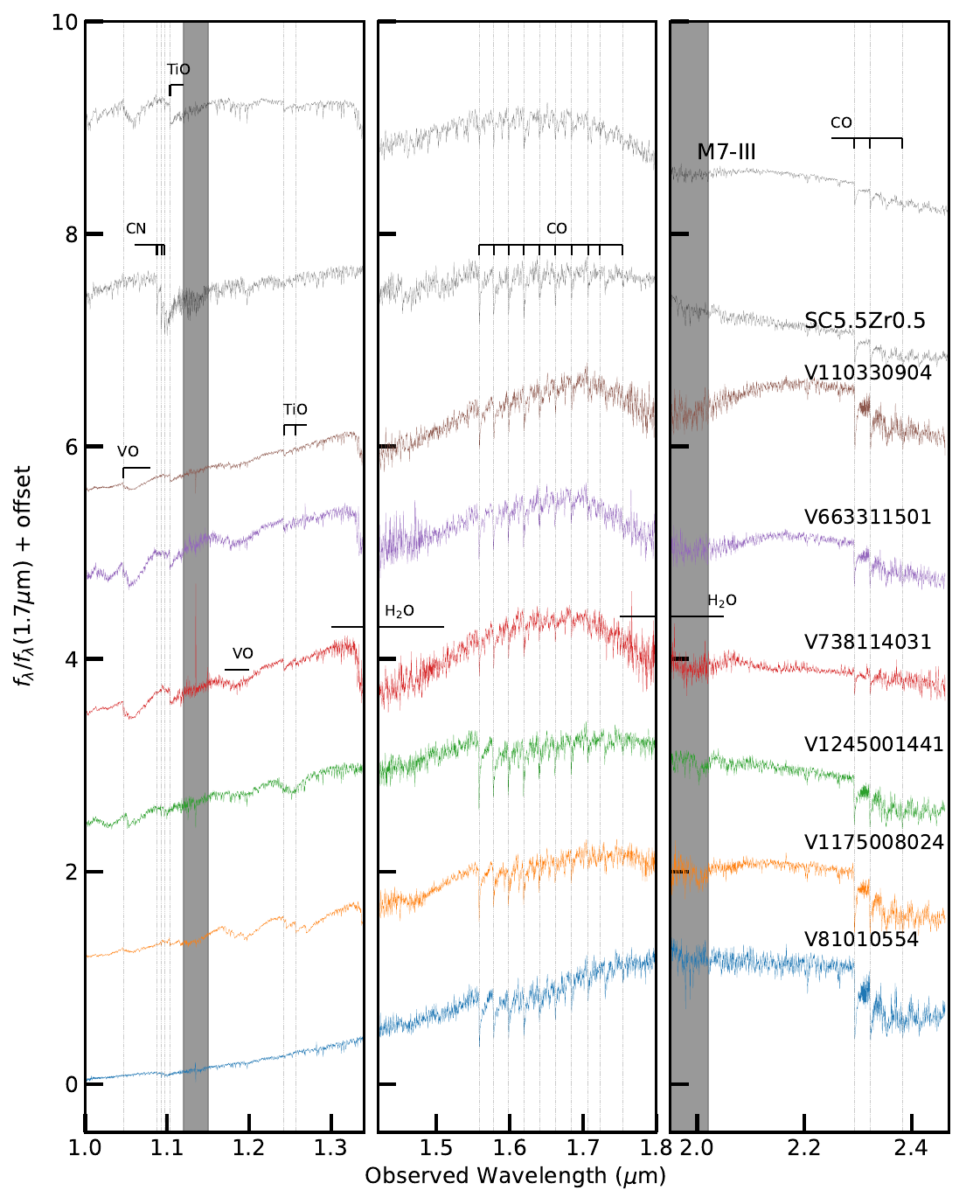}
\caption{LPV spectra, continued from Figure~\ref{fig:lpv_spec}, \ref{fig:ohir0_spec}, and \ref{fig:ohir1_spec}. Spectra in this subsample exhibit enhanced CO absorption in the $H$-band and progressive flattening of the $K$-band continuua from top to bottom, similar to S-type stars (in which C$\sim$O abundances). TiO and VO absorption in the $J$-band are prominent for four of the five sources, and CN absorption begins to appear in V1245001441 and V81010554. Water absorption from $J$ to $H$ and from $H$ to $K$ begins to decrease. Poor atmospheric transmission prevents reliable identification of ZrO at $\lambda<1${\um}, additional markers for S-type stars. This subsample of stars likely constitute bridges between the O- and C-rich AGB sequence. Two reference stars are shown at the top: HD108849 (spectral type M7 III) and HD44544 (SC5.5Zr0.5). }
\label{fig:lpvflatk_spec}
\end{figure*}

\begin{figure*}
\includegraphics[width=\textwidth]{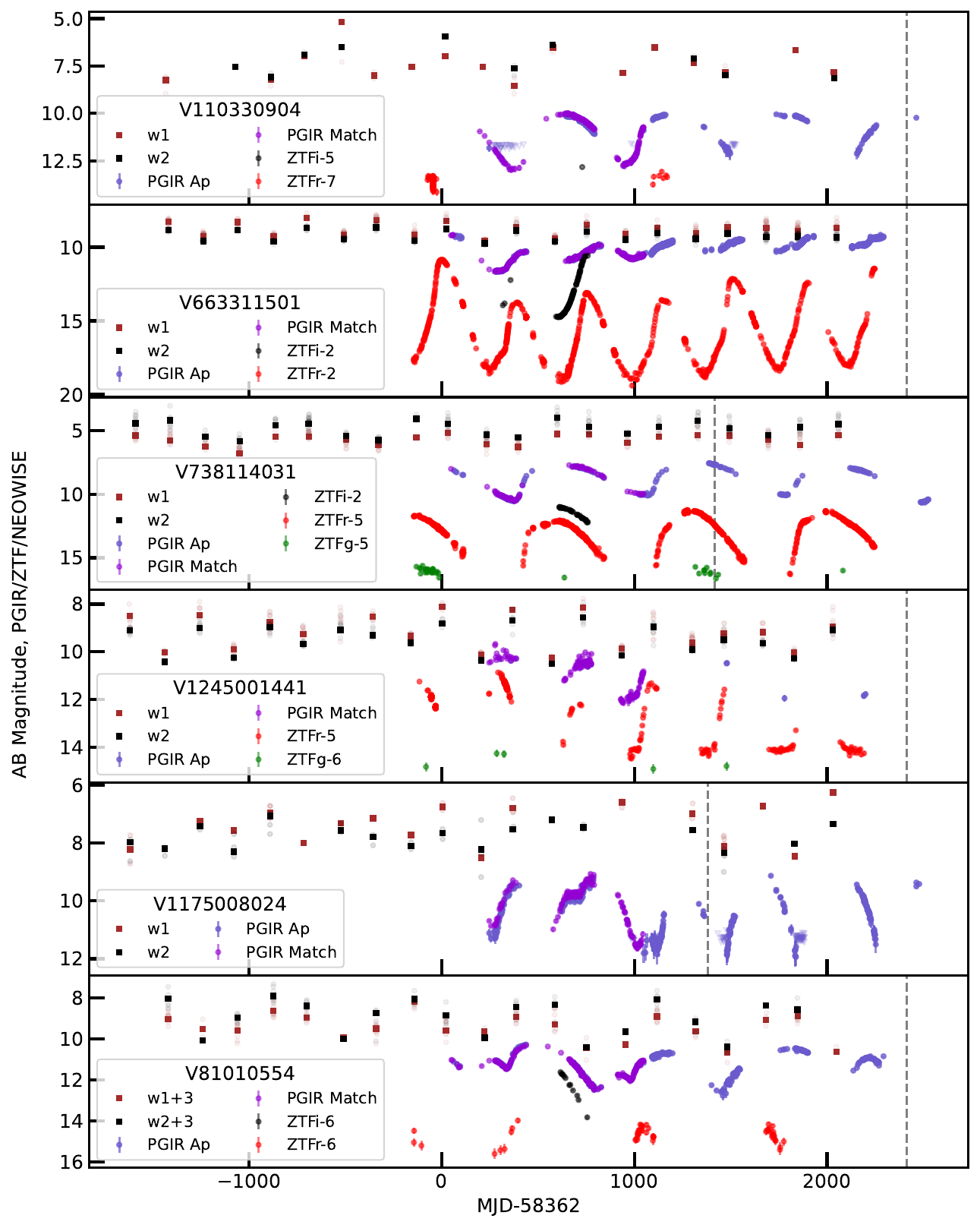}
\caption{LPV lightcurves corresponding to the sources in Figure~\ref{fig:lpvflatk_spec}.}
\label{fig:lpvflatk_lc}
\end{figure*}

\begin{figure*}
\includegraphics[width=\textwidth]{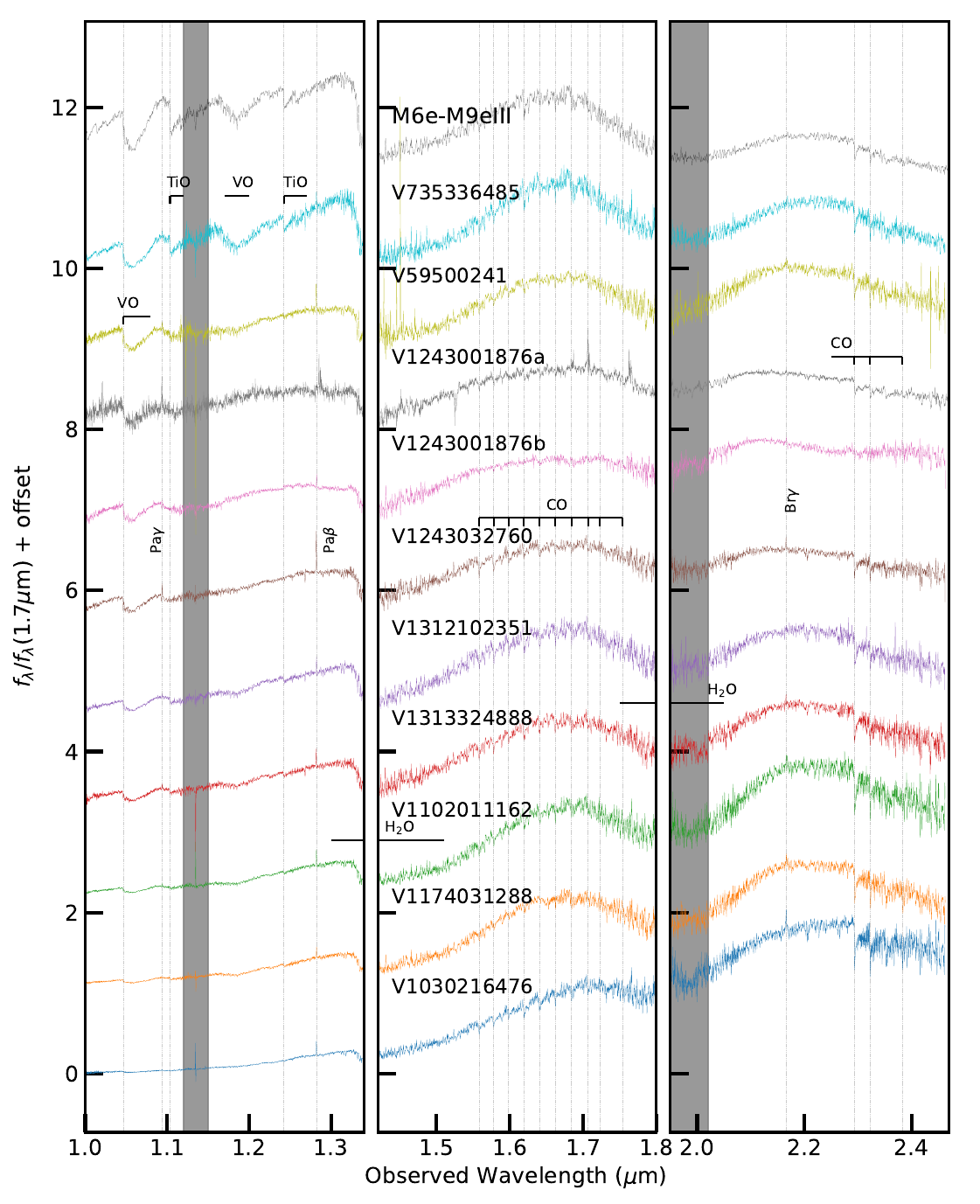}
\caption{O-rich LPV spectra, continued from Figure~\ref{fig:lpv_spec}, \ref{fig:ohir0_spec}, \ref{fig:ohir1_spec}, and \ref{fig:lpvflatk_spec}. Spectra in this subsample exhibit prominent emission lines common for Mira variables such as Pa-$\beta$, along with typical molecular absorption markers such as VO and CO. Two spectra were obtained of V1243001876 more than 700 days apart exhibiting moderate spectral evolution, including weakening CO absorption in $K$-band. The top reference star is HD69243 (M6e-M9eIII).}
\label{fig:lpvline0_spec}
\end{figure*}

\begin{figure*}
\includegraphics[width=\textwidth]{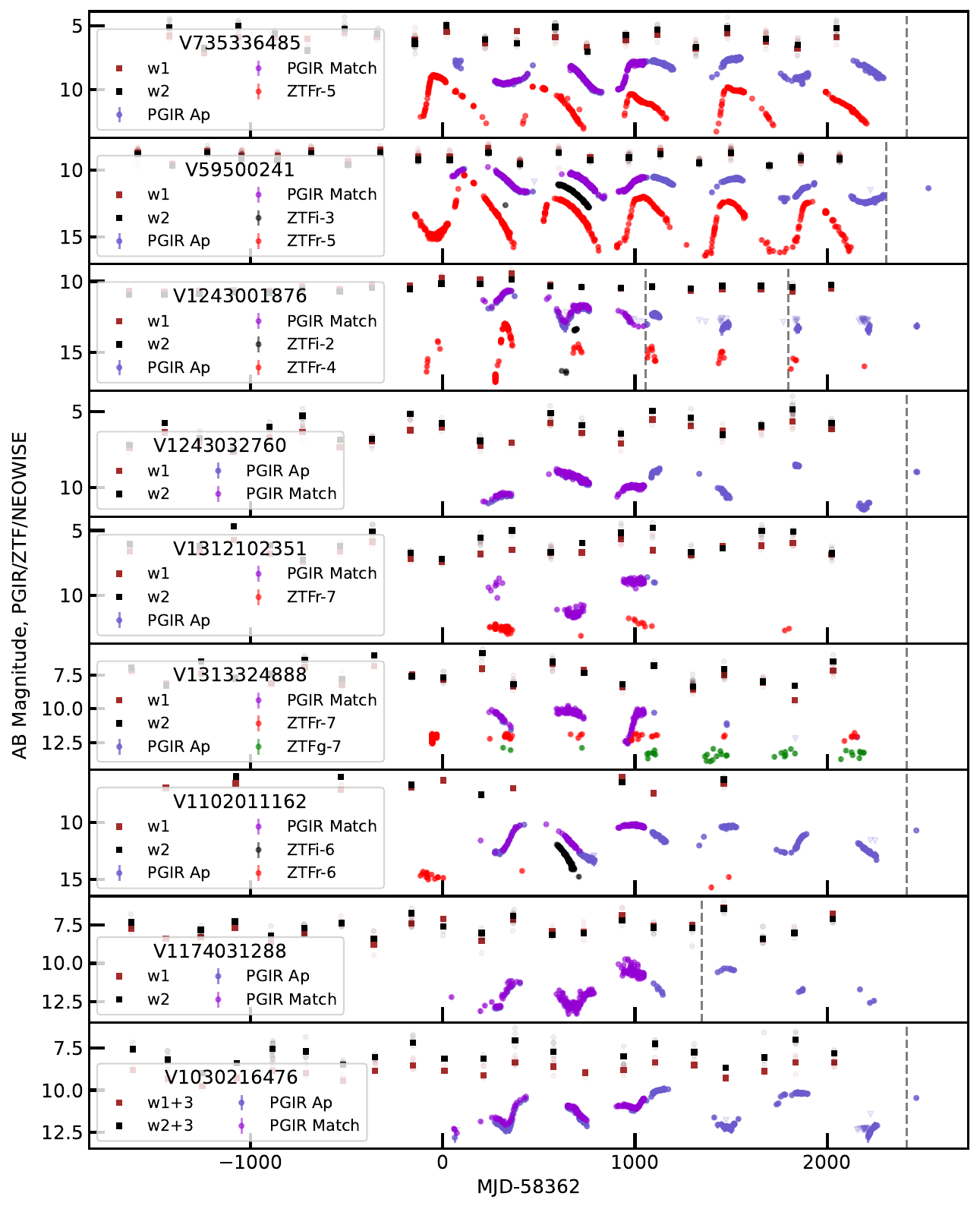}
\caption{O-rich LPV lightcurves corresponding to the sources in Figure~\ref{fig:lpvline0_spec}.}
\label{fig:lpvline0_lc}
\end{figure*}

\begin{figure*}
\includegraphics[width=\textwidth]{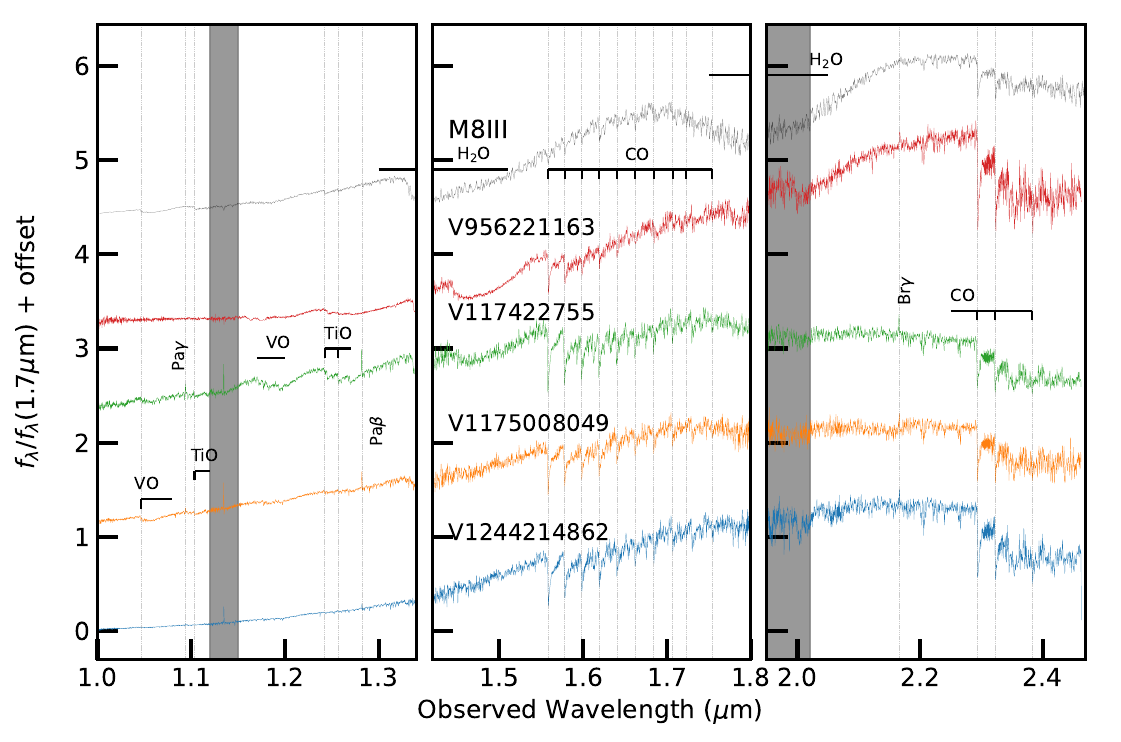}
\includegraphics[width=\textwidth]{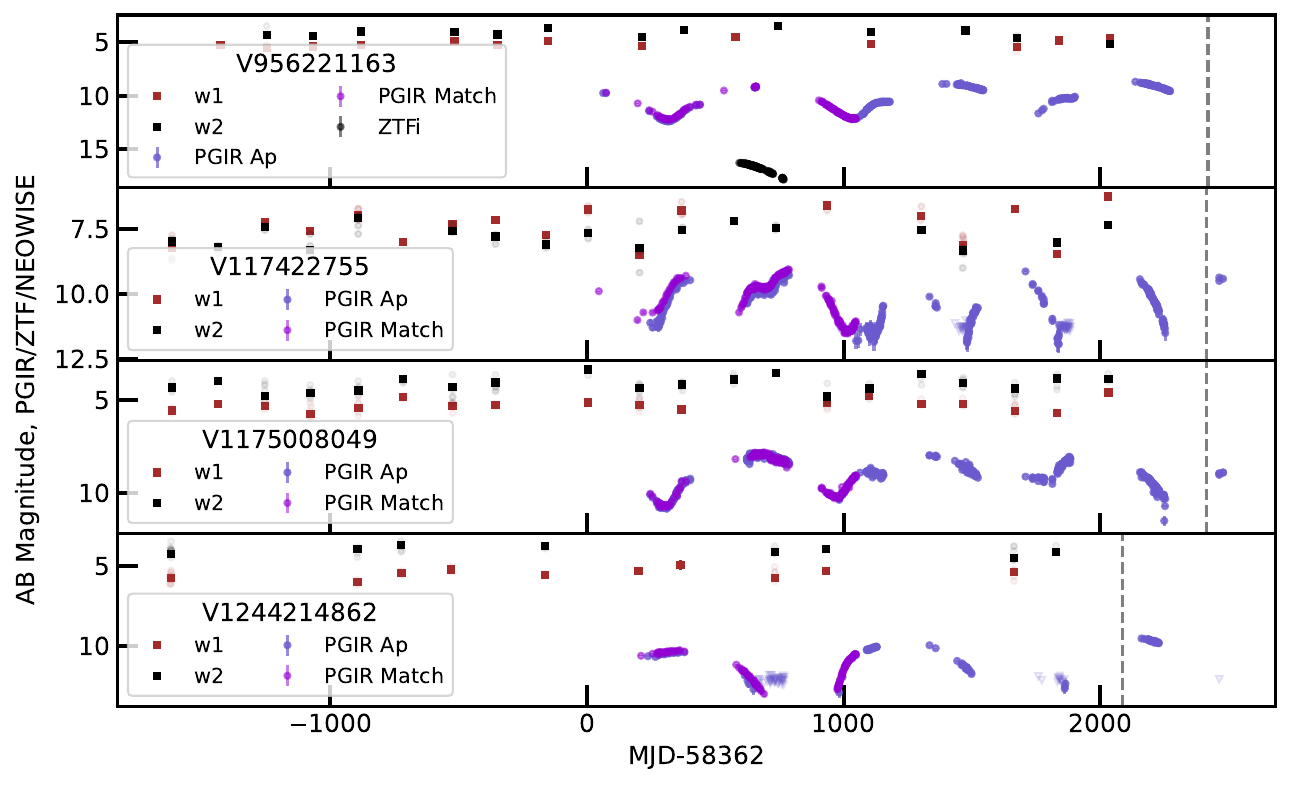}
\caption{Additional O-rich LPV spectra and lightcurves. The spectra exhibit enhanced CO absorption in the $H$-band similar to the spectra in Figure~\ref{fig:lpvflatk_spec} and HI emission lines as in Figure~\ref{fig:lpvline0_spec}. V956221163, a confirmed SiO maser, and V117422755, a known Mira variable, both exhibit a broad absorption feature near 1.5{\um} that contrasts with typical water absorption.}
\label{fig:lpvline1}
\end{figure*}

To limit the sample pursued for spectroscopic identification in this work, strong LPV candidates were largely filtered out by crossmatching sources with the automated PGIR Long Period Variable Catalog \citep{aswin:2024PASP..136h4203S}. These candidates will be spectrally classified in forthcoming works. In general, LPVs are the predominant subpopulation in the PGIR LAV census. Sources satisfying the selection criteria in \S\ref{sec:selection} that were not identified in the catalog account for 65\% of the resulting LAV sample. Many of these evolved pulsators are often not well identified by simple Lomb-Scargle periodogram analyses, as they either show non-sinusoidal variations or have multiple harmonics creating erratic variability in their lightcurves. Moreover, in many existing studies that rely on machine learning algorithms and automated classification, a variety of LAVs are often misclassified as LPVs \citep[e.g.,][]{Chen:2020ApJS..249...18C,2023A&A...674A..14R}

The LPVs broadly consist of a dichotomy between oxygen-rich (O-rich, mostly M-spectral type, Figure~\ref{fig:lpv_spec}---\ref{fig:lpvline1}) and carbon-rich (C-rich, C-spectral type, Figure~\ref{fig:hbandflatk0_spec}---\ref{fig:cstar_lc}) giants. LPVs are a phenomenological class of red giants, supergiants, and thermal pulsating AGB (TP-AGB) stars such as Mira variables which exhibit regular or semiregular photometric variations driven by pulsational modes. For low- and intermediate-mass stars ($0.8 M_\odot \lesssim M \lesssim 8 M_\odot$), post-main-sequence evolution is directed along the red giant and asymptotic giant branches (RGB and AGB, respectively). Following hydrogen shell burning via the CNO cycle and the resulting central He-burning after the He flash, AGBs host an inert C-O core along with H and He shell burning. These AGBs begin as spectral M-types (O-rich, $\rm{C}<\rm{O}$), but can transition into C-types as carbon is dredged up and begins dominating the atmosphere \citep[e.g., see][]{2018A&ARv..26....1H}. The chemistry of the AGB stellar envelopes subsequently dictate wind-driven dynamical evolution, heavy mass-loss, and the enrichment of the interstellar medium with metal-rich elements \citep[e.g.,][]{Wood:1996MNRAS.282..958W,Willson:2000ARA&A..38..573W, Tielens:2005pcim.book.....T}.

Temperatures for the spectral types corresponding to these stellar objects are on the order of 3000\,K, corresponding to blackbody emission peaking at $\sim 1${\um}. NIR surveys are well suited for regular photometric monitoring and spectroscopic classification \citep[e.g.,][which uses IRTF SpeX observations of several Galactic late-type giants and red supergiants]{Messineo21:2021AJ....162..187M}. Among the several spectral features of the O-rich class, these variable M stars are known to harbor characteristic water absorption in the $H$- and $K$-bands ($1.5-2.2${\um}) \citep{2019IAUS..343..309L}. Deep molecular features in the spectra arise due to the extended atmospheres of the pulsating giants driving cool molecule and dust formation \citep{2000A&AS..146..217L}. Near-infrared spectra are marked with the characteristic and prominent metal oxide absorption features such as the TiO$\gamma$ system around 1.24{\um} and the VO$\gamma$ system around 1.05{\um} \citep{1967ApJ...147..117W}. Hydrogen emission lines can also arise as a result of shocks propagating in the dynamic atmosphere, as seen most clearly by the Pa-$\beta$ profiles in  Figures~\ref{fig:lpvline0_spec} and \ref{fig:lpvline1}. As a dominant subpopulation in the PGIR LAV sample, the abundance of O-rich LPVs reflects the oxygen-rich chemistry of the majority of giants in the Galactic Bulge \citep{2013AcA....63...21S}. In the spatial distribution of LAVs in Figure~\ref{fig:census}, O-rich LPVs are largely clustered near the bulge with Galactic longitudes ranging from $0-50^\circ$. 

C-rich LPVs form 27\% of the census and are preferentially distributed in the outer extents of the Galactic disk in the right panel of Figure~\ref{fig:census}. We note the increasing fraction of C-rich to O-rich LPVs as a function of disk radius \citep{2011A&A...534A..79I, 2023MNRAS.521.2745S}. This is consistent with the observed negative metallicity gradient from the inner to outer regions of the disk \citep[e.g.,][]{Hayden:2014AJ....147..116H, Magrini:2023A&A...669A.119M,Willett:2023MNRAS.526.2141W} and/or with an inside-out disk formation scenario \citep{Martig:2016ApJ...831..139M,Frankel:2019ApJ...884...99F}. In the Bulge, the LAV distribution is consistent with the predominance of an old, metal-rich stellar population \citep[e.g.,][]{Zoccali:2003A&A...399..931Z,Bensby:2017A&A...605A..89B}.

Spectrally, due to the consumption of oxygen in the production of CO, C-rich stars are known to harbor strong molecular absorption features such as CN and C$_2$ in contrast to the metal oxide absorption features of the O-rich stars. The spectral markers are shown in Figure~\ref{fig:hbandflatk0_spec}---\ref{fig:cstar_spec}. The spectral and photometric library presented here of 36 C-rich LAVs further increases the size of existing carbon-star libraries \citep[e.g.,][]{2016A&A...589A..36G}, supplementing this class of objects for continued theoretical modeling. 

A set of LPVs exhibit enhanced CO absorption in the $H$-band common for S-type stars in which carbon and oxygen abundances become equivalent along with M-type $J$-band metal oxide markers (Figure~\ref{fig:lpvflatk_spec}). We interpret these stars to be representative of evolutionary links in the sequence between the O- and C-rich giants. We also highlight {\bf V59500241 (IRAS 19518+2033)}, 2.78" from source coordinates of a candidate YSO \citep[MASTER OT J195405.17+204126.6,][]{Gress:2016ATel.9204....1G}. Outside of poor SNR due to low atmospheric transmission blueward of 1.5{\um}, the spectrum and lightcurves in Figure~\ref{fig:lpvline0_spec} and \ref{fig:lpvline0_lc} strongly indicate the source is an O-rich LPV with HI emission features. 

We consider in further detail two flavors of long-period-variables in the census: (i) oxygen-rich variables previously identified as OH/IR stars or sources with counterpart maser emission and (ii) carbon-rich variables with the 1.53{\um} absorption feature.

\subsubsection{OH/IR stars and oxygen-rich masers}
\label{sec:maser}
Oxygen-rich giants expel a large amount of dust in the latter phases of their evolution \citep[mass-loss rates between $10^{-8}-10^{-4} \textrm{ M}_\odot \textrm{ yr}^{-1}$,][]{deBeck:2010A&A...523A..18D}, 
which can fully enshroud the stars with optically thick circumstellar shells, thereby obscuring them at optical wavelengths. Molecular OH maser emission at 1612\,MHz can commonly arise due to pumping in the expanding shell. Other maser emission lines can include SiO and H$_2$O. Because of the optically thick circumstellar dust environment, these stars are often only detectable in the infrared and radio \citep[e.g.,][]{ Wilson:1972A&A....17..385W,Herman:1985A&AS...59..523H, Baud:1981A&A....95..156B}. While OH/IR stars are known to be a natural extension of the class of Mira variables in their spectral properties and variability, a large fraction of O-rich AGB stars do not exhibit a 1612\,MHz maser signal \citep{1983A&A...124..123E, Lewis:1992ApJ...396..251L, 2003A&A...412..481T, 2006A&A...455..645V}. Thus, OH/IR masers can be treated as a subset of the sequence of O-rich LPVs with extreme mass-loss \citep{Baud:1983A&A...127...73B}. 

Nineteen LPVs in the census, shown in Figure~\ref{fig:ohir0_spec}--\ref{fig:ohir1_lc}, are infrared counterparts to known OH/IR sources -- most cataloged in previous studies of OH/IR sources based on the detection of the OH maser line at 1612 MHz or SiO and H$_2$O \citep{1990ApJS...74..911B,1994A&AS..103..541B, 2004PASJ...56..765D,2013AJ....145...22K,2017ApJS..232...13C, Messineo:2018A&A...619A..35M}. {\bf V39901735 (IRAS 20542+3631)} is a confirmed OH/IR star with suppressed OH maser emission, driven by the presence of a companion star disrupting the velocity coherence in the expanding shell \citep{Lewis:1987AJ.....94.1025L}.

All targets in the subsample host prominent water absorption in the {\it H}-band and minimal CO absorption in close comparison with other O-rich LPVs. VO and TiO absorption is visible and the spectra are visually similar to the IRTF M spectral type templates. PGIR and {\it NEOWISE} photometry demonstrate semiregular variability that compares closely with LPV lightcurves. Outside of limited detection in the $r$- and $i$-bands for three of the sources, the variables are mainly detected in IR photometry. The subsample consists primarily of fundamental mode pulsators, hosting periods $\gtrsim500-700$ days, consistent with the peak in the period distribution of OH/IR stars \citep{jimenez:2021MNRAS.505.6051J}. {\bf V664317076 (IRAS 19291+2012)} \citep{Lewis:1994ApJS...93..549L} and {\bf V3383041 (OH 83.4--0.9)} \citep{Bowersa:1978A&AS...31..127B,Bowersb:1978A&A....64..307B} stand as notable exceptions with periods $\sim 1000-1500$ days, in the regime of ``extreme" OH/IR stars which live in the tail of the OH/IR star period distribution 
\citep{Engels:2024IAUS..376..328E}. These contrast with the other Mira variables hosting $\sim 100$ day periods. 

SiO masers include {\bf V1312335822 (RAFGL 6903S/IRAS 17515+2407)}, originally misclassified as a carbon star by a low-resolution spectrum with IRAS \citep{Chen:2003ChJAA...3..551C}. The medium-resolution NIR spectrum in Figure~\ref{fig:ohir1_spec} exhibits definitive O-rich markers. In addition, {\bf V956221163 (IRAS 18441-0325/BAaDE ad3a-13063)}, a SiO maser \citep{Dike:2021AJ....161..111D}, and {\bf V117422755 (V3923 Sgr)} a Mira variable \citep{Maffei:1975IBVS..985....1M} with no confirmed history of counterpart maser detection, share a broad absorption feature near 1.5{\um}, bluer than an absorption feature near 1.53{\um} in C-rich LPVs, but contrasting with the typical water absorption in the $H$-band of other O-rich LPVs (Figure~\ref{fig:lpvline1}).

\subsubsection{1.53{\um} absorbers}
\begin{figure*}
\includegraphics[width=\textwidth]{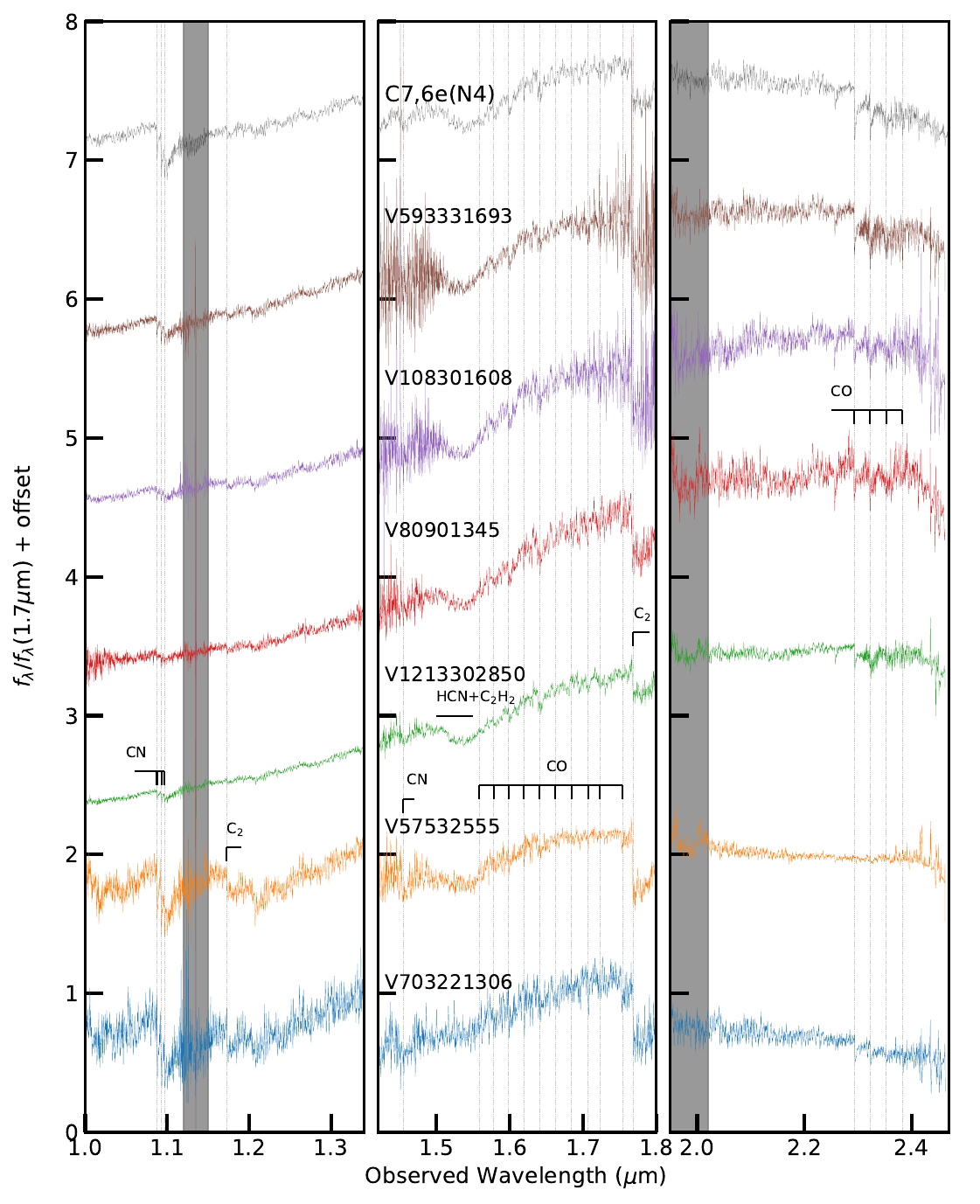}
\caption{C-rich LPV spectra. Oxygen is primarily bound up in the formation of CO, leaving an abundance of carbon-rich molecular absorption features in place of the metal oxide markers common to the O-rich LPVs. Most sources exhibit a broad HCN+C$_{2}$H$_2$ absorption feature near 1.5{\um}. Other strong absorption are attributable to CN and C$_2$. The reference spectrum at the top corresponds to the carbon star R Lep (HD 31996, spectral type C7,6e (N4)).}
\label{fig:hbandflatk0_spec}
\end{figure*}

\begin{figure*}
\includegraphics[width=\textwidth]{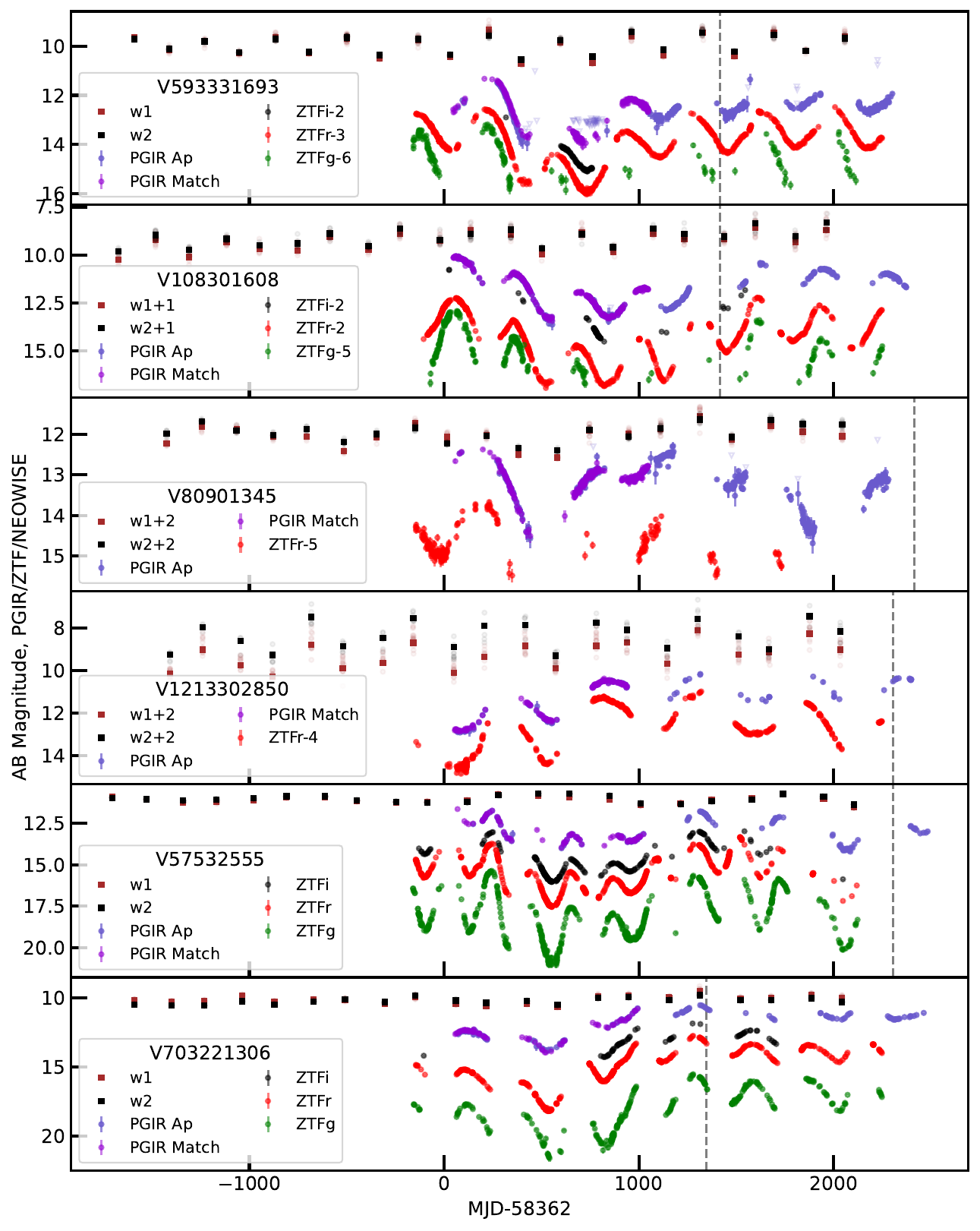}
\caption{C-rich LPV lightcurves corresponding to the sources in Figure~\ref{fig:hbandflatk0_spec}.}
\label{fig:hbandflatk0_lc}
\end{figure*}

\begin{figure*}
\includegraphics[width=\textwidth]{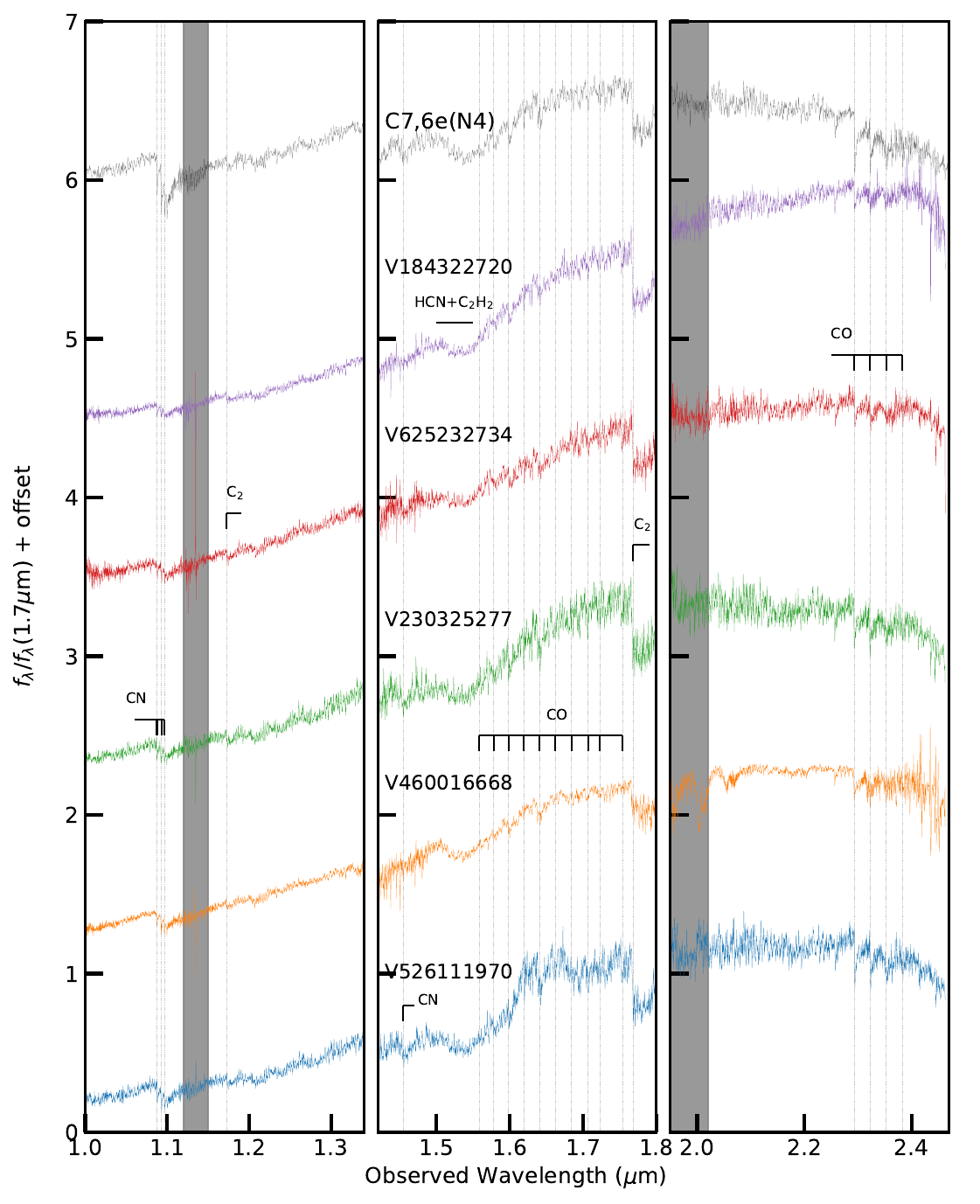}
\caption{C-rich LPVs continued from Figure~\ref{fig:hbandflatk0_spec}.}
\label{fig:hbandflatk1_spec}
\end{figure*}

\begin{figure*}
\includegraphics[width=\textwidth]{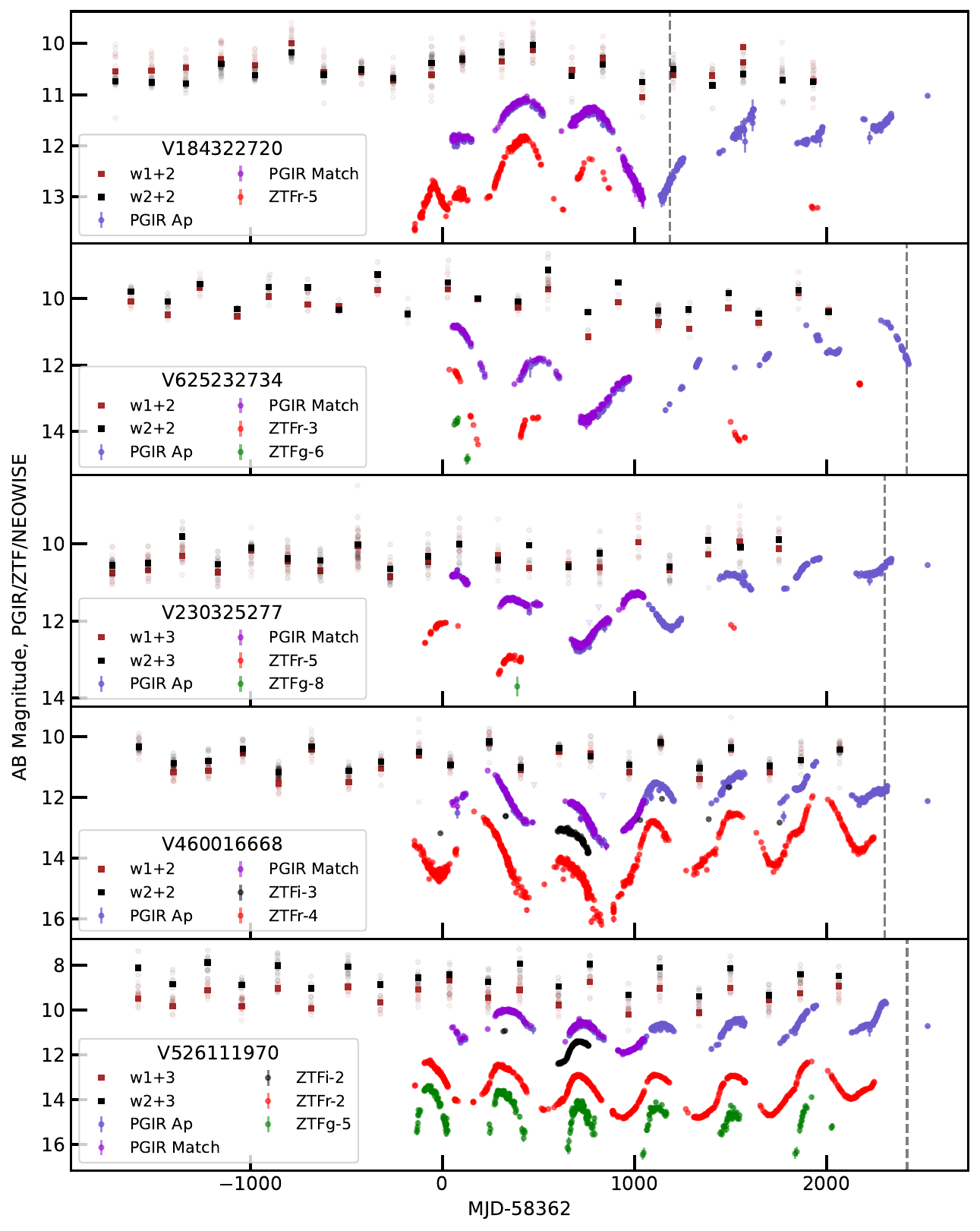}
\caption{C-rich LPV lightcurves corresponding to the sources in Figure~\ref{fig:hbandflatk1_spec}.}
\label{fig:hbandflatk1_lc}
\end{figure*}

\begin{figure*}
\includegraphics[width=\textwidth]{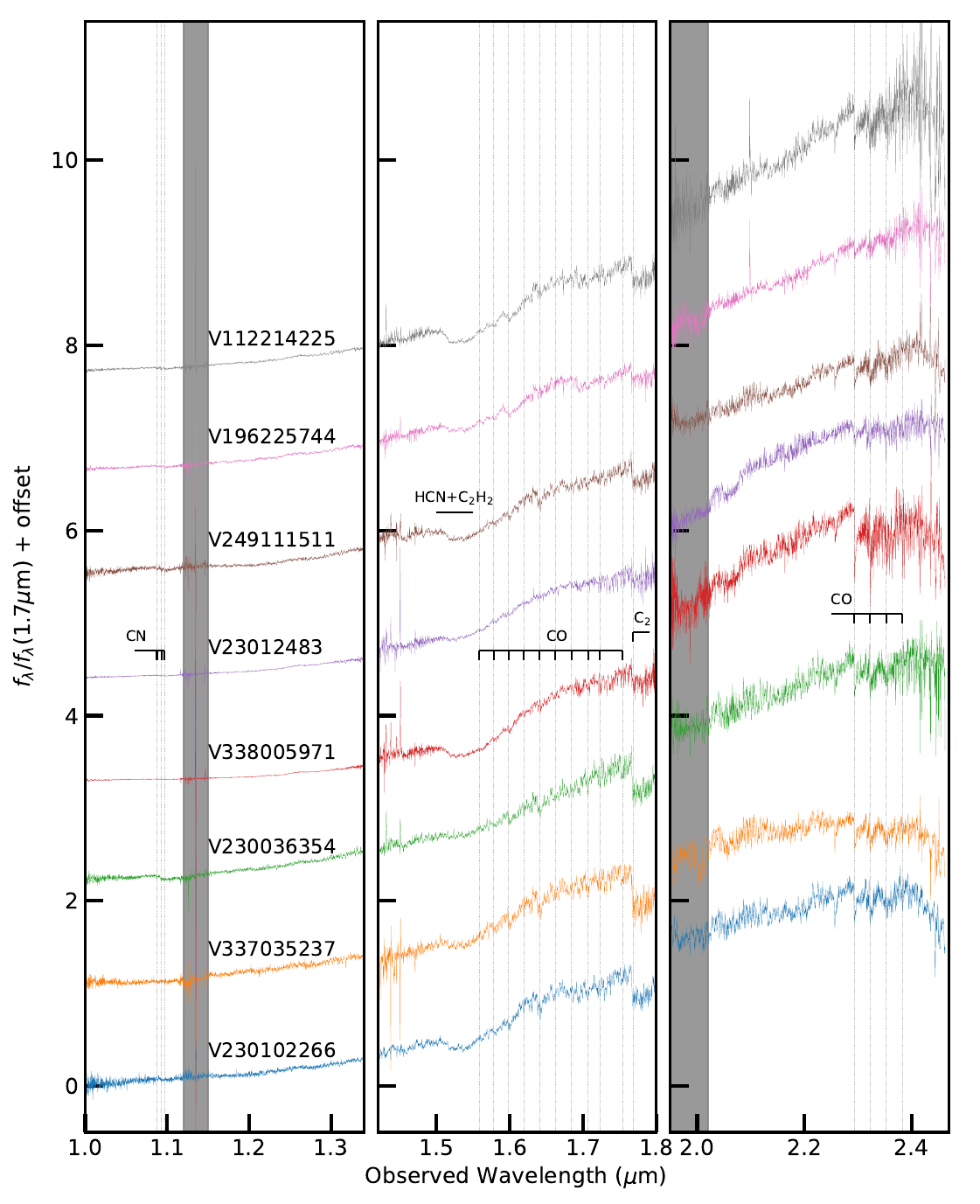}
\caption{A spectral subset of the C-rich LPVs are red and dusty variables with steeply rising $K$-band continua.}
\label{fig:monsterhband0_spec}
\end{figure*}

\begin{figure*}
\includegraphics[width=\textwidth]{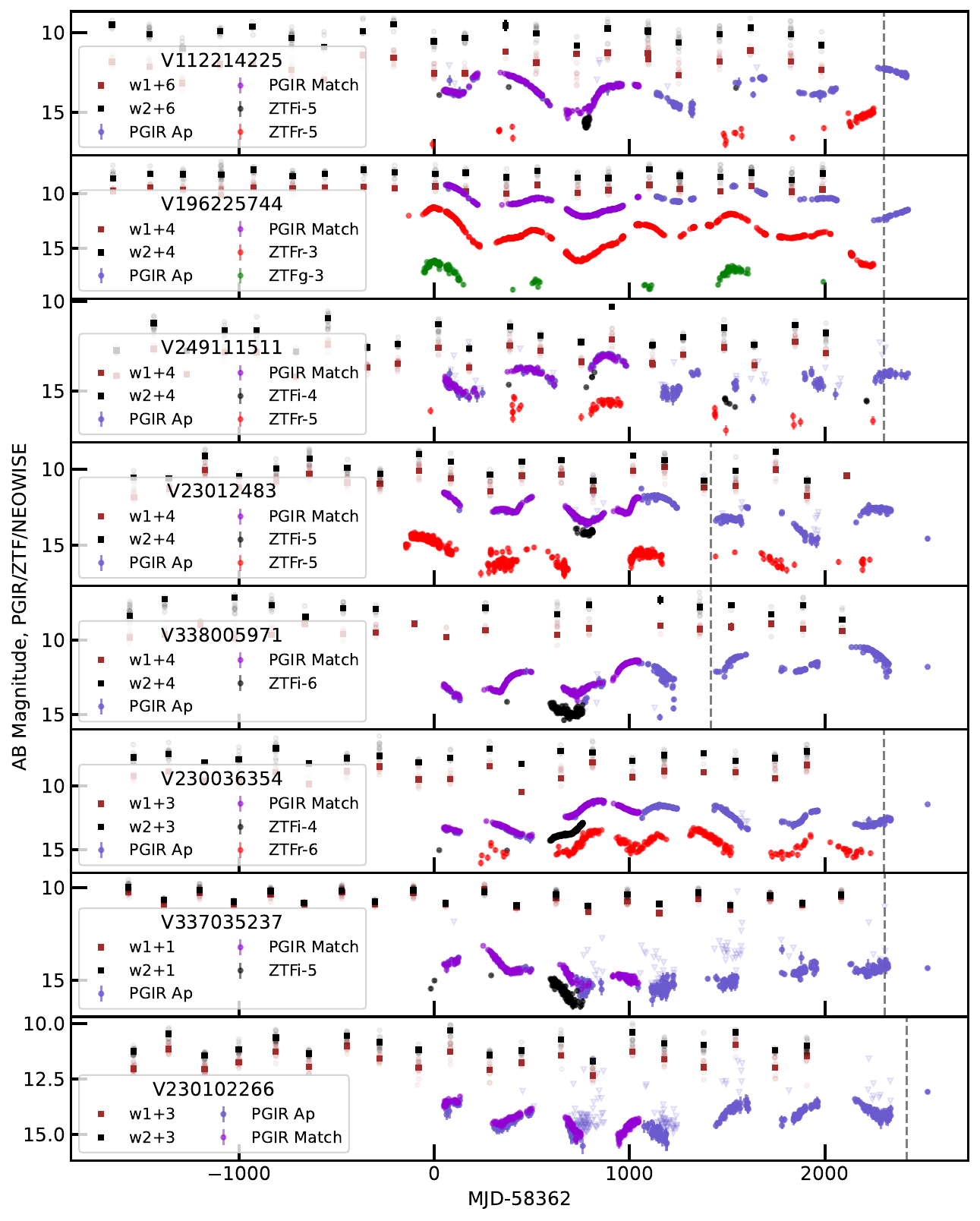}
\caption{C-rich LPV lightcurves corresponding to the sources in Figure~\ref{fig:monsterhband0_spec}. The red C-rich stellar lightcurves exhibit broad-band semiregular variability analogous to Mira variables.}
\label{fig:monsterhband0_lc}
\end{figure*}

\begin{figure*}
\includegraphics[width=\textwidth]{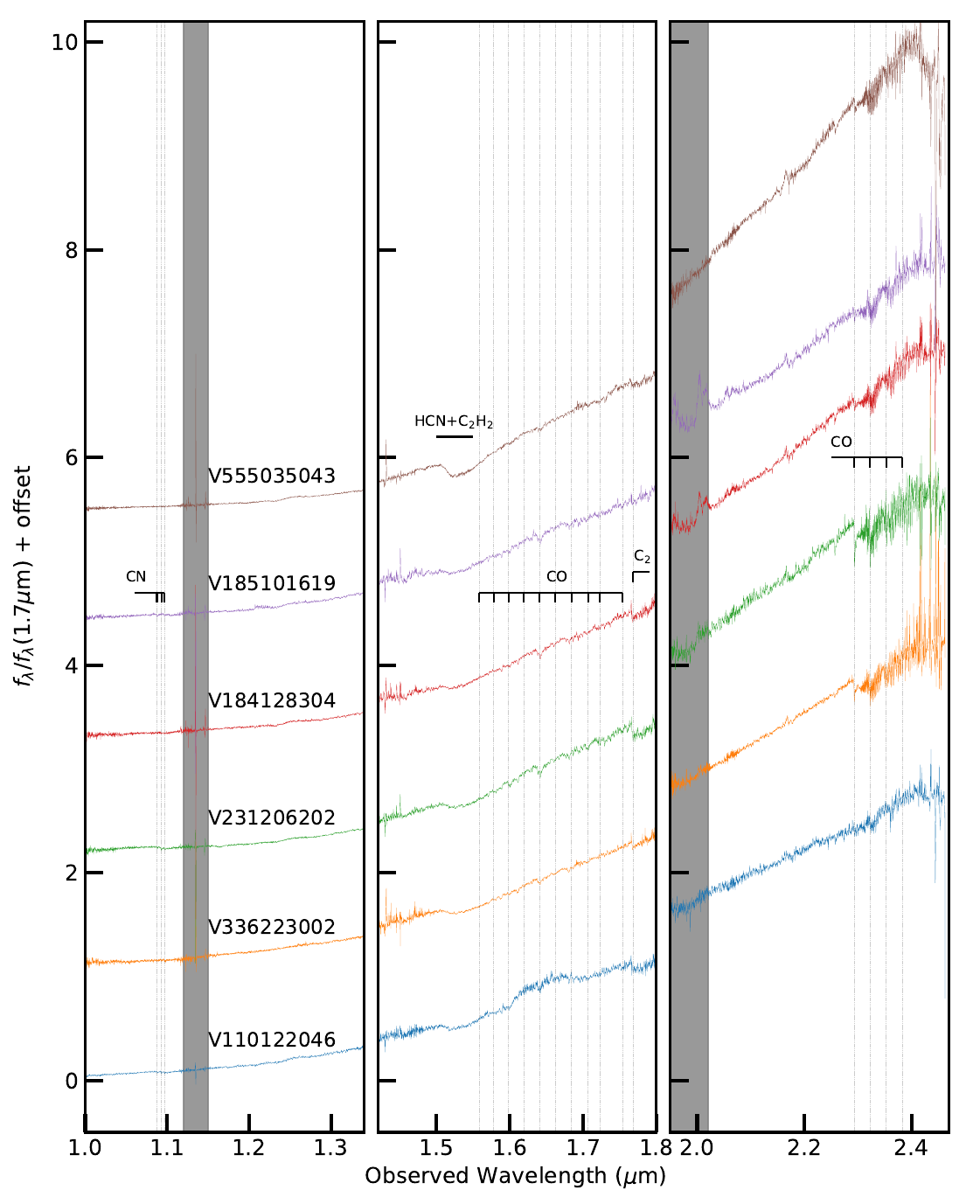}
\caption{C-rich LPVs with steeply rising $K$-band continua, continued from Figure~\ref{fig:monsterhband0_spec}.}
\label{fig:monsterhband1_spec}
\end{figure*}

\begin{figure*}
\includegraphics[width=\textwidth]{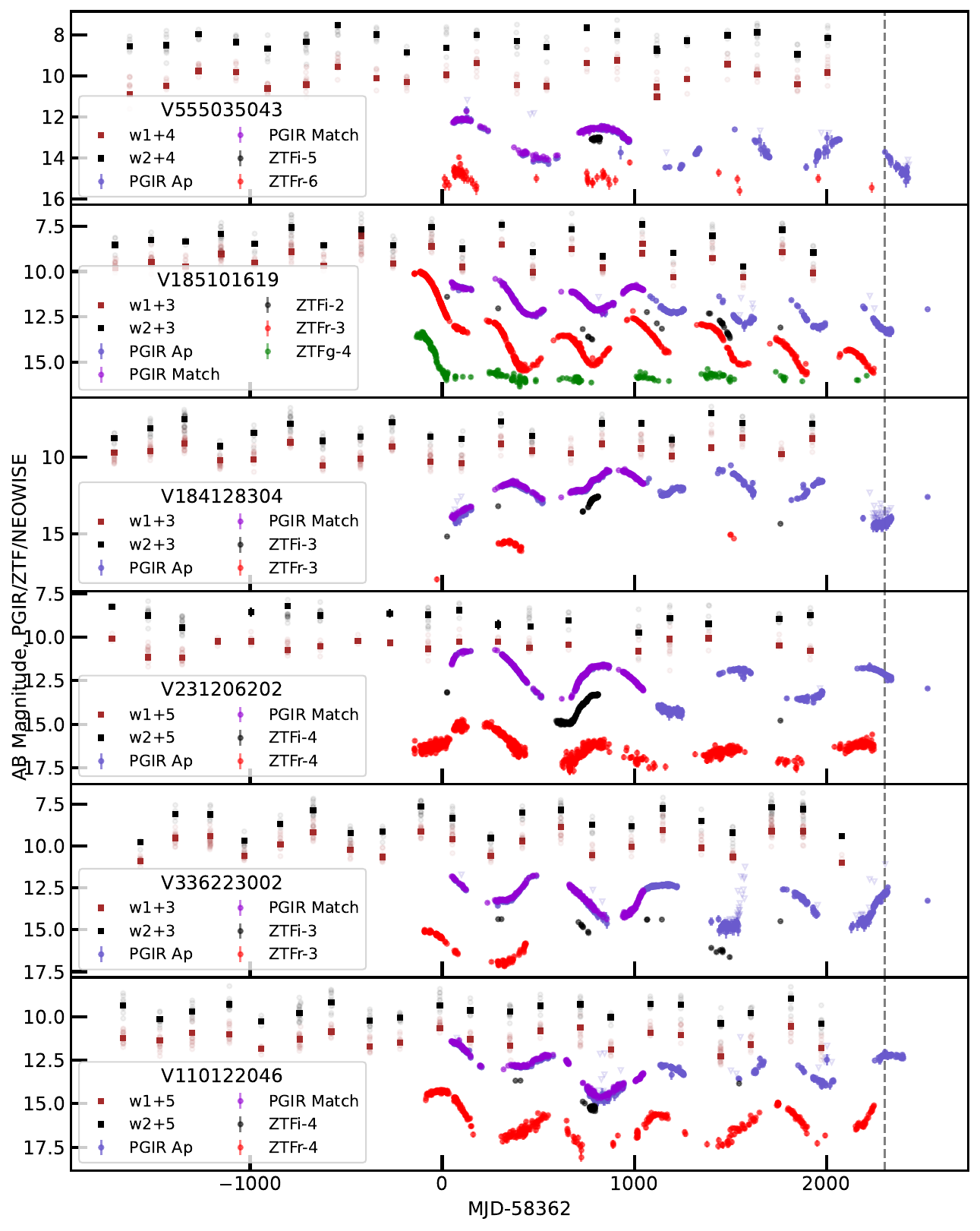}
\caption{C-rich LPV lightcurves corresponding to the sources in Figure~\ref{fig:monsterhband1_spec}.}
\label{fig:monsterhband1_lc}
\end{figure*}

\begin{figure*}
\includegraphics[width=\textwidth]{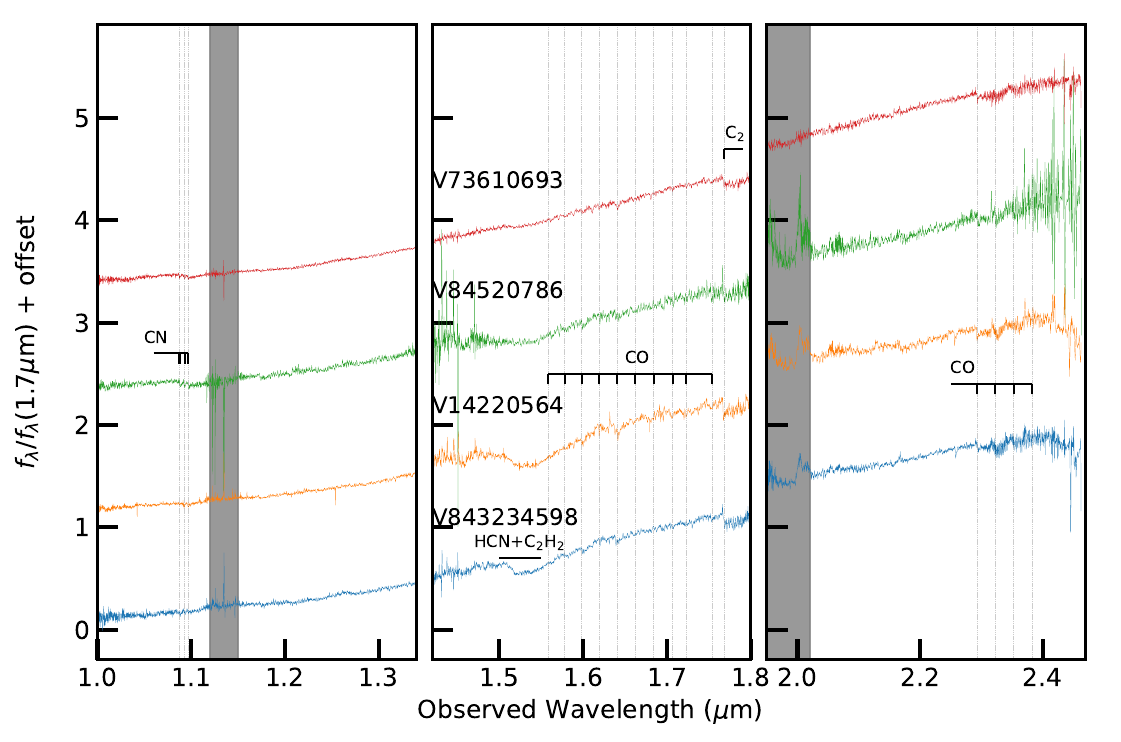}
\includegraphics[width=\textwidth]{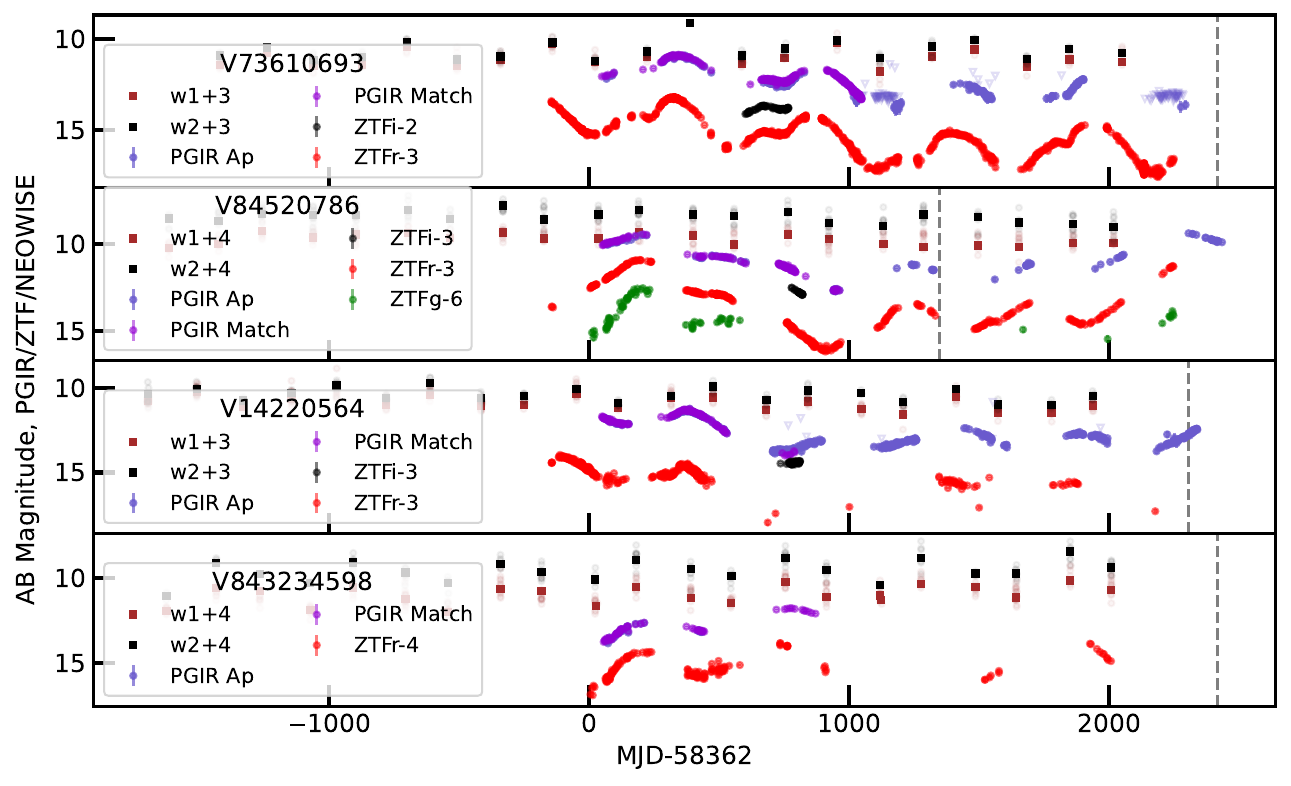}
\caption{C-rich LPVs continued from Figure~\ref{fig:monsterhband1_spec}, sorted by increasing 1.53{\um} absorption from top to bottom.}
\label{fig:monsterhbandlinerise}
\end{figure*}

\begin{figure*}
\includegraphics[width=\textwidth]{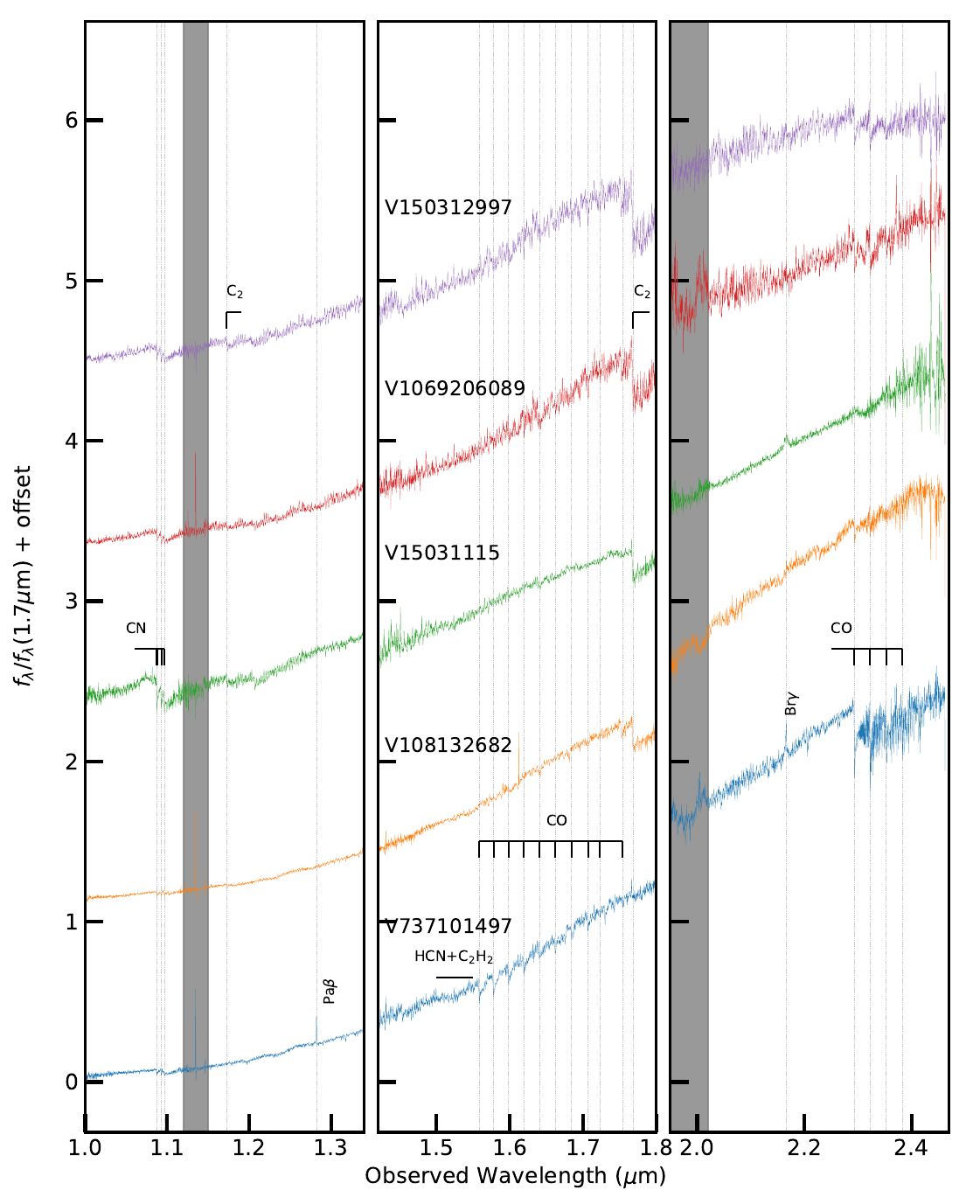}
\caption{Red C-rich LPVs without prominent HCN+C$_2$ absorption. V737101497 hosts enhanced CO absorption in the $H$-band as well as HI emission lines in $J$ and $K$ and an absence of C$_2$ at 1.77{\um}. A subtle 1.53{\um} absorption feature is highlighted in V737101497.}
\label{fig:cstar_spec}
\end{figure*}

\begin{figure*}
\includegraphics[width=\textwidth]{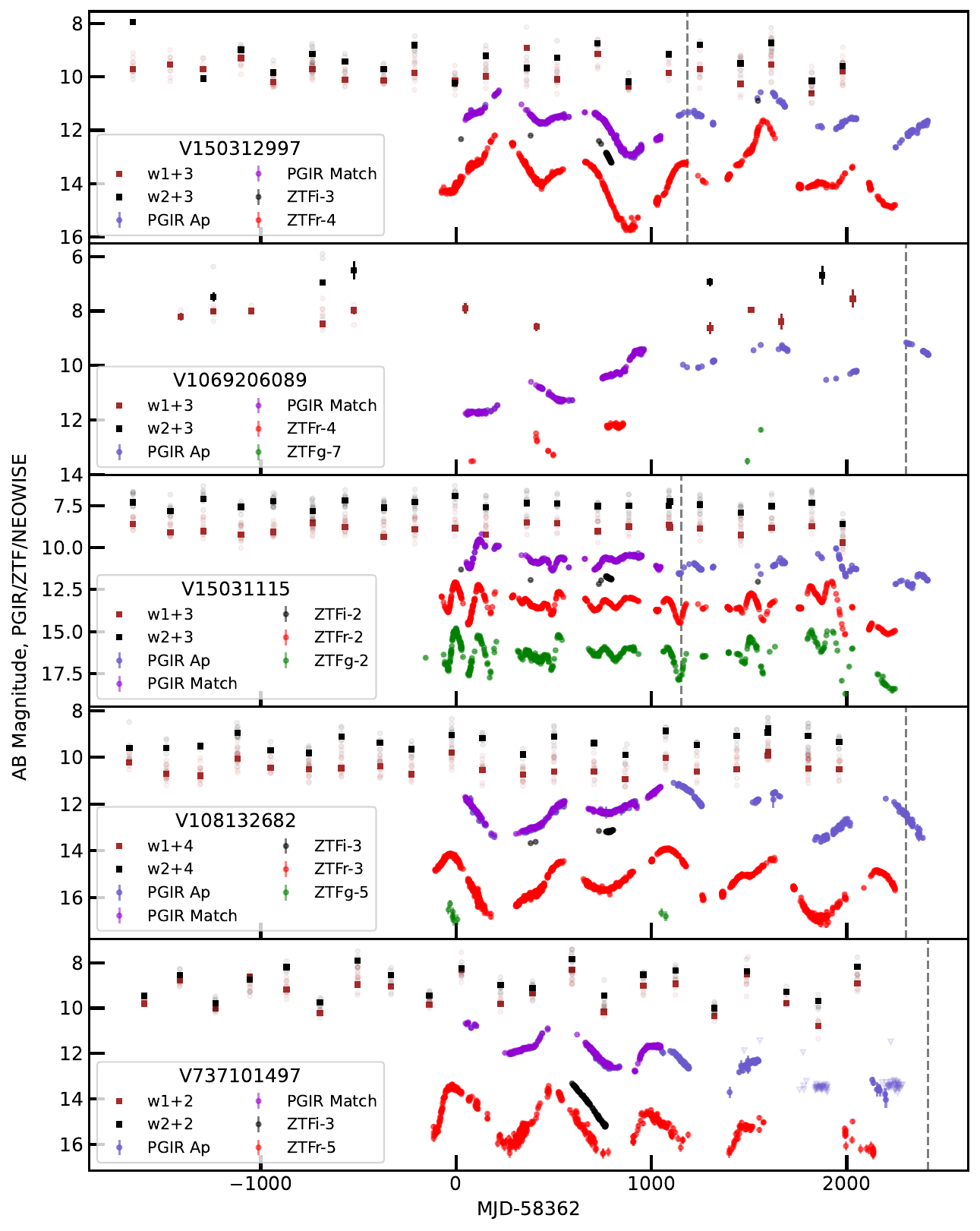}
\caption{C-rich LPV lightcurves corresponding to the sources with diminished HCN+C$_2$H$_2$ features in Figure~\ref{fig:cstar_spec}. Spectra were acquired at various phases, suggesting the presence of HCN+C$_2$H$_2$ in C-rich circumstellar envelopes is not strictly phase-dependent.}
\label{fig:cstar_lc}
\end{figure*}

Most C-rich LPVs include sources with common absorption features in their spectrum with short wavelength edges near 1.53{\um}, in close comparison with the C-rich Mira and IRTF template source R Lep \citep[HD\,31996, spectral type C7.6e(N4),][]{2009ApJS..185..289R}. These sources are presented in Figures~\ref{fig:hbandflatk0_spec}---\ref{fig:monsterhbandlinerise}. Their optical/IR (OIR) lightcurves demonstrate regular or semi-regular variability, consistent with other C-rich Miras. Moreover, spectra of the carbon stars were acquired at various phases. This spectral characteristic was first suggested to arise from HCN and/or C$_2$H$_2$ in cool ($T \sim 2000-3000$ K) carbon star atmospheres \citep{1981ApJ...246..455G, 1998AJ....115.2059J}. Synthetic spectral modeling of carbon stars with strong winds at phases of minimum light have demonstrated that the 1.53{\um} feature can similarly arise due to a combination of HCN, C$_2$H$_2$, as well as C$_2$ \citep{2004A&A...422..289G}. Even for carbon rich LPVs at their photometric minima such as {\bf V15031115 (NIKC 2-59)} in Figure~\ref{fig:cstar_spec} and \ref{fig:cstar_lc}, the 1.53{\um} feature is not universally detected. The feature in {\bf V23012483 (IRAS 21135+5149)}, shown in Figure~\ref{fig:monsterhband0_spec} and \ref{fig:monsterhband0_lc}, was not observed in previous spectral studies \citep[][ referred therein as ERSO 113]{2009MNRAS.400.1413W}, furthering the interpretation of the feature as phase-dependent. However, the relation between this spectral feature and carbon star large-amplitude variability in terms of phase-dependence remains to be fully settled. We do not detect a clear relation between the presence of the 1.53{\um} absorption feature and the variable's phase.

Several of the LAVs which host the strongest 1.53{\um} absorption features are characterized with steeply rising $K$-band continua, which we interpret to be dominated by thermal dust emission. Suppressed flux in the $J$-band, along with weakened CN absorption bandheads, suggests that the stellar photospheres are heavily obscured by a circumstellar dust environment. Presence of the 1.53{\um} feature has been shown to be sufficient, yet not necessary, to the presence of large-amplitude variability \citep[i.e., while all sources in previous studies hosting the 1.53{\um} absorption feature were LAVs, the existence of large-amplitude variability did not imply 1.53{\um} absorption,][]{2016A&A...589A..36G,2017A&A...601A.141G}. Similarly, we find that while the majority of C-rich LAVs host a strong 1.53{\um} absorption signature, a small subset does not (Figure~\ref{fig:cstar_spec}). In total, the 1.53{\um} absorption feature may be an important probe of the chemistry in C-rich circumstellar envelopes---detailed dust-driven modeling is required to determine the exact line carriers and dynamics.

The semiregular lightcurves in Figures~{\ref{fig:hbandflatk0_lc}}, \ref{fig:hbandflatk1_lc}, \ref{fig:monsterhband0_lc}, \ref{fig:monsterhband1_lc}, and \ref{fig:monsterhbandlinerise} suggest against any singular, dramatic dust formation episodes outside of typical pulsational activity for TP-AGBs. In the following sections, other classes of sources exhibiting periods of dust formation exhibit more erratic variability in their timeseries data.

\subsection{R Coronae Borealis stars}
\label{sec:rcb}
\begin{figure*}[htbp]
    \centering
    \includegraphics[width=\textwidth]{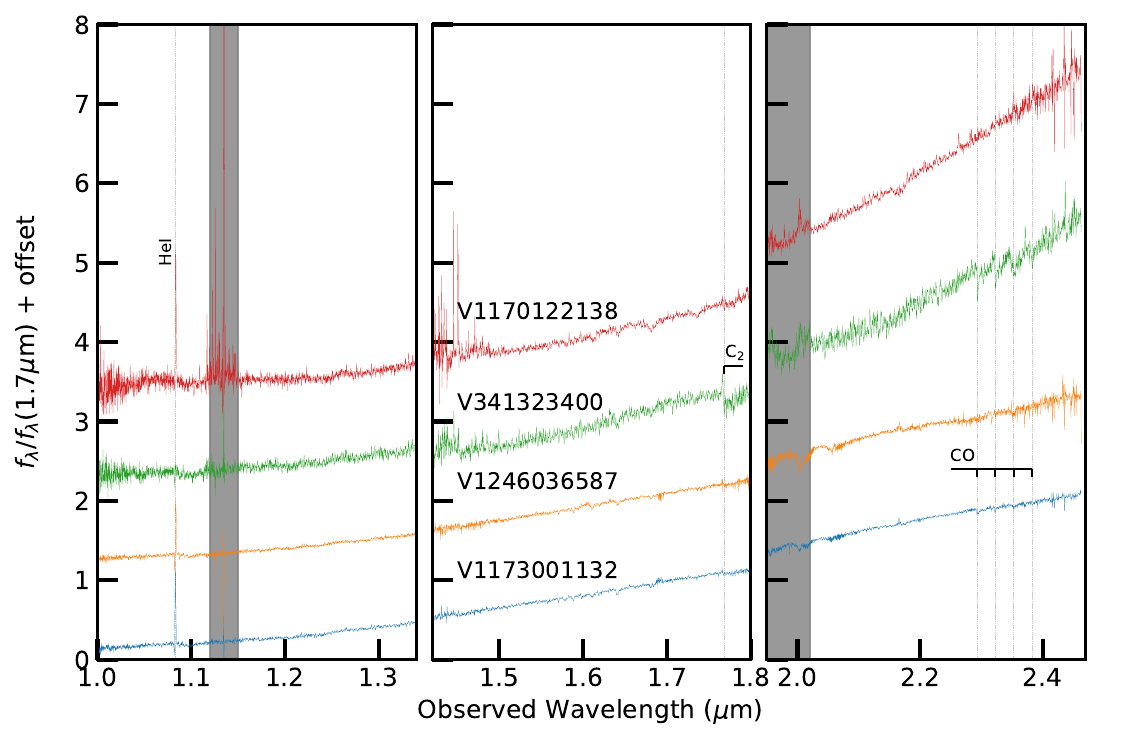}
    \includegraphics[width=\textwidth]{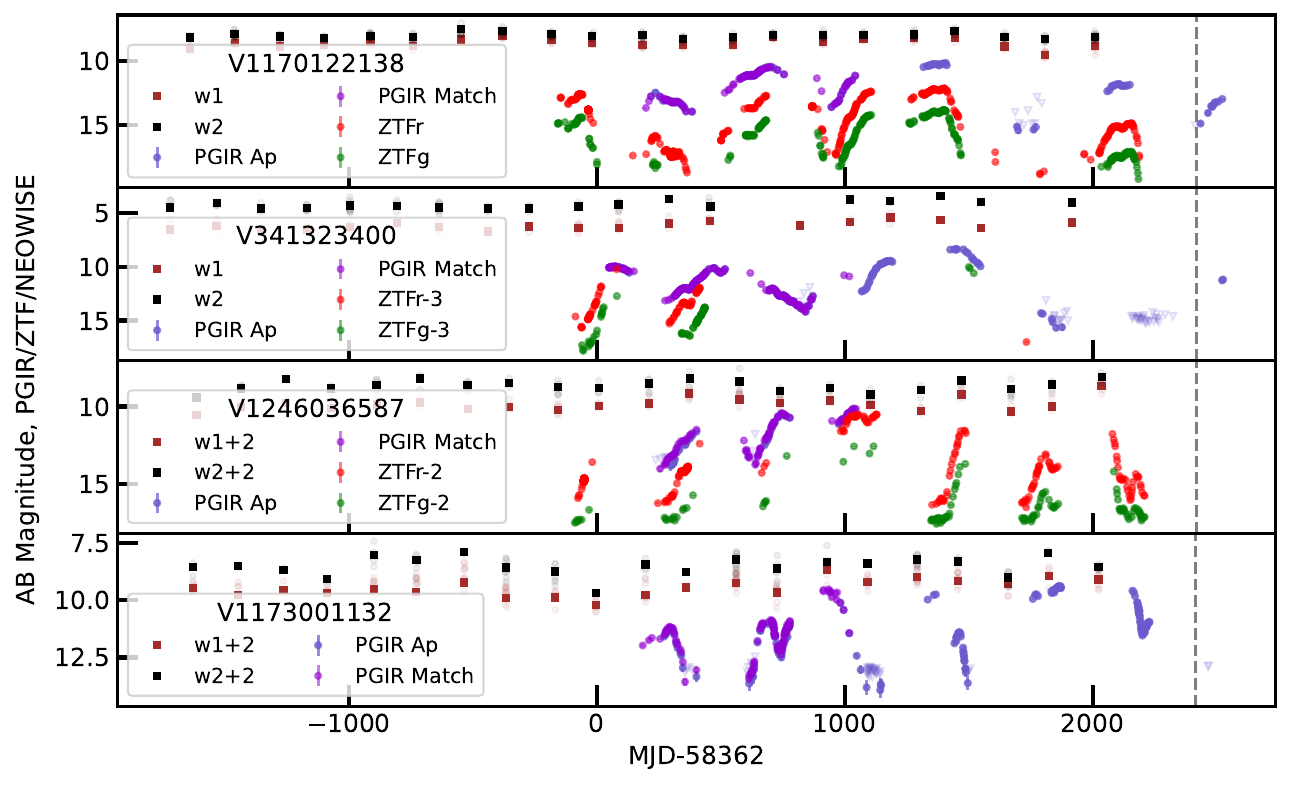}
    \caption{RCBs with steeply rising $K$-band continua. While not universal amongst RCBs, all sources in this spectral subset exhibit HeI-10833 either in emission or absorption. V341323400 (V381 Lac) exhibits a blue-shifted HeI-10833 absorption feature ($\sim -250$\,km\,s$^{-1}$) as well as enhanced C$_2$ absorption in the $H$-band. The bottom two spectra exhibit a P Cygni HeI-10833 profile.}
    \label{fig:rcbrisek}
\end{figure*}

\begin{figure*}[htbp]
    \centering
    \includegraphics[width=\textwidth]{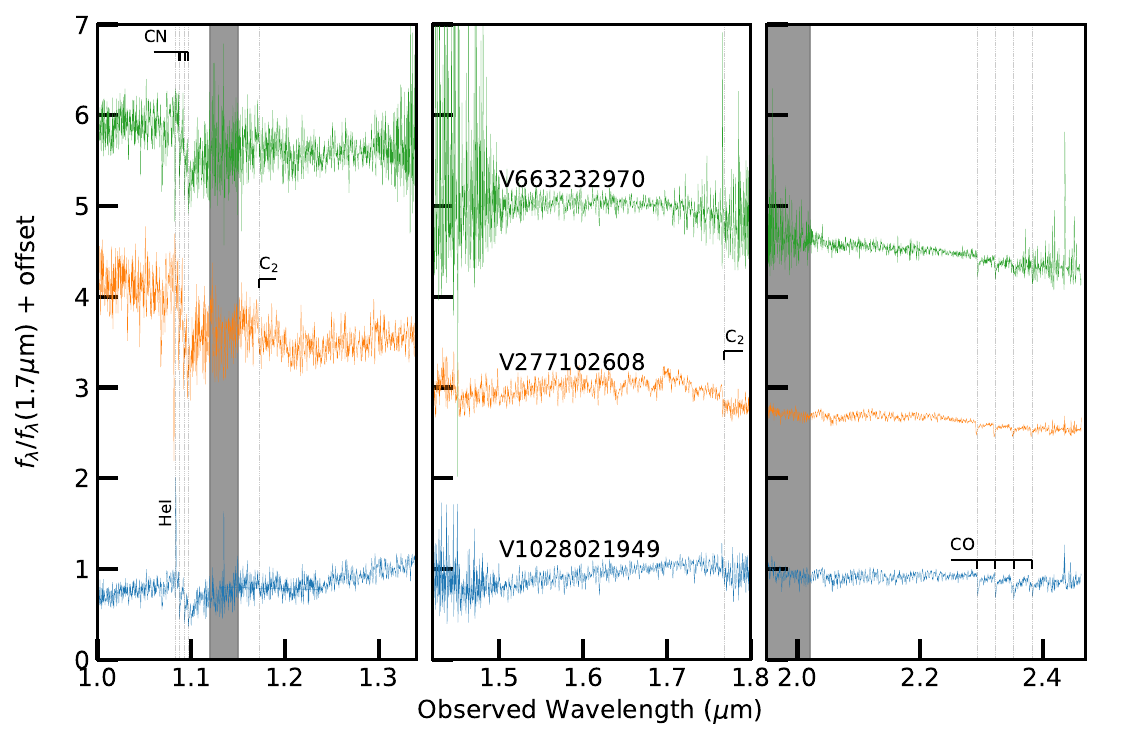}
    \includegraphics[width=\textwidth]{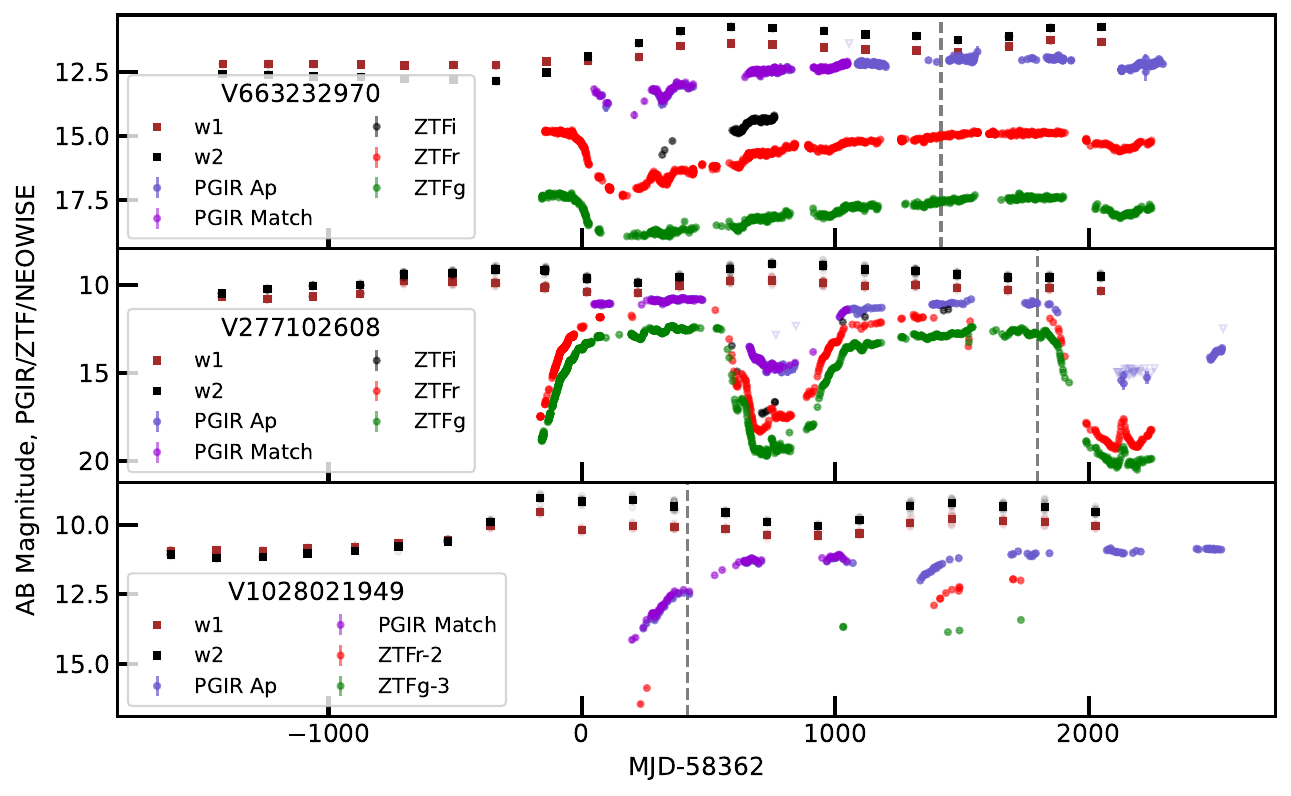}
    \caption{RCBs at or near maximum light, with flattened $K$-band continua, strong HeI-10833 profiles, and familiar C-rich variable absorption features.}
    \label{fig:rcbflatk}
\end{figure*}

R Coronae Borealis stars (RCBs) are a rare class of helium- and carbon-rich, hydrogen-deficient supergiants exhibiting large-amplitude erratic variability \citep{Clayton:2012JAVSO..40..539C, Clayton:1996PASP..108..225C, Asplund:2000A&A...353..287A}. Due to their unique chemistry, RCBs are known to be the products of C-O white-dwarf and He-white dwarf mergers. While the mechanisms are still ill-defined, large-amplitude variability is often driven by these sources' expulsion of dust at maximum light. 

The set of spectral and photometric data for all of these sources have been detailed in other studies \citep[e.g.,][and references therein]{Karambelkar:2021ApJ...910..132K,viraj:2024PASP..136h4201K}. We present spectra for 7 out of 16 census sources to compare with other classes. All RCBs that satisfy the census criteria are provided in Appendix~\ref{sec:appendix_observation}. 

\subsection{Young stellar and pre-main-sequence objects}
\label{sec:yso}
\begin{figure*}
\includegraphics[width=\textwidth]{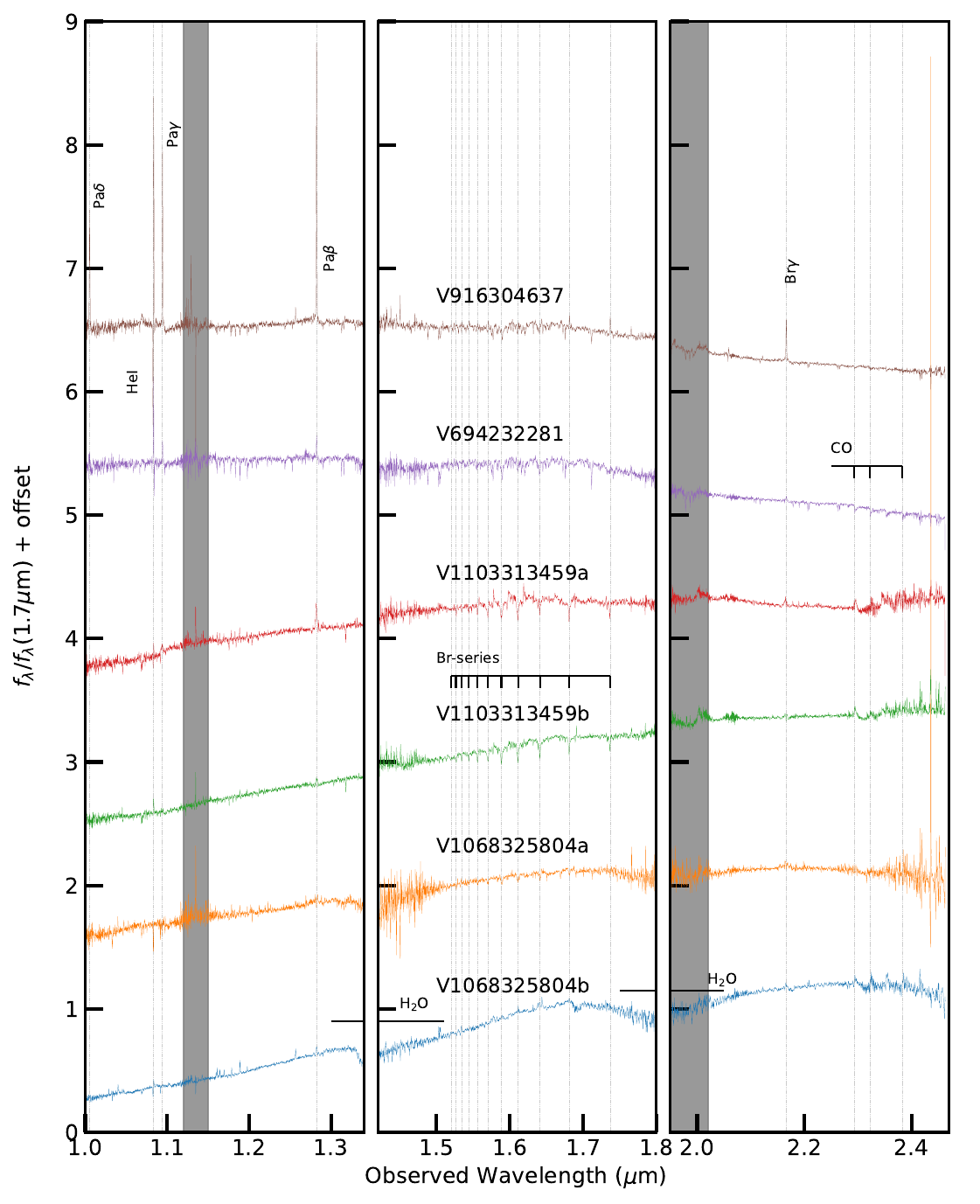}
\caption{YSO candidates. Spectra of V1103313459 were acquired 5 days apart, showing rapid evolution in He, H, and CO line profiles as well as its K-band continuum. Spectra of V1068325804 were acquired 45 days apart, with enhanced water absorption in the later spectrum and He absorption evolving into a P Cygni profile.}
\label{fig:yso_spec}
\end{figure*}

\begin{figure*}
\includegraphics[width=\textwidth]{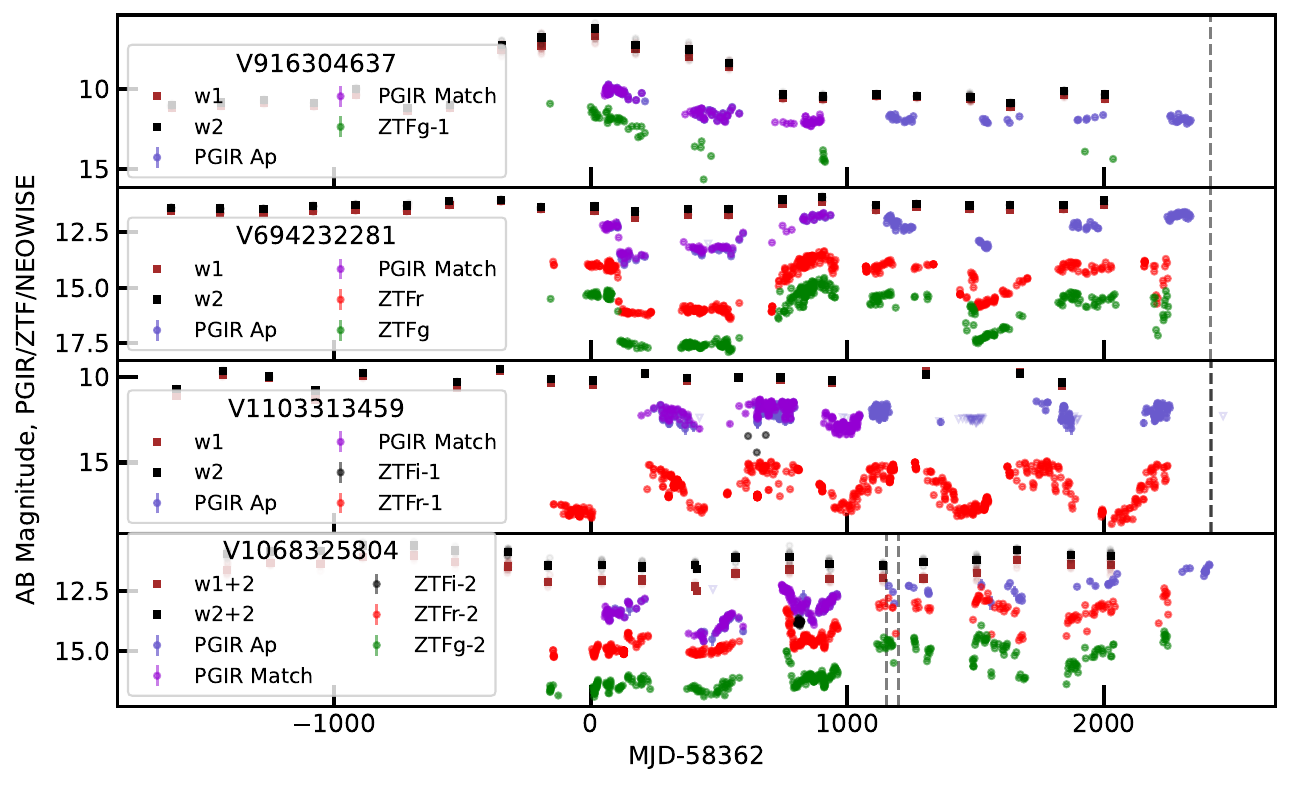}
\caption{Lightcurves of YSO candidates from Figure~\ref{fig:yso_spec}. Two spectra were acquired for V1103313459 5 days apart and for V1068325804 45 days apart.}
\label{fig:yso_lc}
\end{figure*}

\begin{figure}
\epsscale{1}
\includegraphics[width=\columnwidth]{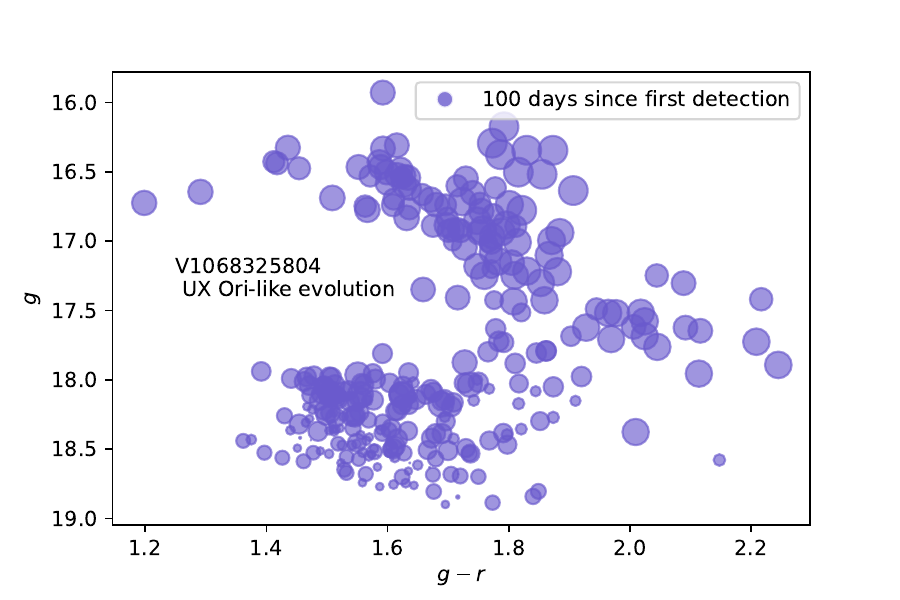}
\caption{The observed optical color-magnitude diagram for V1068325804 (IRAS 07069-1026) indicates a UX Ori-type morphology, in which early coverage shows the source following a reddening vector and a subsequent color-reversal at its minimum brightness. The size of the marker is proportional to the time since the first detection with ZTF.}
\label{fig:uxori}
\end{figure}

\begin{figure*}[htbp]
    \centering
    \includegraphics[width=\textwidth]{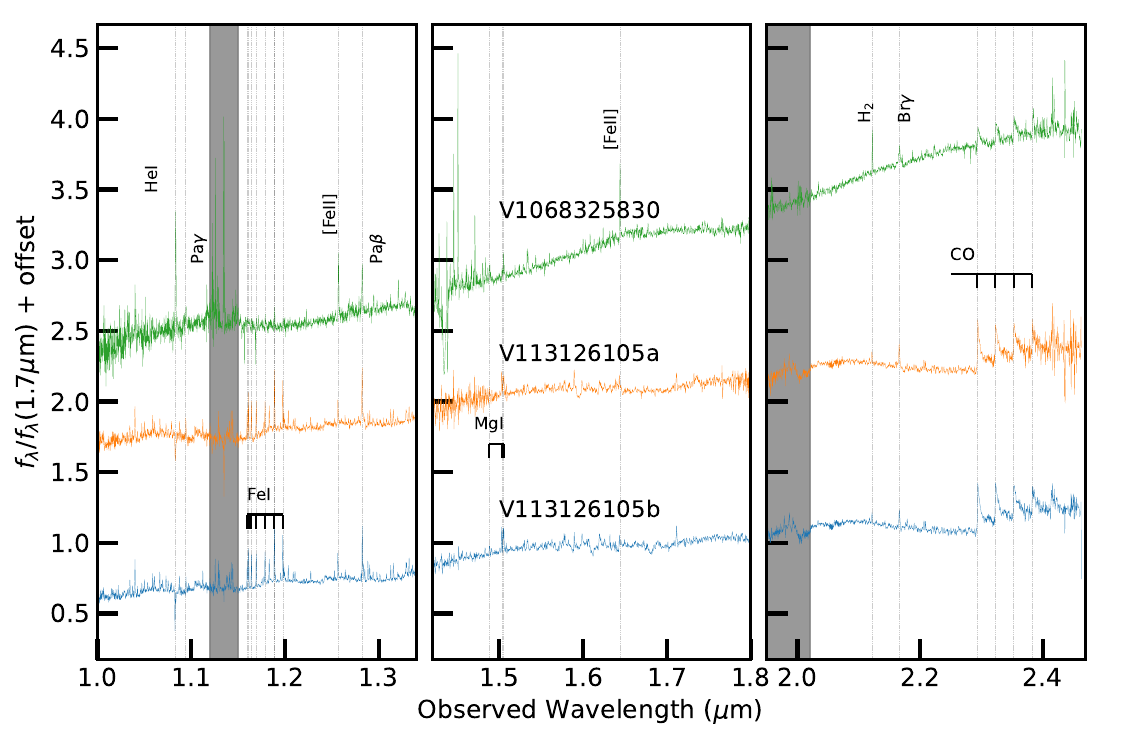}
    \includegraphics[width=\textwidth]{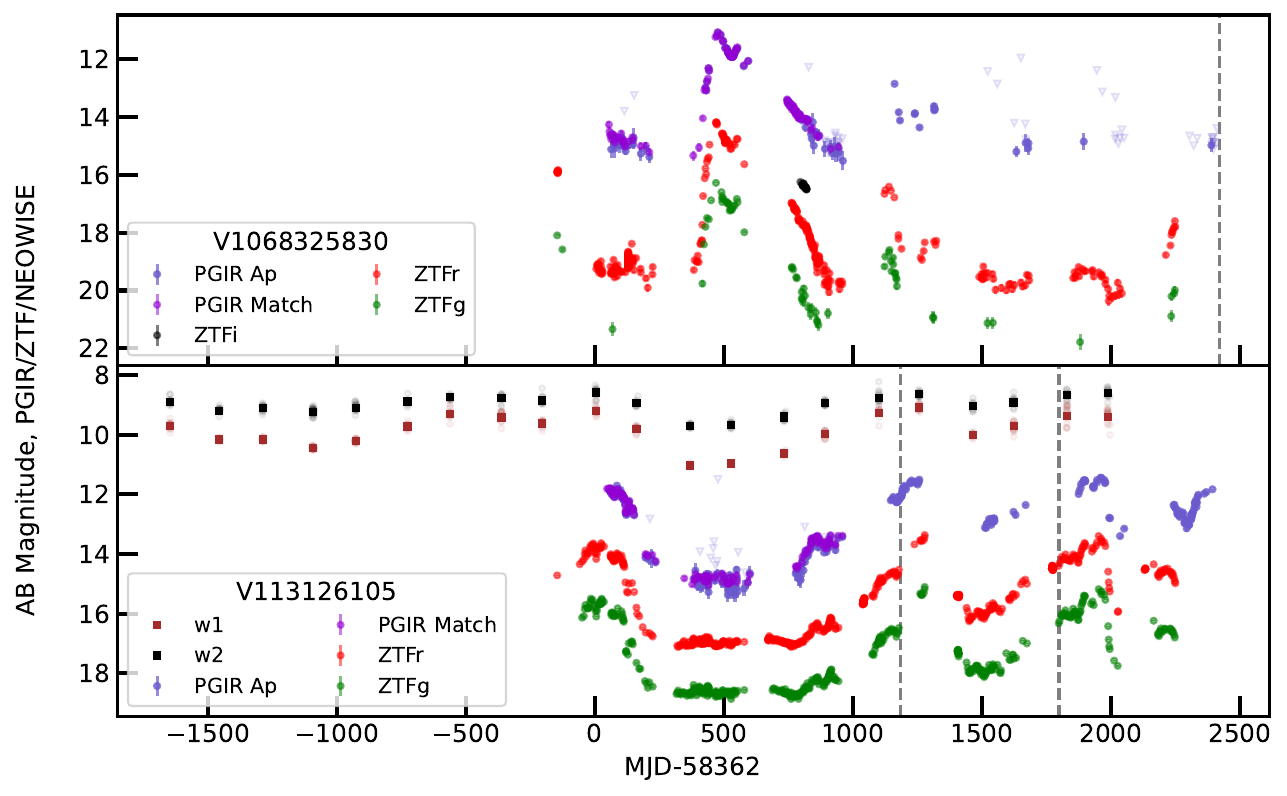}
    \caption{YSOs identified with EXor-like behavior characterized by repetitive outbursts in the timeseries data. HeI-10833 is seen in emission in V106832830 and absorption in spectra collected 613 days apart of V113126105. Spectral features features include molecular CO and H$_2$O; atomic MgI, SiI, NaI, and CaI; permitted and forbidden (particularly for V1068325830) Fe lines; and deep wind lines in HI and HeI.}
    \label{fig:EXor}
\end{figure*}

\begin{figure}[htbp]
    \centering
    \includegraphics[width=\columnwidth]{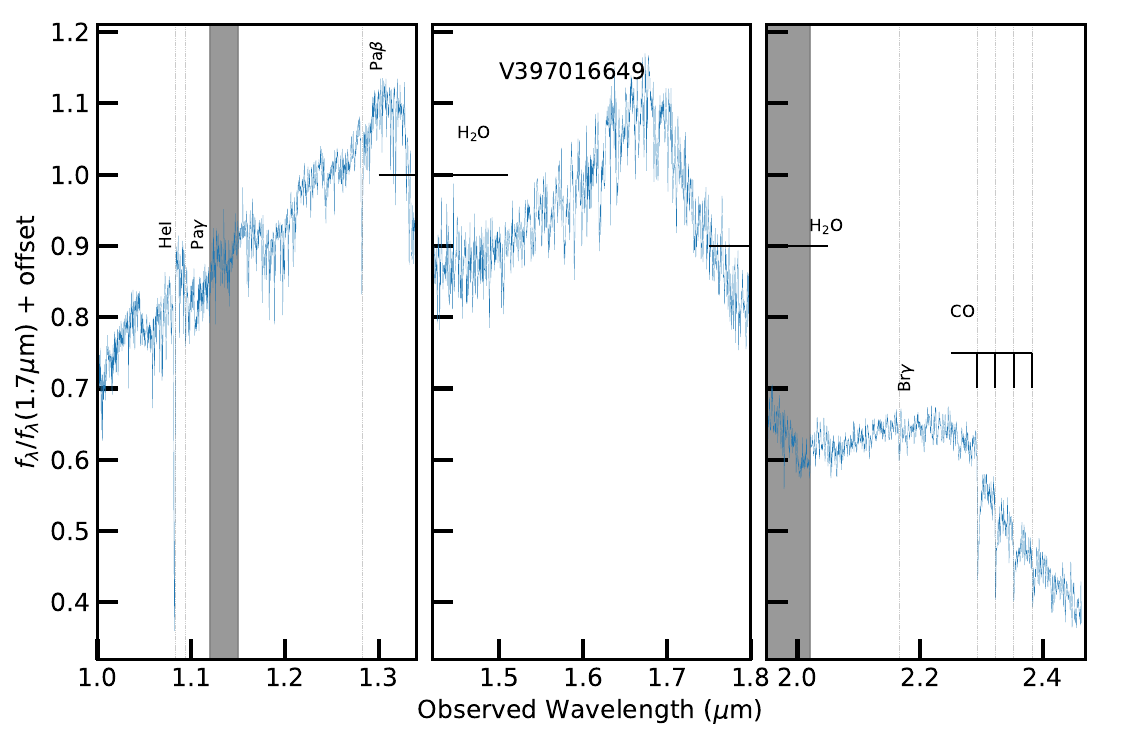}
    \includegraphics[width=\columnwidth]{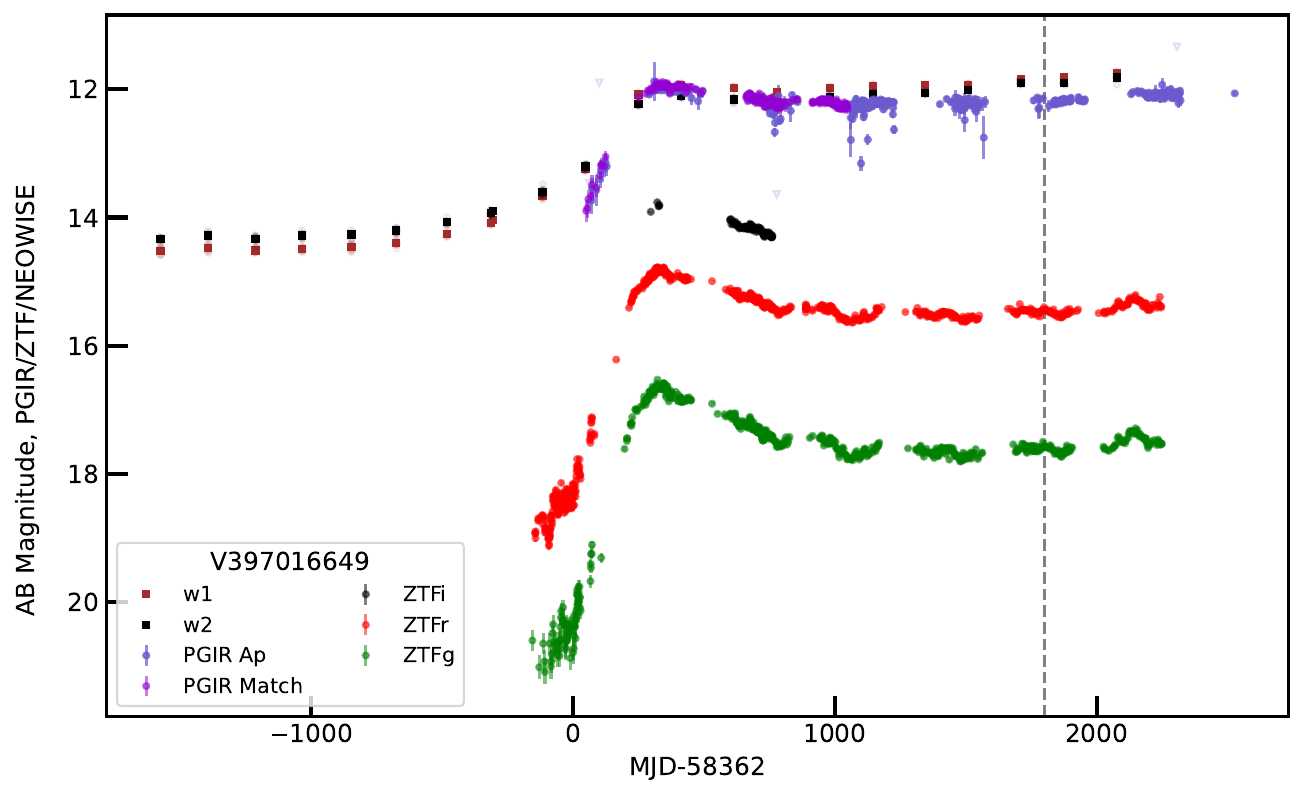}
    \caption{Gaia 18dvy (V397016649) is the only confirmed FUOr in the census. The post-eruption spectrum is marked with a blueshifted HeI-10833 absorption component, as well as HI and cool molecular absorption features.}
    \label{fig:fuor}
\end{figure}

\begin{figure*}[htbp]
    \centering
    \includegraphics[width=\textwidth]{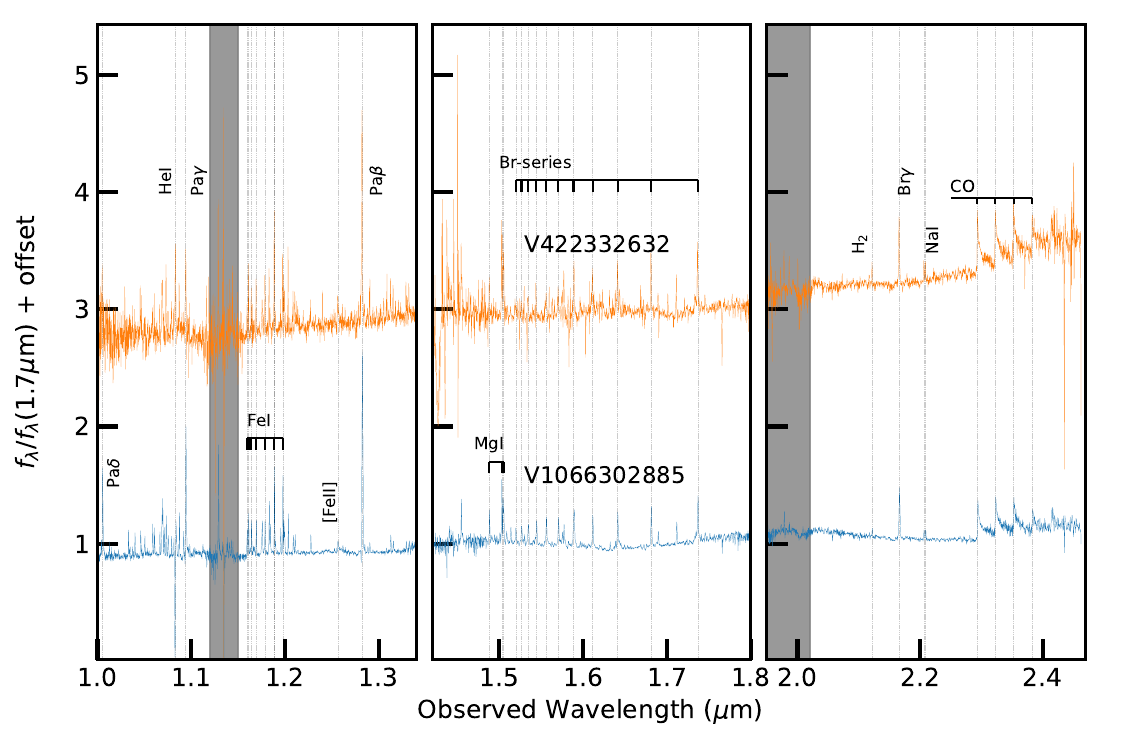}
    \includegraphics[width=\textwidth]{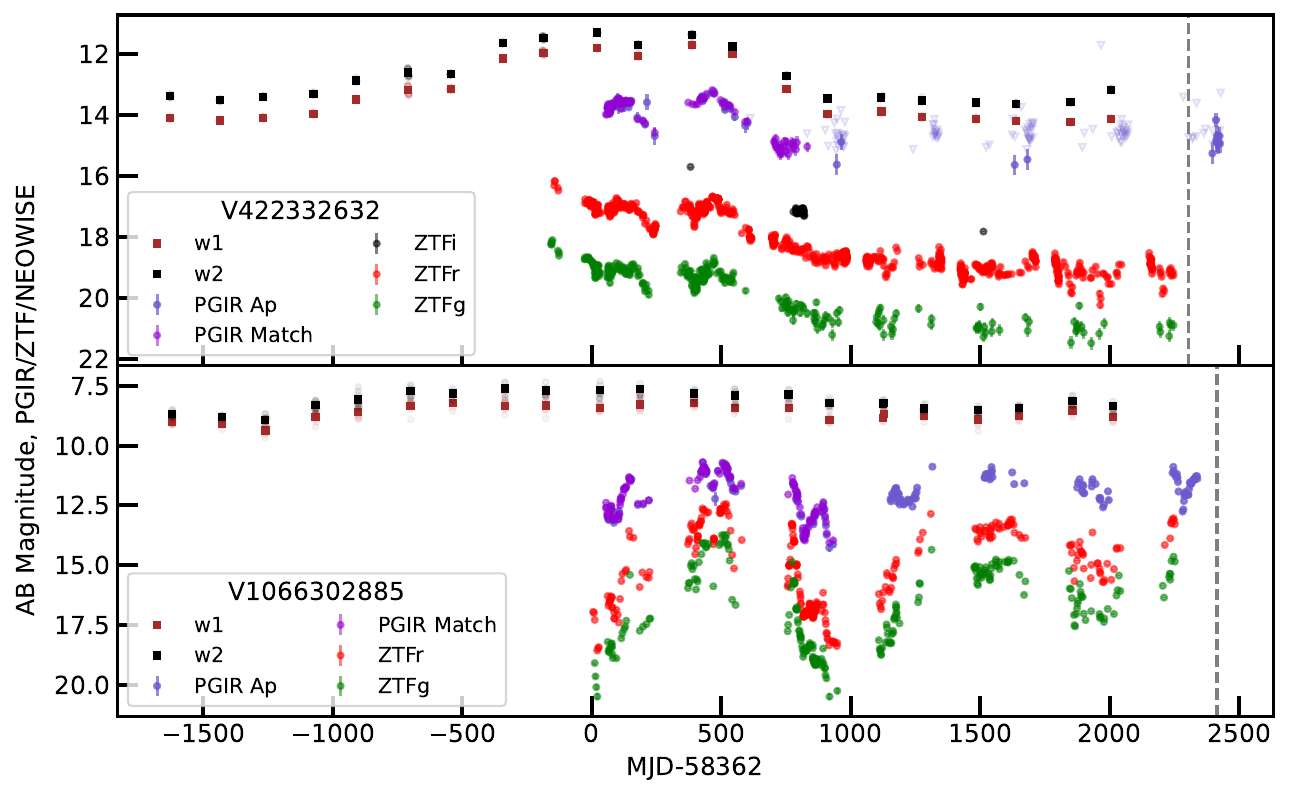}
    \caption{V1066302885 (SS 43/PDS 23) is a confirmed Herbig Ae/Be star and V422332632 is a candidate. Both exhibit similar line-dominated NIR spectra and erratic variability in their multiband photometry.}
    \label{fig:HAeBe}
\end{figure*}

\begin{figure*}[htbp]
    \centering
    \includegraphics[width=\textwidth]{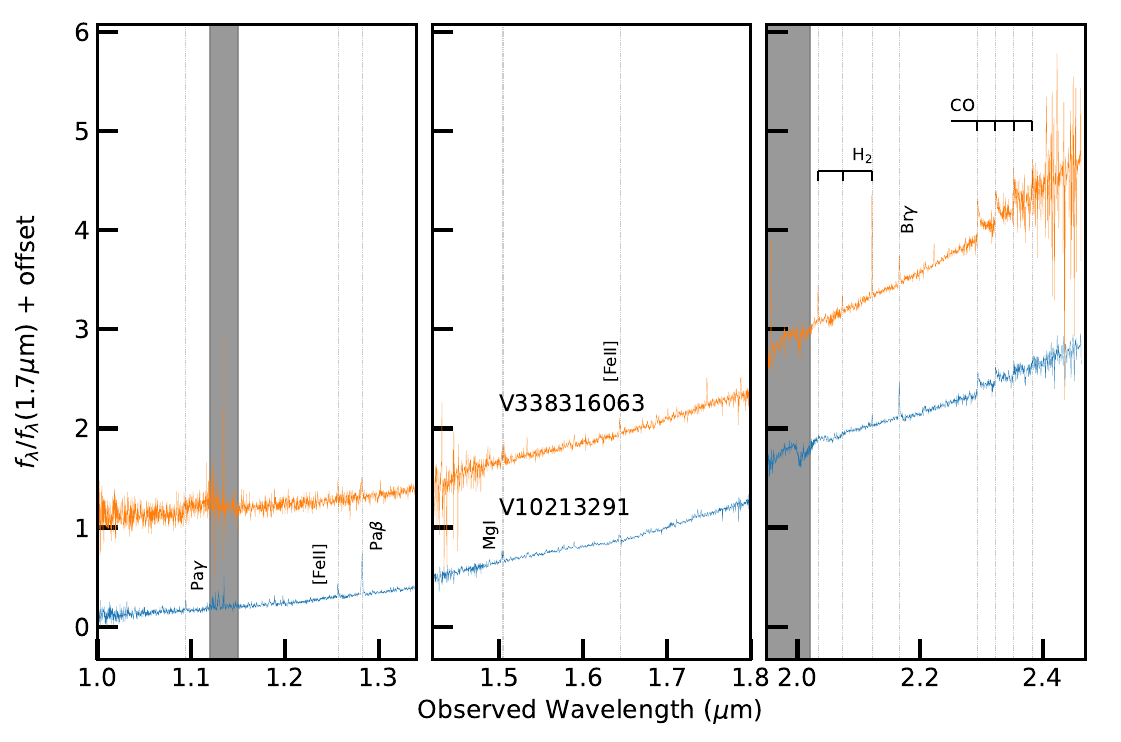}
    \includegraphics[width=\textwidth]{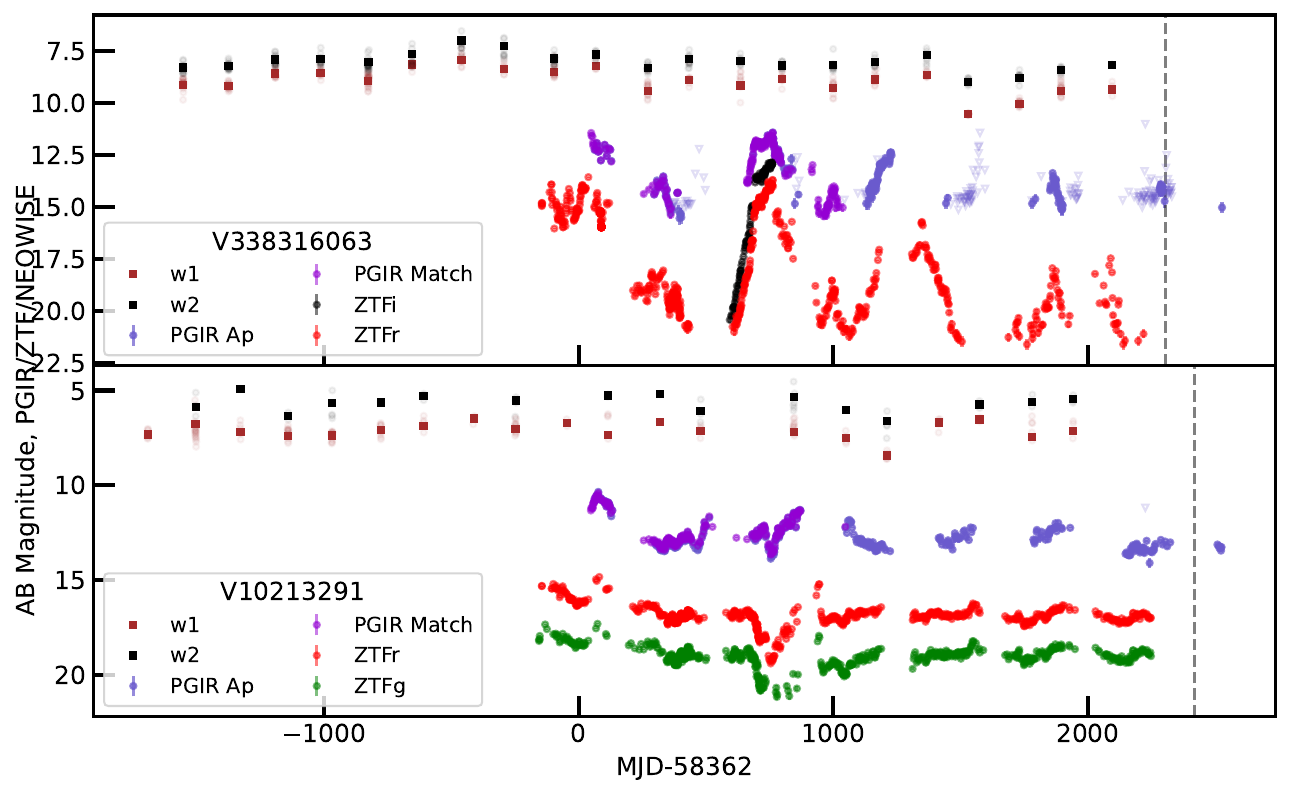}
    \caption{The dust-embedded YSOs PTF10nvg (V338316063) and PV Cep (V10213291). Their steeply rising $K$-band continua contrast with the other YSOs in the census.}
    \label{fig:ysoexotic}
\end{figure*}

Accretion processes and/or extinction variations in young stellar objects (YSOs) power large-amplitude variability in their timeseries data. The diversity of YSOs and the physical mechanisms behind their processes remain to be fully defined. While most sources are identified based on SEDs from photometry and machine learning algorithms \citep[e.g.,][]{2016MNRAS.458.3479M}, through our systematic higher-resolution spectroscopic follow-up of LAVs, PGIR has unveiled a zoo of pre-main-sequence objects and eruptive events. Sources in this sample are located near star-forming regions, host high-energy excitation lines and erratic variability, and/or have been proposed as candidate YSOs or pre-main-sequence objects in prior works in contrast to the Mira variables in \S\ref{sec:lpv}

{\bf V916304637 (Haro 5-94/V523 Ori)} is shown in Figure~\ref{fig:yso_spec} and Figure~\ref{fig:yso_lc} and is a well-studied Class II YSO \citep[e.g.,][]{Koenig:2014ApJ...791..131K, Cottle:2018ApJS..236...27C}. Strong emission lines include Pa$\delta$, $\gamma$, and $\beta$ as well as HeI-10833 and Br$\gamma$. V916304637 is located on the Gaia extinction-corrected de-reddened CaMD with $M_G=3.37$ and $BP-RP-E(BP-RP)=0.55$.

{\bf V694232281 (Haro 6-68/V460 Ori)} is shown in Figure~\ref{fig:yso_spec} and Figure~\ref{fig:yso_lc} and is a long-claimed member of the $\lambda$ Ori star-forming region \citep{Duerr:1982ApJ...261..135D}. It was classified as a YSO with spectral type K in a low-resolution optical spectrum with LAMOST \citep{Zhang:2023ApJS..267....7Z}. The NIR spectrum exhibits strong HeI-10833 and Pa$\beta$ emission.

{\bf V1103313459 (2MASS J18362900-0922384)} is shown in Figure~\ref{fig:yso_spec} and Figure~\ref{fig:yso_lc} and exhibits a similar spectrum as V1068325804. Due to halo contamination in $W3$ and $W4$, it is not included in the WISE color-color diagrams in Figure~\ref{fig:colors}. Prominent spectral features include a P Cygni HeI-10833 line evolving into purely HeI-10833 emission. Strong H emission features populate the $H-$band alongside CO emission in the $K$-band. The source lacks cool molecular features, hosts HeI and HI emission lines, and an evolving $K$-band continuum. The source has previously been identified as a red source in Spitzer Galactic midplane observations \citep{Robitaille:2008AJ....136.2413R} where it was suggested to be a standard AGB star rather than YSO due to the color selection $[8.0]-[24.0] = 2.04 < 2.5$. Given the risk of contamination in the simple color selection cut, we suggest the source is not a standard AGB in the sample, evidenced by the differences between the spectrum and the AGB spectral properties shown in Figure~\ref{fig:lpv_spec}$-$\ref{fig:lpvline1}. This object has been flagged in the optical due to its periodic ZTF $r$-band photometry \citep[ZTF J183629.01-092238.3,][]{Chen:2020ApJS..249...18C}. 

{\bf V1068325804 (IRAS 07069-1026)} is shown in Figure~\ref{fig:yso_spec} and Figure~\ref{fig:yso_lc} and is a YSO candidate in Canis Majoris (CMa) \citep{Sewilo:2019ApJS..240...26S, 2023ApJS..267...46I}. We acquired two spectra, 45 days apart, one in a local peak in magnitude and the other at the trough in a dimming event of 0.8 mags in the {\it J}-band. Notable spectral evolution is observed between the timepoints. Whereas HeI-10833{\AA} absorption is observed, the signature evolves into a P Cygni profile in the subsequent spectrum. Moreover, while no signatures of the CO bandheads are seen in the {\it K}-band in the first observation, CO molecular band emission arises in the later spectrum. The general continuum in the {\it H}-band evolves as well, suggesting enhanced water absorption between $1.5-1.65${\um}. Early Brackett series absorption lines in the {\it H} band also appear to diminish or vanish in the later spectrum.

In the ZTF $g$ vs. $g-r$ color-magnitude diagram in Figure~\ref{fig:uxori}, V1068325804 exhibits UX Ori-like behavior in which the star follows a reddening vector prior to turning blueward due to scattered light in the circumstellar dust, driven by the presence of an edge-on disk or thick envelope \citep{1988SvAL...14...27G,1997ApJ...491..885N}. The amplitude of UX Ori-type variability ranges from $2-3$ mags and is proposed to be driven by clumpy accretion from a protostellar cloud or the circumstellar medium in the vicinity of a pre-main-sequence star \citep{2022ApJ...930..111D, 2024arXiv240320065G}. Radio observations with ALMA \citep{olmi:2023MNRAS.518.1917O} indicate the emission from the source (referred therein as HG2728) originates from a shell expanding radially outwards.

Two sources exhibit repeated outbursts in their timeseries data, similar to  EX Lup-type (EXor) YSOs. EXor-like outbursts, are eruptive T Tauri stars exhibiting repetitive outbursts lasting months or years due to magnetospheric-accretion instabilities \citep{Herbig:2008AJ....135..637H, Fischer:2023ASPC..534..355F}. 

{\bf V1068325830 (iPTF\,15afq/Gaia\,19fct)}, shown in Figure~\ref{fig:EXor}, is a confirmed EXor star located in CMa \citep{Miller:2015ATel.7428....1M, Hillenbrand:2019ATel13321....1H, Park:2022ApJ...941..165P}. The source has been identified in previous outburst events dating from 2015 with the intermediate Palomar Transient Factory (iPTF) and in 2018 with ZTF and Gaia. Optical spectroscopy 6 months following the discovery outburst presented metal-rich absorption features along with a host of emission lines common for accreting T Tauri stars. PGIR coverage in 2019 captured a $\sim4$ mag months-long outburst in $J$-band, followed by a subsequent decline over $\sim 500$ days. The NIR spectrum presented herein extends the spectroscopic coverage, revealing prominent HeI, HI, and [FeII], H$_2$, and CO emission lines alongside a reddened continuum. The NIR spectrum compares well with the spectrum of the dust-embedded YSO PTF10nvg, shown in Figure~\ref{fig:ysoexotic} and described in further detail at the end of this section.

{\bf V113126105 (2MASS J03580766+6244253)} is shown in Figure~\ref{fig:EXor} and was observed at timepoints 600 days apart. A spectrum was obtained 100 days prior to a precipitous $3-4$ mag decline in the optical and then the following spectrum obtained at approximately the same magnitude as the first timepoint during the subsequent rise. It has been identified as a candidate EXor in the American Association of Variable Star Observers (AAVSO) International Variable Star Index VSX \citep[AAVSO UID 000-BMT-403,][]{2006SASS...25...47W}, supported by the recurring outbursts in the optical-IR timeseries. In contrast to the other targets, there is no Brackett emission in the {\it H} band. While there are minimal differences in the spectra taken at the different times, Brackett absorption appears in the second spectrum. Prominent spectral features include Pa-$\beta$ and CO emission in the {\it K} band, along with a host of emission lines in the {\it J} band. It is unclear upon visual inspection whether He-10833{\AA} or Si-10827{\AA}, or both, are responsible for the relevant absorption signature in both spectra.

{\bf V397016649 (Gaia 18dvy)}, is shown in Figure~\ref{fig:fuor} and is a confirmed FUOr in the Cygnus OB3 Association \citep{2020ApJ...899..130S}. We obtained a spectrum four years following the spectrum presented in \citet{2020ApJ...899..130S}, well into the outburst stage. Several absorption lines are observed including Pa-$\beta$, no longer exhibiting a P Cygni profile. Consistent with the earlier NIR spectrum, we observe broad water absorption at the onset of 1.3{\um} and strong CO bandheads in the {\it K}-band. 

Two sources are confirmed and candidate Herbig Ae/Be (HAeBe) variables. These stars are pre-main-sequence intermediate mass ($2 M_\odot< M < 10 M_\odot$) objects, which provide an evolutionary link between the low- and high-mass stellar formation pathways \citep{Herbig:1960ApJS....4..337H, Hillenbrand:1992ApJ...397..613H}.

{\bf V422332632 (IPHAS J055702.42+335534.3)} is shown in Figure~\ref{fig:HAeBe} and has been identified as a pre-main-sequence candidate or potential planetary nebula (PN). The latter classification has been supported by the measurement of high $r-H\alpha$ colors in the INT photometric H-alpha survey (IPHAS) catalog \citep[$r-H\alpha=1.403$,][]{Viironen:2009A&A...504..291V,Vioque:2020A&A...638A..21V}. The forest of NIR emission lines compare well with the other HAeBe star in Figure~\ref{fig:HAeBe} (V1066302885), yet PNs are known to host similar nebular emission lines. Due to the absence of reliable extinction $A_G$ and reddening $E(BP-RP)$ estimates in DR3, V422332632 (Gaia DR3 3451755127193604480) is not included in the CaMD in Figure~\ref{fig:colors}. However, without such corrections, $M_G = 6.1^{+1.6}_{-1.1}$ and $BP-RP=2.2\pm0.2$, on the border of the delineation between HAeBe and PN by \citet{Vioque:2020A&A...638A..21V}. 

{\bf V1066302885 (SS 43/PDS 23)} is shown in Figure~\ref{fig:HAeBe} and is a confirmed Herbig Ae/Be variable \citep{Vieira:2003AJ....126.2971V}. The NIR spectrum hosts a forest of emission lines with prominent HI and CO emission features. A blueshifted HeI-10833 absorption component ($\sim - 250$\,km\,s$^{-1}$) may also be present, yet higher resolution observations are required for reliable identification.

Two sources are dust-embedded YSOs with steeply rising NIR continua. {\bf V338316063 (V2492 Cyg/PTF10nvg)} is shown in Figure~\ref{fig:ysoexotic} and was first characterized by the Palomar Transient Factory (PTF) in 2010 as a rare and unusual Class I YSO undergoing a brightening event \citep{Covey:2011AJ....141...40C}. Rather than accretion instabilities exclusively driving the large-amplitude variability as for other events such as FUOrs or EXors, semi-periodic variability is driven by cyclical extinction variations, obscuring then revealing the inner disk. Close to 15 years removed from discovery, PTF10nvg has not settled into a quiescent state; rather, the optical and IR lightcurves have repeatedly exhibited outbursts with dampening peaks in contrast to EXor-style outbursts such as V113126105. The NIR spectrum exhibits similar characteristics as the spectra collected near maximum brightness in \citet{Hillenbrand:2013AJ....145...59H}, with common molecular and atomic emission features. We detect enhanced H$_2$ emission at 2.1218{\um} ($1-0$ S(1)) in comparison to spectra taken at maximum light in 2010.
    
{\bf V10213291 (PV Cep)} is shown in Figure~\ref{fig:ysoexotic} and is an extensively studied large-amplitude YSO with significant differences between other common periodically eruptive objects such as EXors \citep[e.g.,][]{Lorenzetti:2011ApJ...732...69L, pvcep:2013A&A...554A..66C}. The spectra of PV Cep and PTF10nvg stand apart from the other YSOs in the PGIR census sample. 

\begin{figure}
\includegraphics[width=\columnwidth]{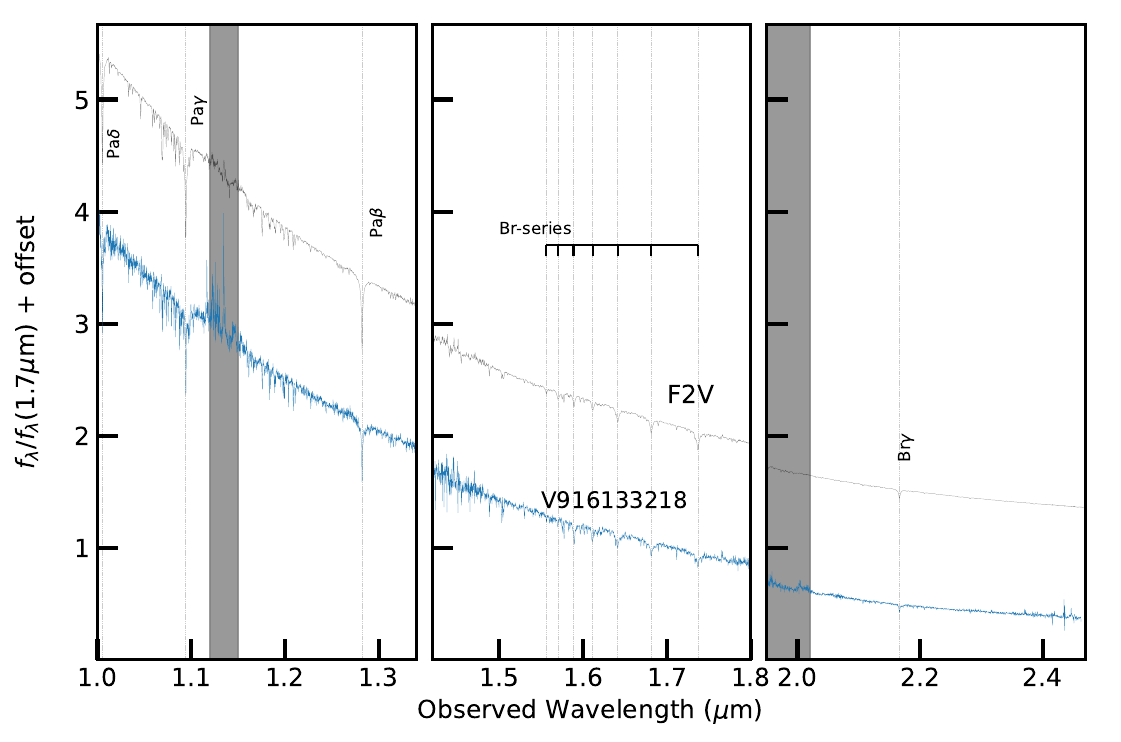}
\includegraphics[width=\columnwidth]{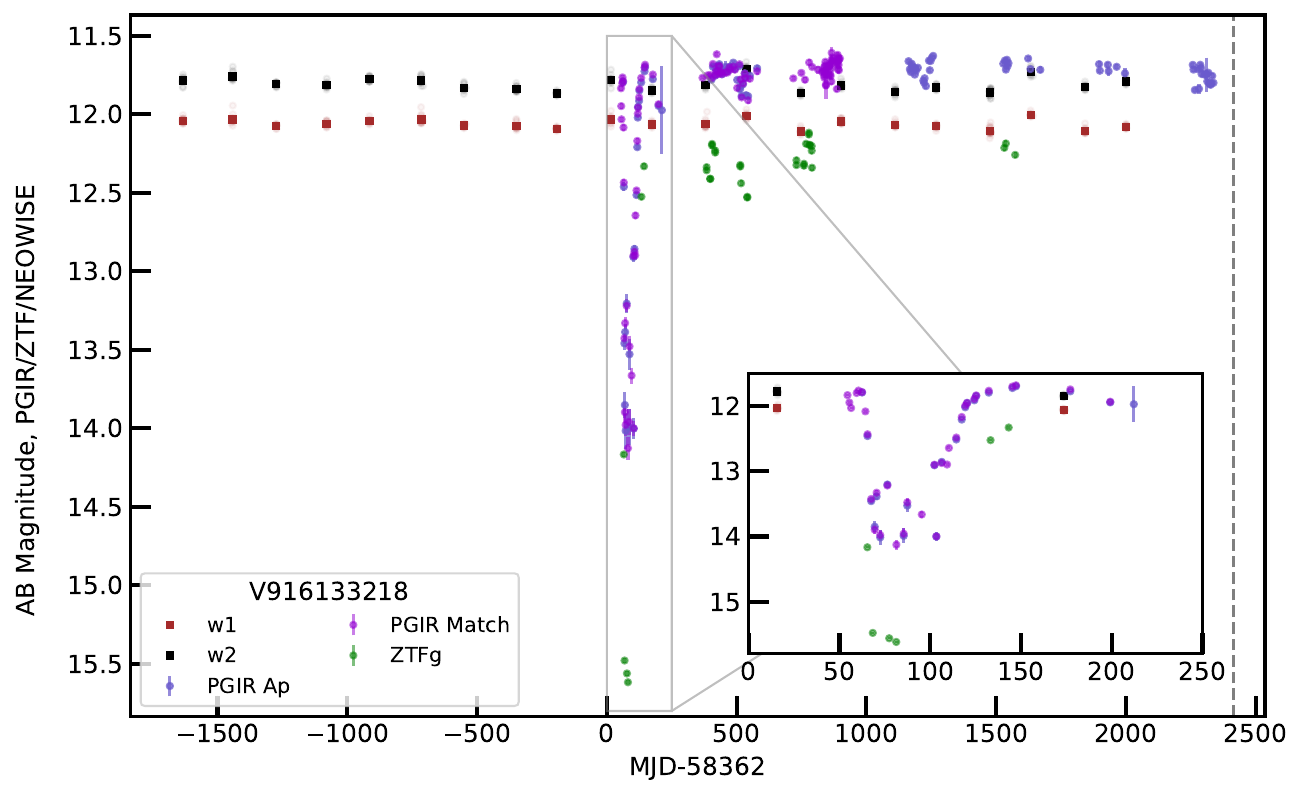}
\caption{Candidate YSO V916133218 (2MASS J05362855+0004456) with an F-type spectral continuum exhibiting $J$- and $g$-band large-amplitude variability. The F-type main sequence star HD113139 is plotted for reference. The inset shows that the source underwent a precipitous $\gtrsim 2$ magnitude decline in $J$ within the first 150 days of PGIR coverage.}
\label{fig:ttauri}
\end{figure}

{\bf V916133218 (2MASS J05362855+0004456)}, presented in Figure~\ref{fig:ttauri},  is a YSO candidate first identified by \citet{Sanchez:2014A&A...572A..89S} based on UV excess and has been included in several spectroscopic and rotation studies of the greater Orion star-forming complex \citep[e.g.,][]{Jayasinghe:2018MNRAS.477.3145J}. Its SED is representative for Class II YSOs with large infrared excesses. The source exhibited $> 2 $ mag decline and subsequent rebrightening in the $J$-band in the first months of PGIR observations. Only $g$-band observations were conducted with ZTF, similarly reflecting large-amplitude behavior at the onset of NIR coverage. Deep dimming and recovery events over similar timescales are common for YSOs \citep[e.g.,][]{Jiang:2022RNAAS...6..232J,Jiang:2024AJ....167..221J}. The spectrum was acquired nearly 2500 days following the outburst, demonstrating an F-type spectral continuum with He and H absorption lines. V916133218 is located on the Gaia extinction-corrected de-reddened CaMD with $M_G=3.69$ and $BP-RP-E(BP-RP)=0.67$.

\begin{table*}
	\centering
	\caption{Properties of YSOs or PMS objects in the PGIR LAV census}
	\begin{tabular}{llllcccl}
		\hline
		 PGIR ID & RA & Dec & Gaia DR3 ID & RUWE & Distance (pc) & Comments\\
		\hline
            V916133218 & 84.119 & 0.079 & 3220829441656781824 & 1.063 & $401^{+2}_{-3}$ & Class II Orion YSO\\
            V916304637 & 86.122 & -1.371 & 3218175740278049664 & 1.155 & $405\pm3$ & Class II Orion YSO \\
            V694232281 & 83.179 & 12.352 & 3340979816028134144 & 4.912\tablenotemark{a} & $534^{+26}_{-24}$ & $\lambda$ Ori YSO\\ 
            V1066333453 & 97.734 & -9.533 & 3003021861847549568 & 1.812\tablenotemark{a} & $860^{+163}_{-89}$ & Vdb 80 candidate\\
            V422332632 & 89.260 & 33.926 & 3451755127193604480 & 1.089 & $2171^{+2293}_{-861}$  &Candidate HAeBe\\
            V1068325804 & 107.343 & -10.516 & 3046391307734381184 & 6.851\tablenotemark{a} & $792^{+132}_{-117}$ & Candidate UX-Ori in CMa\\
            V113126105 & 59.532 & 62.740 & 475072212426610304 & 2.227\tablenotemark{a} & $906^{+113}_{-87}$ & Candidate (EXor) \\
            V1068325830 & 107.339 & -10.493 & 3046391406515862912 & 1.152 & $2173^{+920}_{-788}$ & iPTF 15afq (EXor in CMa) \\
            V1066302885 & 96.725 & -10.260 & 3002852983733291648 & 2.905\tablenotemark{a} & $810\pm35$ & PDS 23 (HAeBe)\\ 
            V10213291 & 311.475 & 67.961 & 2246924068029363840 & 2.422\tablenotemark{a} & $350^{+5}_{-4}$ & PV Cep\\
            V397016649 & 301.275 & 36.487 & 2059895933266183936 & 1.042 & $11138^{+2371}_{-4662}$ & Gaia 18dvy (FUOr)\\
            V1103313459 & 279.121 & -9.377 & 4155528468317586048 & 1.152 & $4838^{+3831}_{-1664}$ & Candidate\\
            V338316063\tablenotemark{b} & 312.860 & 44.090 & 2066869246454772224 & 2.028\tablenotemark{a} & $774^{+26}_{-21}$ & PTF10nvg\\
            \hline 
            \hline
	\end{tabular}
    \label{tab:yso}
    \tablecomments{All coordinates unless otherwise noted correspond to its 2MASS counterpart in Appendix Table~\ref{tab:tmass_colors}. V397016649 (2MASS ID 20050603+3629135) and V1103313459 (2MASS ID 18362900-0922384) suffer from contamination or poor photometric quality in some or all bands and thus not included in the Appendix Table~\ref{tab:tmass_colors}, but are included here. All Gaia DR3 counterparts are within 0.2" of the 2MASS source coordinates. Distances are from \citet{2021AJ....161..147B}.}
    \tablenotetext{a}{Uncertain astrometric data (RUWE $>1.4$).}
    \tablenotetext{b}{No 2MASS counterpart. Source coordinates are from PGIR.}
\end{table*}

\begin{figure}[htbp]
    \centering
    \includegraphics[width=\columnwidth]{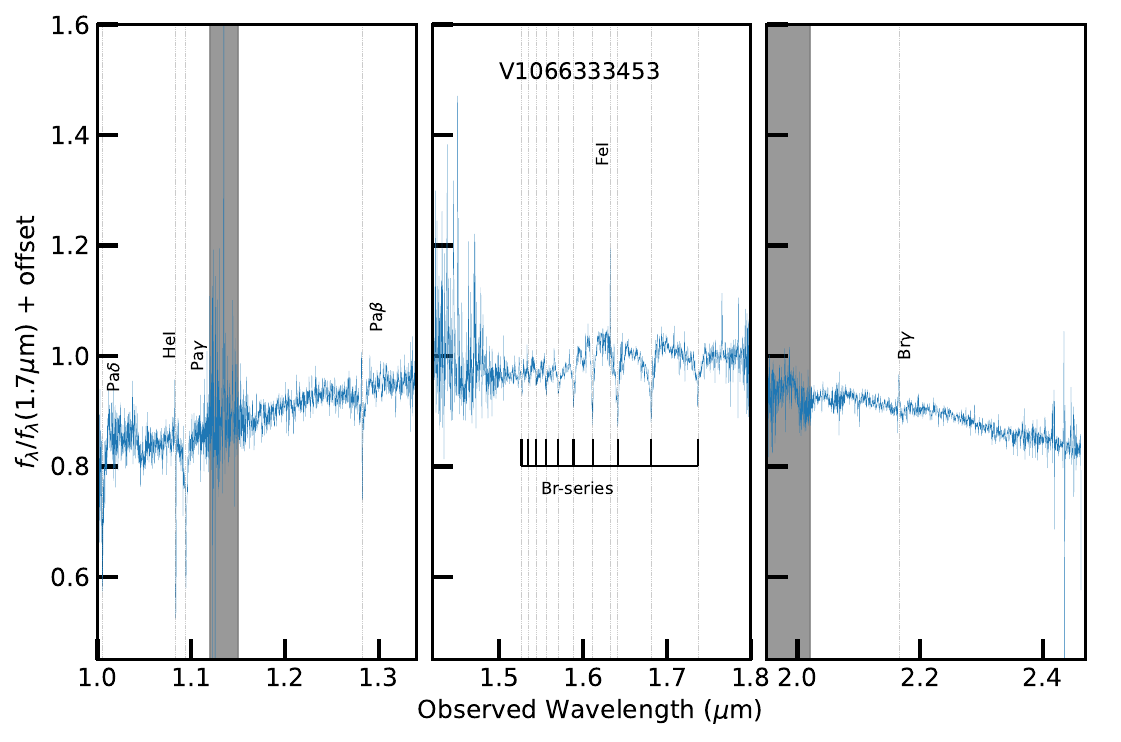}
    \includegraphics[width=\columnwidth]{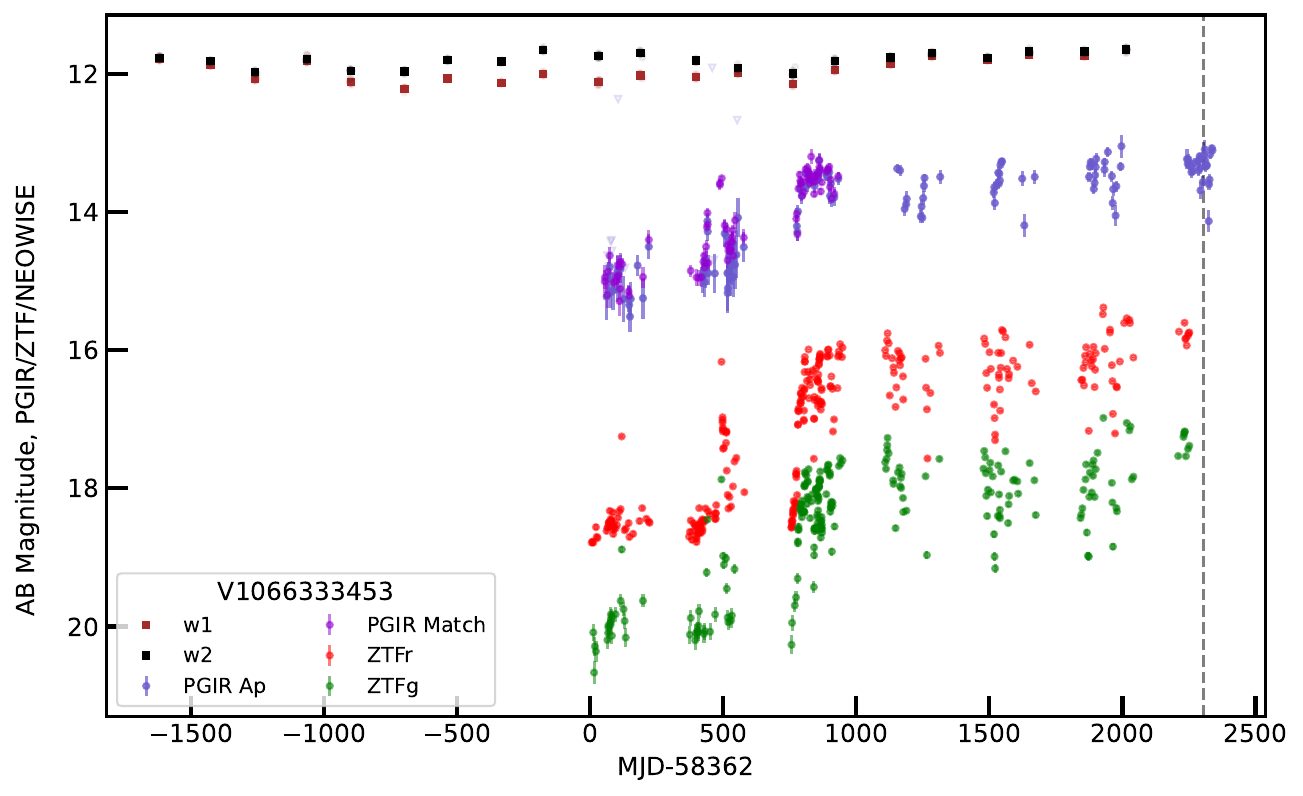}
    \caption{V1066333453 is an ambiguous variable near the star-forming VdB 80 association exhibiting erratic YSO-like variability. Its spectral features include HeI-10833, broad HI, and narrow Fe lines.}
    \label{fig:exotic}
\end{figure}

{\bf V1066333453 (2MASS\,J06305614-0931590)}, has previously been ambiguously classified via photometric data as a candidate RCB \citep{Tisserand:2020A&A...635A..14T}, YSO \citep{2016MNRAS.458.3479M}, or AGN \citep{Edelson:2012ApJ...751...52E, Secrest:2015ApJS..221...12S, Bailer:2019MNRAS.490.5615B}. The NIR spectrum (Figure~\ref{fig:exotic}) demonstrates extremely broad Brackett-series features with superposed narrow Fe lines, Pa$\beta$, $\delta$, and $\gamma$ broad lines, as well as HeI-10833. The source is $\sim 7$ arcmin away from the center of the star-forming Lagotis HII region in the reflection nebula VdB 80 (RA(J200)$=6^{\rm{h}}30^{\rm{m}}53.5^{\rm{s}}$ and Dec(J200)$=-09^\circ39^{'}08.6^{\rm{s}}$, \citet{Bradley:2025PASA...42...32B}). Moreover, its Gaia counterpart (Gaia DR3 3003021861847549568) is located on the extinction-corrected, de-reddened Gaia CaMD with $M_G=5.75$ and $BP-RP-E(BP-RP)=0.43$. The source distance is $860^{+163}_{-89}$pc, consistent with the distance to the VdB 80 association and effectively ruling out its proposed AGN classification. In addition, its SED is typical for other YSOs peaking in the near- to mid-IR. We therefore tentatively include it within the sample of YSOs. 

\subsection{Symbiotic stars and late-type HeI-10833 emitters}
\label{sec:symbiotic}

\begin{figure*}[htbp]
    \centering
    \includegraphics[width=0.95\textwidth]{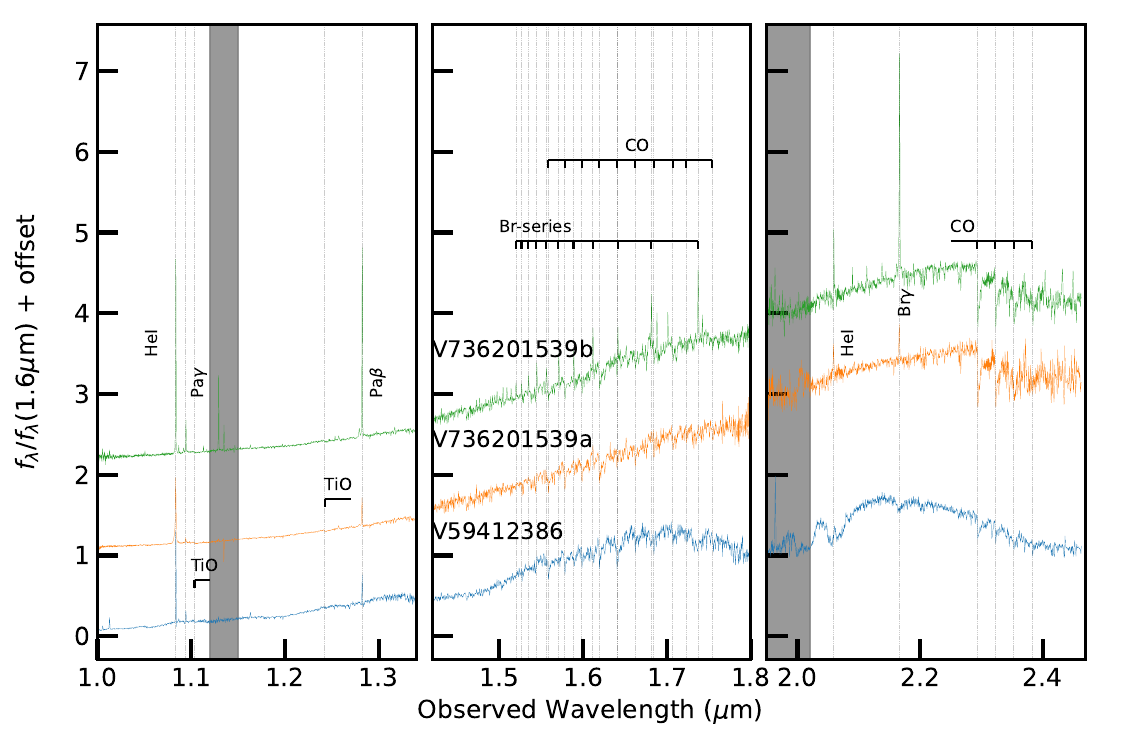}
    \includegraphics[width=0.95\textwidth]{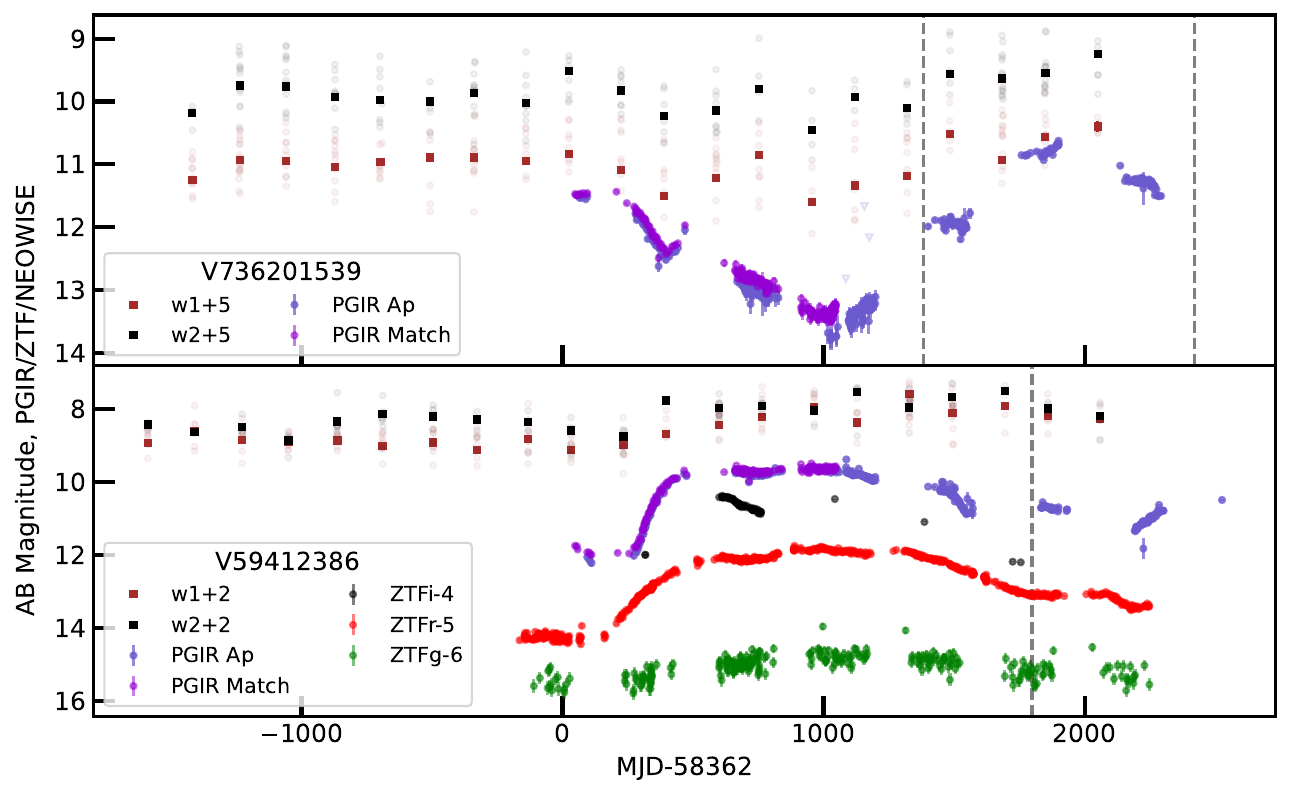}
\caption{V59412386 (IPHAS J193830.62+235438.4) is a confirmed D-type symbiotic star that was recovered in the census. Two spectra were acquired for V736201539 (IRAS 19148+1138). Both spectra exhibit a broad-based HeI-10833 and HI emission line profiles that strengthen 1037 days following the first spectrum, along with cool molecular absorption features in common with M-type giants. In contrast to the confirmed source, the new candidate symbiotic does not host optical detections down to 20.5 AB mag with ZTF.}
\label{fig:symbiotic}
\end{figure*}

\begin{figure*}
\centering
\includegraphics[width=0.95\textwidth]{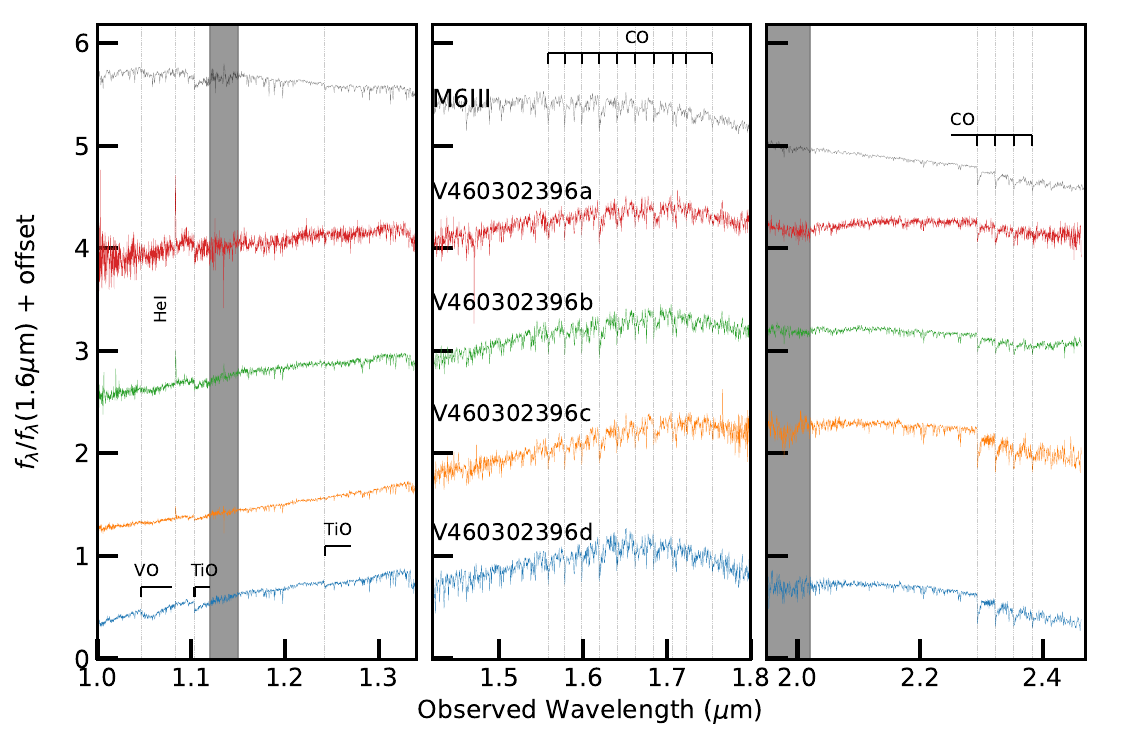}
\includegraphics[width=0.95\textwidth]{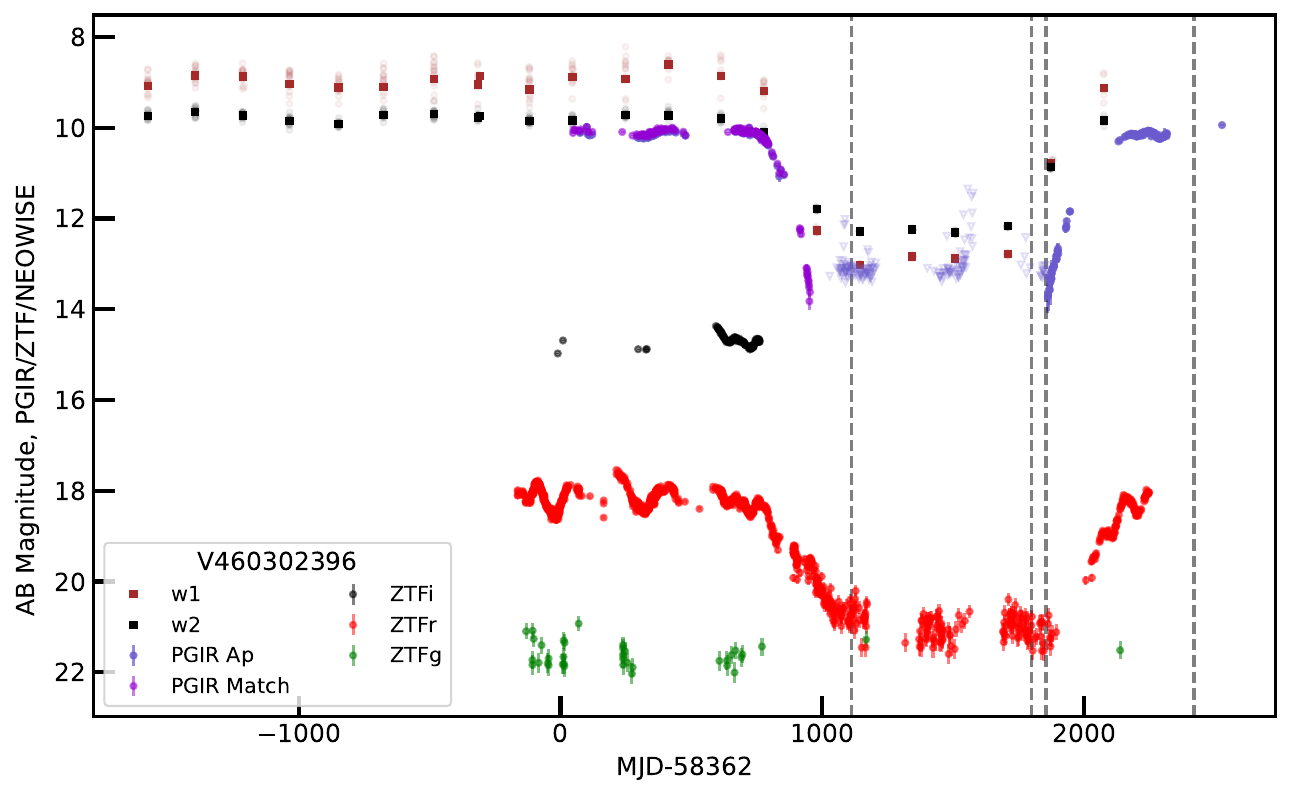}
\caption{Four spectra were obtained of V460302396 (Gaia 21aor) at various points in its evolution. During its low state, HeI-10833 emission weakens in subsequent spectra over time. The most recent spectrum upon return to its quiescent state indicates typical markers of O-rich giants. The topmost reference spectrum corresponds to HD196610, spectral type M6 III.}
\label{fig:He10833em0}
\end{figure*}

\begin{figure*}
\centering
\includegraphics[width=0.95\textwidth]{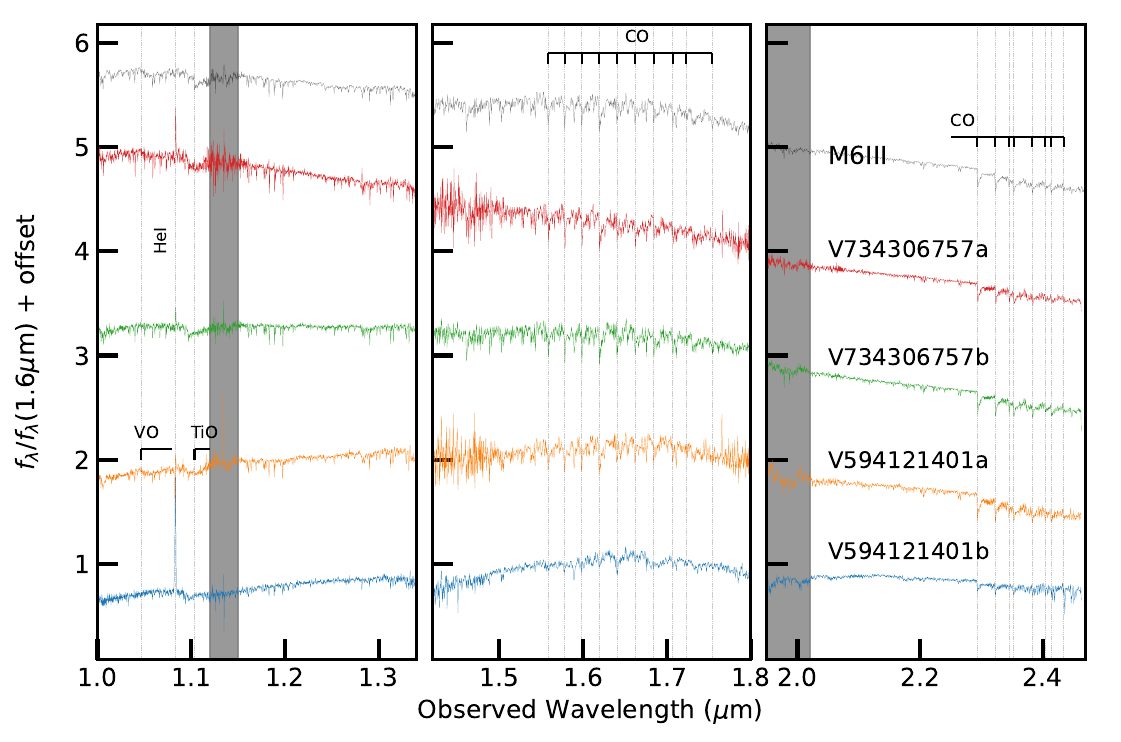}
\includegraphics[width=0.95\textwidth]{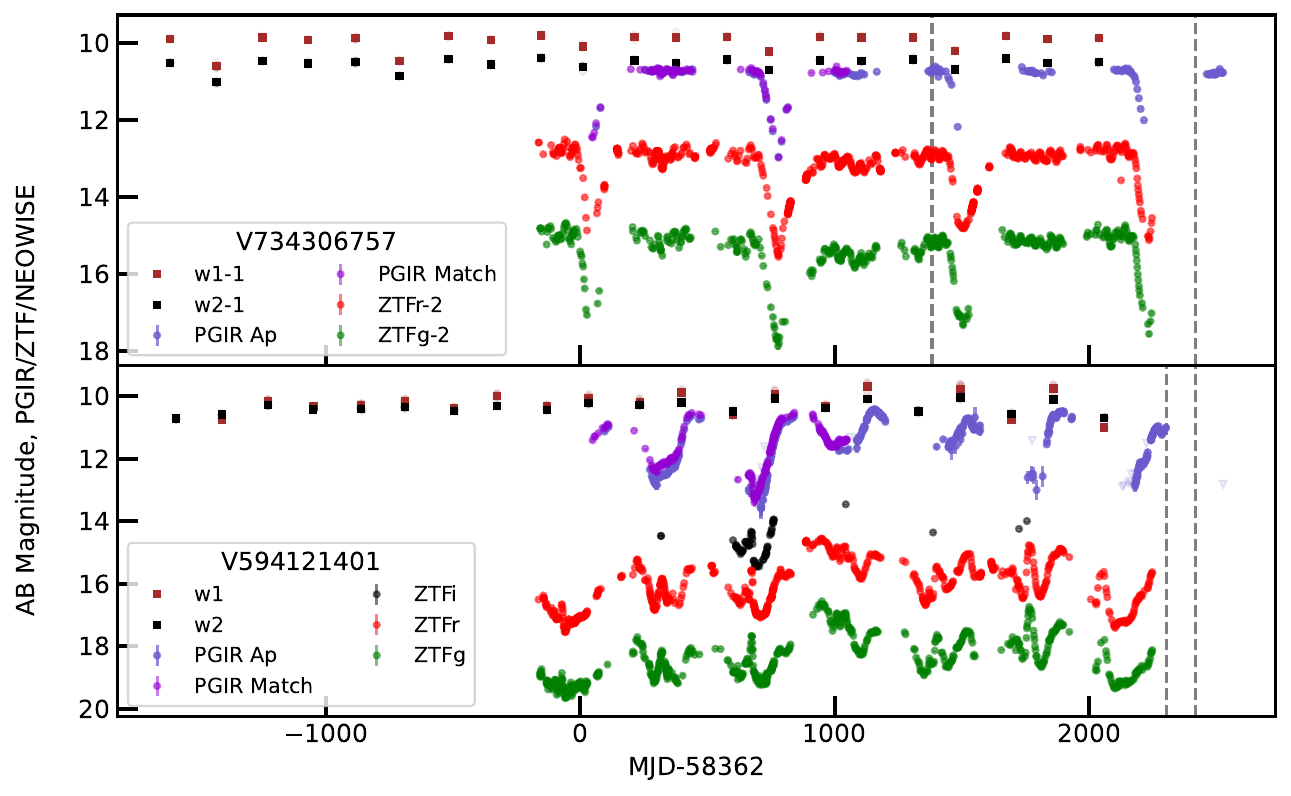}
\caption{Two M-type spectra were obtained of two sources exhibiting multiband periodic variability. Both sources exhibit markers of HeI-10833 emission along with cool molecular absorption features.}
\label{fig:He10833em1}
\end{figure*}

\begin{figure*}
\centering
\includegraphics[width=0.95\textwidth]{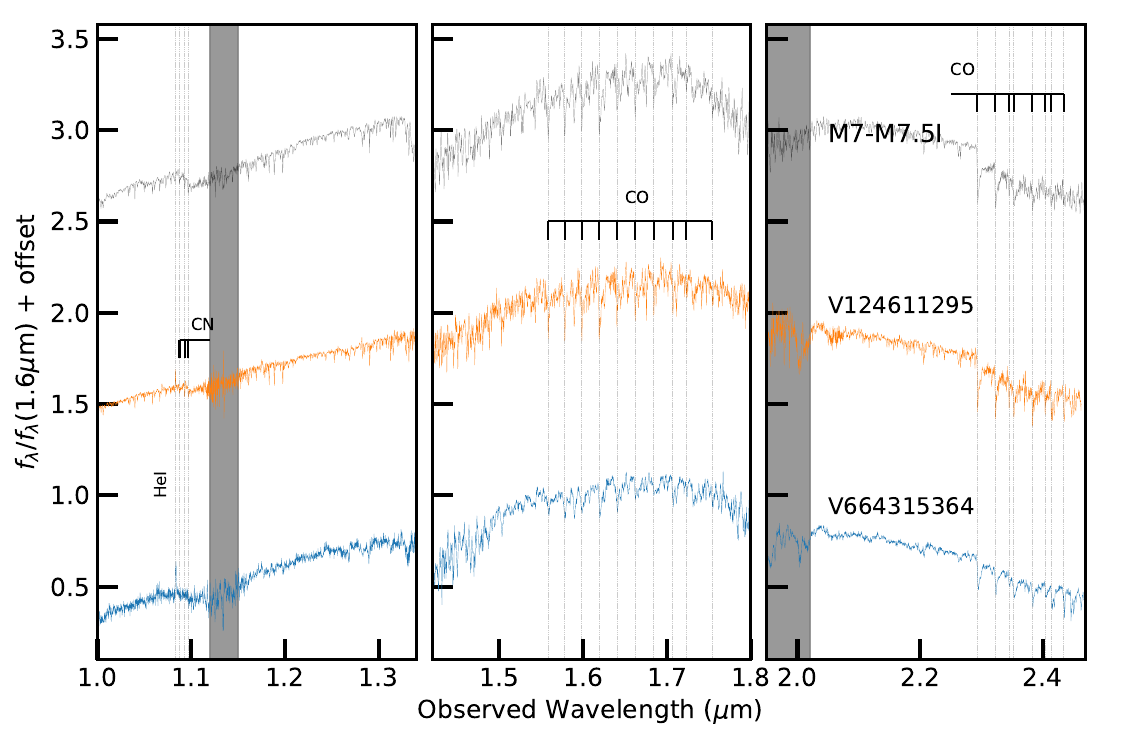}
\includegraphics[width=0.95\textwidth]{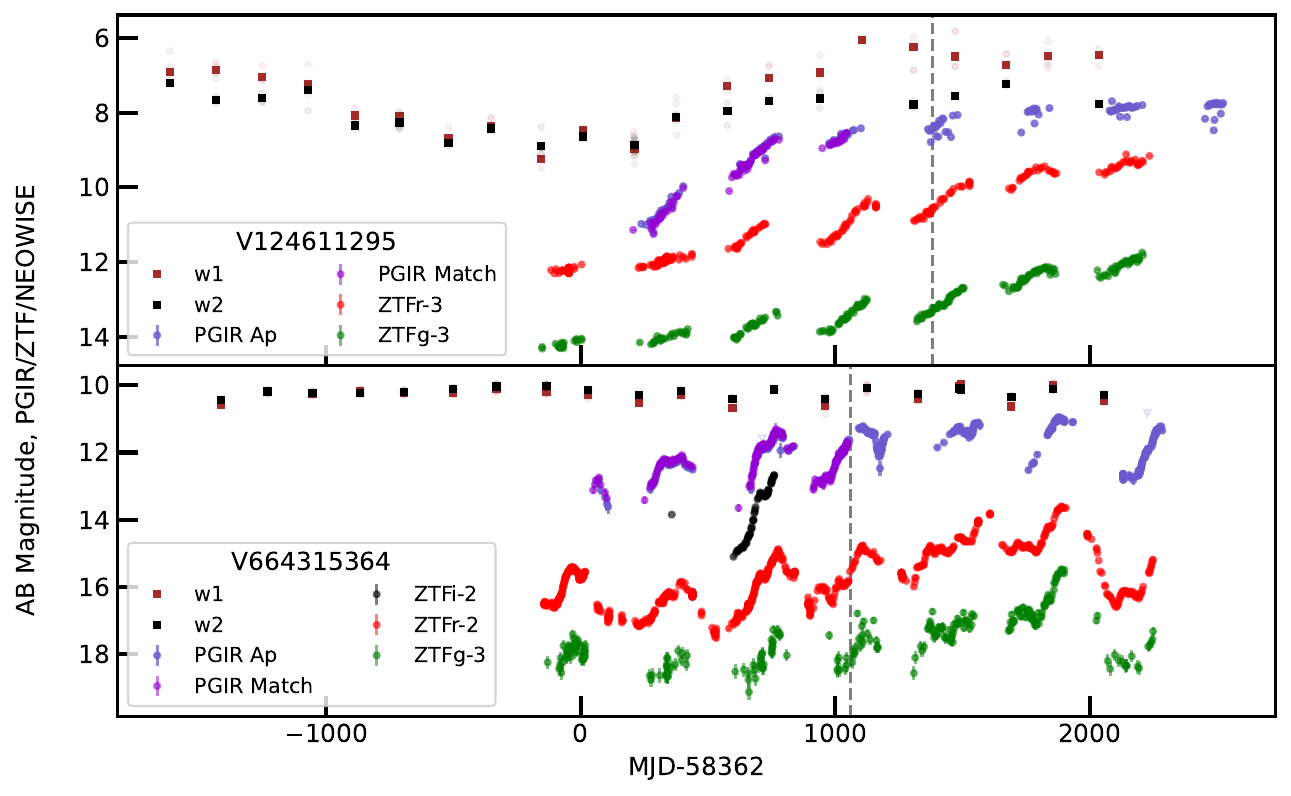}
\caption{Continued from Figure~\ref{fig:He10833em1}. 2 sources exhibit markers of HeI-10833 emission along with cool molecular absorption features similar to the supergiant reference star MY Cep (spectral type M7-M7.5I).}
\label{fig:He10833em2}
\end{figure*}

Symbiotic stars (SySts) are binary systems of evolved giants interacting with compact companions---generally white dwarfs (WD) or neutron stars (NS) \citep{Kenyon:1986syst.book.....K}. For a review, see \citet{2012BaltA..21....5M,Munari:2019arXiv190901389M}; and references therein. While stellar population synthesis models suggest there should be thousands of Galactic symbiotic variables \citep[e.g., $10^3-10^{5}$ depending on varying assumptions of evolutionary pathways,][]{Allen:1984PASA....5..369A,Magrini:2003ASPC..303..539M} only roughly 300 are currently known \citep{Akras:2019ApJS..240...21A,2019RNAAS...3...28M}. The majority of these binaries have thus far been selected from public catalogs (e.g., {\it Gaia}, 2MASS, {\it WISE}) and classified through low-resolution optical spectroscopy. Defining spectral features have historically included high-excitation emission lines (e.g., [FeVII] $\lambda6087$), strong H$\alpha$, and/or the OVI $\lambda\lambda6830,7088$ Raman-scattered lines in conjunction with molecular absorption features common to the photospheres of red giants (e.g., TiO, VO, and CN) \citep{Belczy:2000A&AS..146..407B,Akras:2019ApJS..240...21A}. 
Yet censuses premised on optical surveys can also be biased against the dustiest systems suffering from heavy extinction. Moreover, exclusively IR color-based selection criteria for SySts can suffer from contamination from semi-regular and long-period variable AGB stars or red giants \citep{Akras:2019MNRAS.483.5077A}.
NIR lightcurve-based large-amplitude selection criteria and prolonged NIR spectroscopic follow-up provide alternative and complementary discovery avenues. Many symbiotic O-rich Miras undergo sporadic, intense dust formation episodes producing erratic variations ($\geq 1$ mag) in their light curves \citep[e.g.,][]{Munari:1988A&A...200L..13M,De:2022ApJ...935...36D}. This higher-resolution NIR spectroscopic sample builds upon previous atlases \citep[e.g.,][]{SchulteLadbeck:1988A&A...189...97S} and is required for a detailed accounting of evolving spectral features that could be obscured at shorter wavelengths. 

We obtained spectra for 7 sources with erratic variability, which exhibit NIR signatures of high-excitation emission lines along with cool molecular absorption features (generally, deep CO bands and other markers of evolved giants such as TiO, VO, CN, or H$_2$O). All sources exhibit M-type spectra similar to the LPVs in Figure~\ref{fig:lpv_spec}---\ref{fig:lpvline1}, along with a prominent emission line coincident with the HeI--10833{ \AA} triplet at any point in our spectral coverage\footnote{The line corresponds to the $2^3\rm{P}_{2,1,0}-2^3S_1$ transition, with strong components at 1.0833217{\um} and 1.0833306{\um}, as well as a weaker component at 1.0832057{\um}, as measured in vacuo \citep[e.g., see][]{Geballe:2009ApJ...698..735G,NIST_ASD}. The line is also commonly referred to as HeI-10830 due to its lab frame measurements in air.}. For several sources, we obtained spectra at multiple timepoints to assess the time-evolution of the HeI signature. In a forthcoming work, we will detail modeling and additional follow-up for sources within this sub-sample.

While being a strong marker of hot stellar chromospheres in later G- and K-type stars as well as early type emission line and Wolf Rayet stars, the HeI--10833{\,\AA} emission line is rarer to similarly arise in evolved M stars due to the high photospheric temperatures needed for excitation \citep[$\sim 20$\,eV,][]{1968ApJ...152..123V, 1976ApJ...208..414Z}. However, in binary systems, HeI-10833{\,\AA} can arise in which the secondary is an M-type star \citep{1986ApJS...62..899O}. The HeI feature has also been attributed to collisional excitation via high-velocity winds around evolved giants \citep[albeit with smaller equivalent widths than for those in symbiotic systems,][]{Dupree:1992ApJ...387L..85D} as well as in disks and polar winds in young stars \citep[][evidenced by spectra presented in \S\ref{sec:yso}]{Edwards:2006ApJ...646..319E}. RCBs (\S\ref{sec:rcb}) are also known to host HeI--10833{\,\AA} signatures during periods of dust formation, in which the circumstellar media is shocked by the expanding dust shell which excites the transition \citep{Clayton:2013AJ....146...23C,Eyres:1999MNRAS.307L..11E}.
In all spectra within this subset of candidate symbiotic stars however, with the possible exception of V124611295 and V664315364 in Figure~\ref{fig:He10833em2}, there is an absence of carbon-rich molecular features such as CN and C$_2$ and a general preponderance of late M spectral types regardless of phase. The spectra within this class are generally distinct from the RCBs with HeI signatures in Figure~\ref{fig:rcbrisek}--\ref{fig:rcbflatk}.

The NIR spectra and OIR lightcurves are presented in Figure~\ref{fig:symbiotic}--\ref{fig:He10833em2}, and various properties for the corresponding sources are presented in Table~\ref{tab:syst}. Collectively, the photometric data of these targets are irregular and erratic. With the exception of one confirmed variable, the sources in this sample are not known in current symbiotic catalogs \citep{2019RNAAS...3...28M}. 

\begin{figure*}
\centering
\includegraphics[width=\columnwidth]{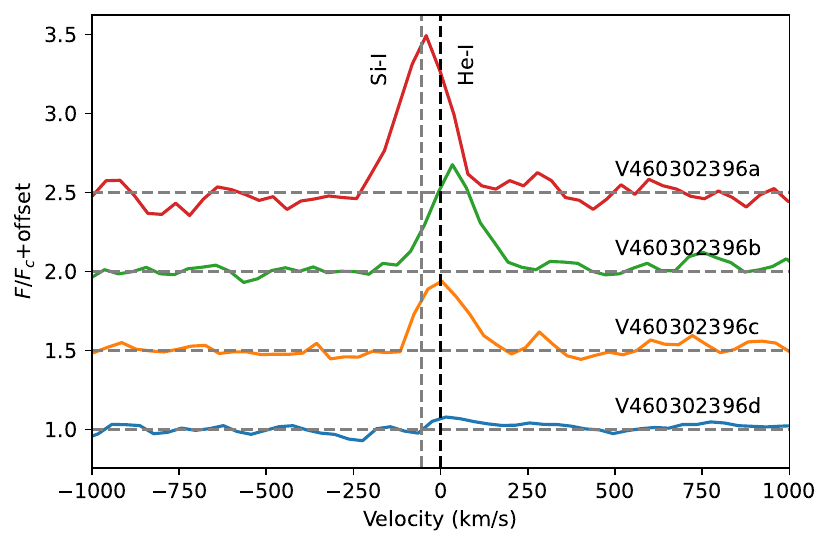}
\includegraphics[width=\columnwidth]{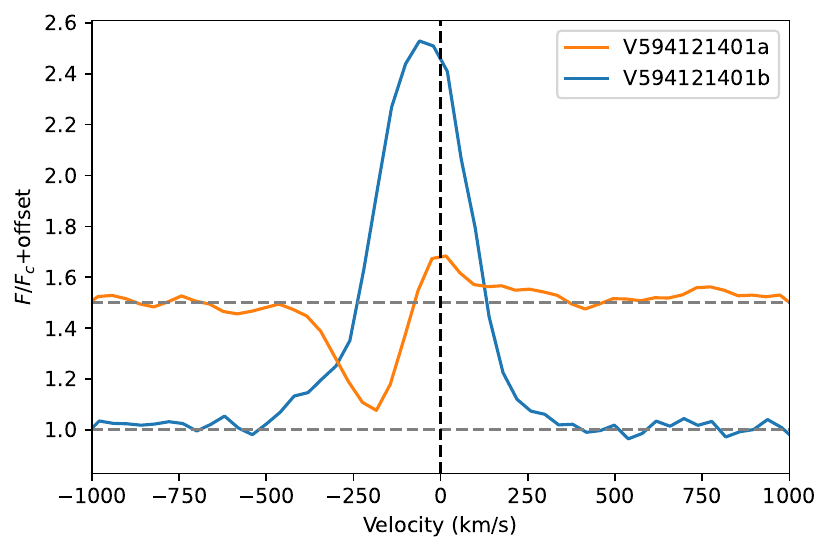}
\includegraphics[width=\columnwidth]{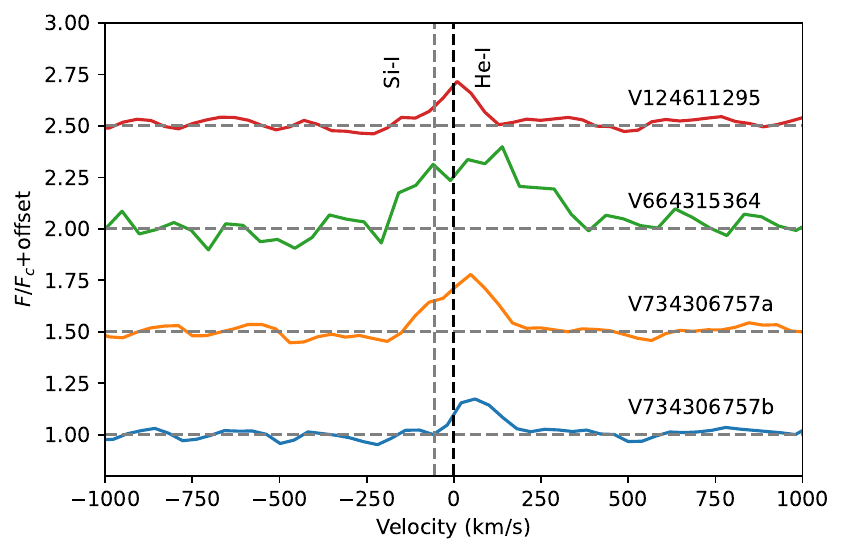}
\includegraphics[width=\columnwidth]{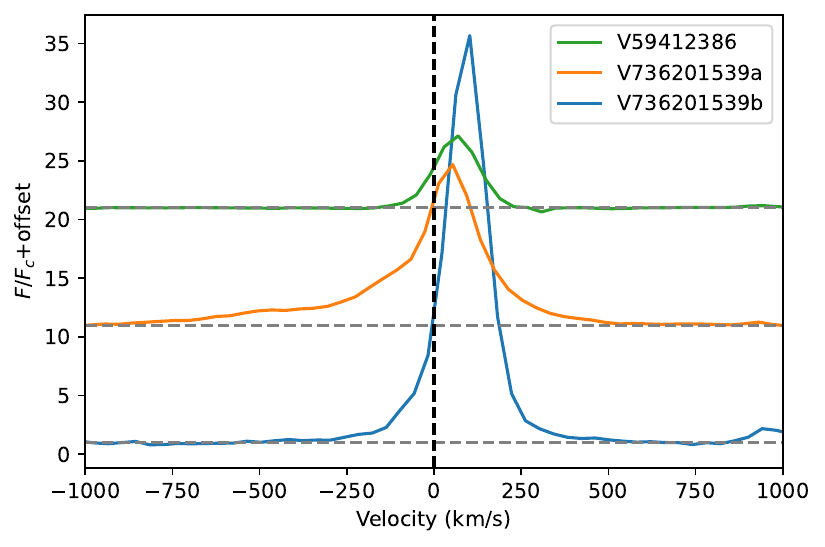}
\caption{The HeI-10833 line in velocity space, normalized with respect to the continuum flux $F_c$ for each of the candidate symbiotic sources in the census. {\it Top left:} A weak and evolving low-velocity ($<50$\,km\,s$^{-1}$) yet broad potential HeI signature is seen in V460302396 (Gaia 21aor). Four spectra were acquired at various timepoints spanning 1300 days, from the onset of its photometric decline to its return to quiescence. The common SiI-10831 line is also indicated. {\it Bottom left:} Same as top left, but for the other candidate low-velocity HeI-10833 emitters. Two spectra were acquired for the eclipsing binary V734306757. {\it Top right:} Two spectra of V594121401 were acquired 115 days apart. The HeI P Cygni profile evolves into a broadened emission line {\it Bottom right:} V59412386 is a known D-type symbiotic star. Two spectra were acquired for V736201539, a new high-probability candidate symbiotic star. The strong broadened profile in the first spectrum with its asymmetric base is suggestive of a strong stellar wind and mass transfer.}
\label{fig:heline}
\end{figure*}

In Figure~\ref{fig:heline}, we plot the HeI-10833 line in the observer rest-frame for each source, correcting for barycentric motion.
From the medium resolution spectra alone, we cannot definitively conclude whether a lower velocity HeI component is present without contributions from the SiI-10831 line, especially for the weaker signatures in the left panels of Figure~\ref{fig:heline}. Line formation via low-velocity winds is feasible given the escape velocities of red giants ($\sim 50$\,km\,s$^{-1}$), however the ionization mechanism is unclear. Ultimately, higher-resolution observations in the near-infrared are necessary to identify the source of the emission line. Spectral coverage in the optical could also reveal additional symbiotic-like signatures including the OVI Raman-scattered line. In the rightmost panels of Figure~\ref{fig:heline}, however, sources host exceptionally strong and broad signatures, with HeI line velocities of $\gtrsim 100$ km\,s$^{-1}$, and evolving P Cygni profiles. 

\begin{table*}
	\centering
	\caption{Properties of symbiotic stars or evolved HeI emitters}
	\label{tab:syst}
	\begin{tabular}{llllcccl}
		\hline
            PGIR ID & RA & Dec & Gaia DR3 ID & RUWE & Distance (pc) & Comments\\
		\hline
            V736201539 & 289.304 & 11.736 & 4312912322197813248 & 1.31 & 4769$^{+2026}_{-656}$ &  Candidate D-type SySt\\
            V59412386  & 294.628 & 23.911 & 2020286134278200704 & 1.0 & 1773$^{+545}_{-352}$ &  D-type SySt\\
            V460302396 & 301.940 & 32.792 & 2055104643862870400 & 1.101 & 4580$^{+615}_{-661}$ & Evolving HeI emitter\\ 
            V594121401 & 294.447 & 23.683 & 2020275860715618560 & 0.938 & 4769$^{+601}_{-730}$ & Candidate D-type SySt\\
            V734306757 & 278.687 & 14.321 & 4509023694336515712 & 1.22 & 6456$^{+562}_{-551}$ & HeI emitter\\
            V664315364 & 291.810 & 19.849 & 4515694706300129536 & 1.053 & 3496$^{+818}_{-618}$ & HeI emitter\\
            V124611295 & 278.758 & -19.591 & 4092442000143998080 & 1.395 & 2302$^{+977}_{-428}$ & HeI emitter\\
            \hline 
            \hline
	\end{tabular}
    \tablecomments{Same properties as presented in Table~\ref{tab:yso}. All sources host quality 2MASS photometry. Candidate D-type symbiotic stars are located in the D-type locus in the 2MASS $J-H$ vs. $H-K_s$ color diagram from \citet{Akras:2019MNRAS.483.5077A}, whereas the other sources lie closer to the border between S-type and D-type SySts.}
\end{table*}

We describe the individual sources in detail:

{\bf V59412386 (IPHAS J193830.62+235438.4)} is shown in Figure~\ref{fig:symbiotic} and is a D-type symbiotic star first detected in 2014 with IPHAS \citep{2014A&A...567A..49R, 2019ApJS..240...21A}. An optical spectrum is presented in \citet{2014A&A...567A..49R}; we extend the spectral coverage into the near-infrared. It is the only LAV in the census confirmed to be a symbiotic in known catalogs \citep{2019RNAAS...3...28M}. The spectrum hosts strong and broadened HeI emission.

{\bf V736201539 (IRAS 19148+1138)} is shown in Figure~\ref{fig:symbiotic} and hosts an exceptionally strong, asymmetric HeI-10833 emission profile with a broad base to the signature. The line profile suggests the presence of a strong wind and active mass transfer. The equivalent width of the feature exceeds $-180$\,{\AA}, in contrast to other sources in this subset hosting profiles on the order of $-10$\,{\AA}, which already exceed equivalent widths of the HeI profile in typical evolved giants \citep{Dupree:1992ApJ...387L..85D} by several orders of magnitude. It has been resolved in radio observations with the Very Large Array Sky Survey \citep[VLASS,][]{Lacy:2020PASP..132c5001L} and detected to integrated flux densities ($2-4$\,GHz) of $S_\nu=2.4\pm0.3$\,mJy \citep[VLASS1QLCIR J191713.00+114409.3,][]{Gordon:2021ApJS..255...30G}. 
At a distance of $d=4.8^{+2.0}_{-0.7}$\,kpc \citep[Gaia DR3 4312912322197813248,][]{2021AJ....161..147B}, the radio luminosity $L_\nu = 4\pi d^2 S_\nu =0.7^{+1.5}_{-0.4}\times10^{20}$\,erg\,s$^{-1}$\,Hz$^{-1}$, typical for optically thick free-free emission in SySts \citep[e.g.,][]{Seaquist:1990ApJ...349..313S,Ivison:1995MNRAS.273..517I,Mikolajewska:2002AdSpR..30.2045M,Dickey:2021ApJ...911...30D}. 
Radio emission at this level is known to arise as a result of the photoionization of the cool secondary stellar wind by the hot primary \citep{Taylor:1984ApJ...286..263T}. While CO absorption is prominent, V736201539 does not host similarly strong metal oxide absorption markers in its $J$-band spectrum. This could be attributable to contamination from an accretion disk that enhances blue continuum emission, reducing the apparent depth of the TiO feature. Such an effect is well-documented in symbiotic systems \citep[e.g.,][]{Chakrabarty:1997ApJ...489..254C,Corradi:2010A&A...509L...9C,Sonith:2023MNRAS.526.6381S}. Given the presence of the strong counterpart radio source, broad and intense HeI signature along with cool molecular absorption features, V736201539 is a strong symbiotic candidate, which has not been detailed before.

{\bf V460302396 (Gaia 21aor)}, shown in Figure~\ref{fig:He10833em0} and reported in the Gaia alert stream in 2021 \citep{2021TNSTR.317....1H}, faded nearly 1000 days ago by at least 3 mags in the infrared, bounded by upper limits, and rebrightened to become detectable again nearly two years later. Four spectra were acquired, one during the onset of the deep fade, two slightly before PGIR's re-detection, and the last at the onset of the rebrightening. The spectral evolution charted demonstrates a weakening of the HeI-10833{ \AA} emission line (top left, Figure~\ref{fig:heline}) and increased CO absorption over the course of $\sim 3.5$ years. The source corresponds to Gaia DR3 2055104643862870400 with extinction-corrected $BP-RP = 3.21$ and absolute magnitude of $G=-2.58$, presuming distances of 4.58\,kpc, consistent with the reddest AGB stars. 2MASS source coordinates are 0.166" from Gaia source coordinates.

{\bf V734306757 (2MASS J18344488+1419162)}, shown in Figure~\ref{fig:He10833em1}, is a source hosting a 100 day duration dimming event, with a period of $\sim 700$ days. The regularity of the dimming events across the optical to infrared strongly suggests that the source is an eclipsing binary. A weak potential HeI-10833 is seen in emission along with cool CO absorption bandheads and an M-spectral type in two spectra taken 1000 days apart.

{\bf V594121401 (2MASS J19374724+2340578)}, shown in Figure~\ref{fig:He10833em1}, was misidentified as an LPV in \citet{2024PASP..136j4501M} \citep[i.e., it is not included in the PGIR LPV catalog of][]{aswin:2024PASP..136h4203S}. The spectrum hosts a HeI-10833 P Cygni profile which evolves into a broad ($\Delta v\sim 500$\,km\,s$^{-1}$) emission feature 115 days later (top right, Figure~\ref{fig:heline}). 

{\bf V664315364 (ATO J291.8096+19.8492)}, shown in Figure~\ref{fig:He10833em2}, was discovered by the Asteroid Terrestrial-impact Last Alert System \citep[ATLAS][]{Tonry:2018PASP..130f4505T} in 2018 and identified as an LPV in its first catalog of variable stars \citep{Heinze:2018AJ....156..241H}. The multiband lightcurve exhibits semiregular variability and the spectrum exhibits typical markers of late M-type stars in addition to a broad emission feature coincident with HeI-10833.

{\bf V124611295 (IRAS 18320-1937)}, shown in Figure~\ref{fig:He10833em2}, exhibits a smoothly varying lightcurve, with gradual large-amplitude brightening in PGIR and ZTF photometry. Archival NEOWISE data shows the source's period exceeds 2000 days, well in excess of typical LPVs. A weak and low velocity HeI-10833 line is seen in emission. The source has been previously characterized as a carbon star \citep{Abia:2022A&A...664A..45A}. While it exhibits CN bandheads in $J$-band, which late M-type stars can also host, the NIR spectrum does not share other similarities with the C-rich LPVs in Figures~\ref{fig:hbandflatk0_spec}---\ref{fig:cstar_spec}, such as C$_2$ and the 1.53{\um} absorption feature.

In addition to this class of confirmed and candidate symbiotic binaries, we include the classical nova V282302182 \citep[V2891 Cyg/PGIR19brv,][]{De:2021ApJ...912...19D}. We do not present additional spectroscopy or photometry for this known source.

\subsection{Erratic giants and RV Tauri supergiants}
\begin{figure*}[htbp]
    \centering
    \includegraphics[width=\textwidth]{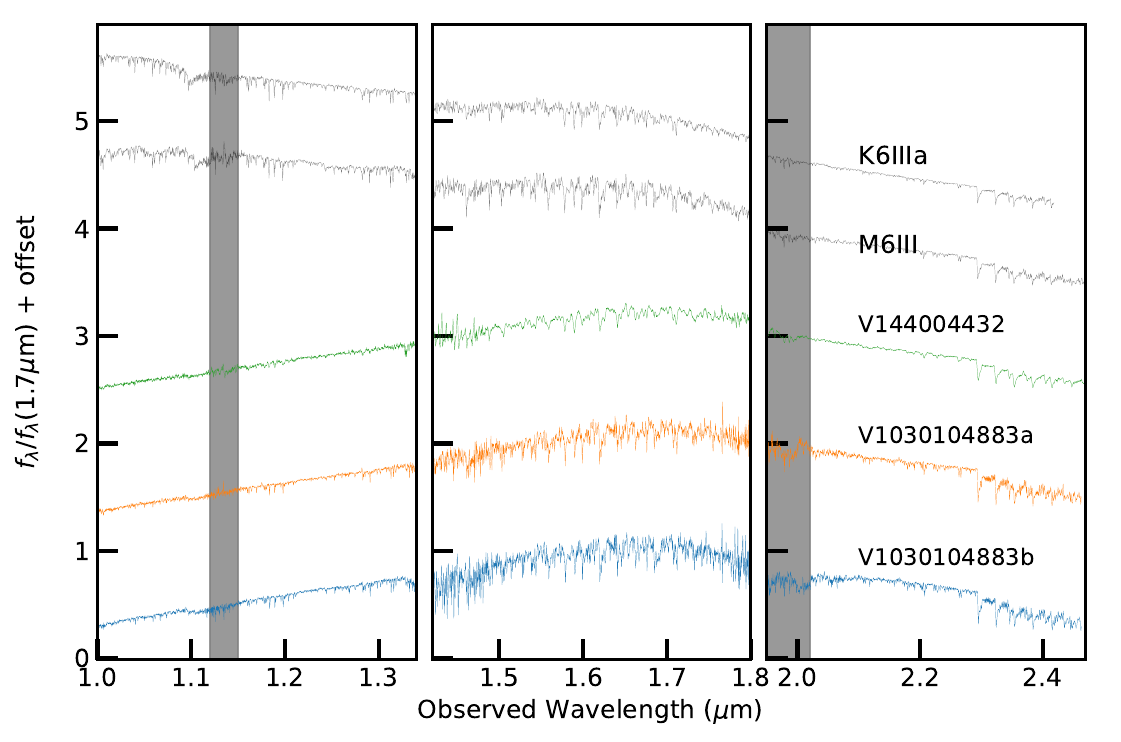}
    \includegraphics[width=\textwidth]{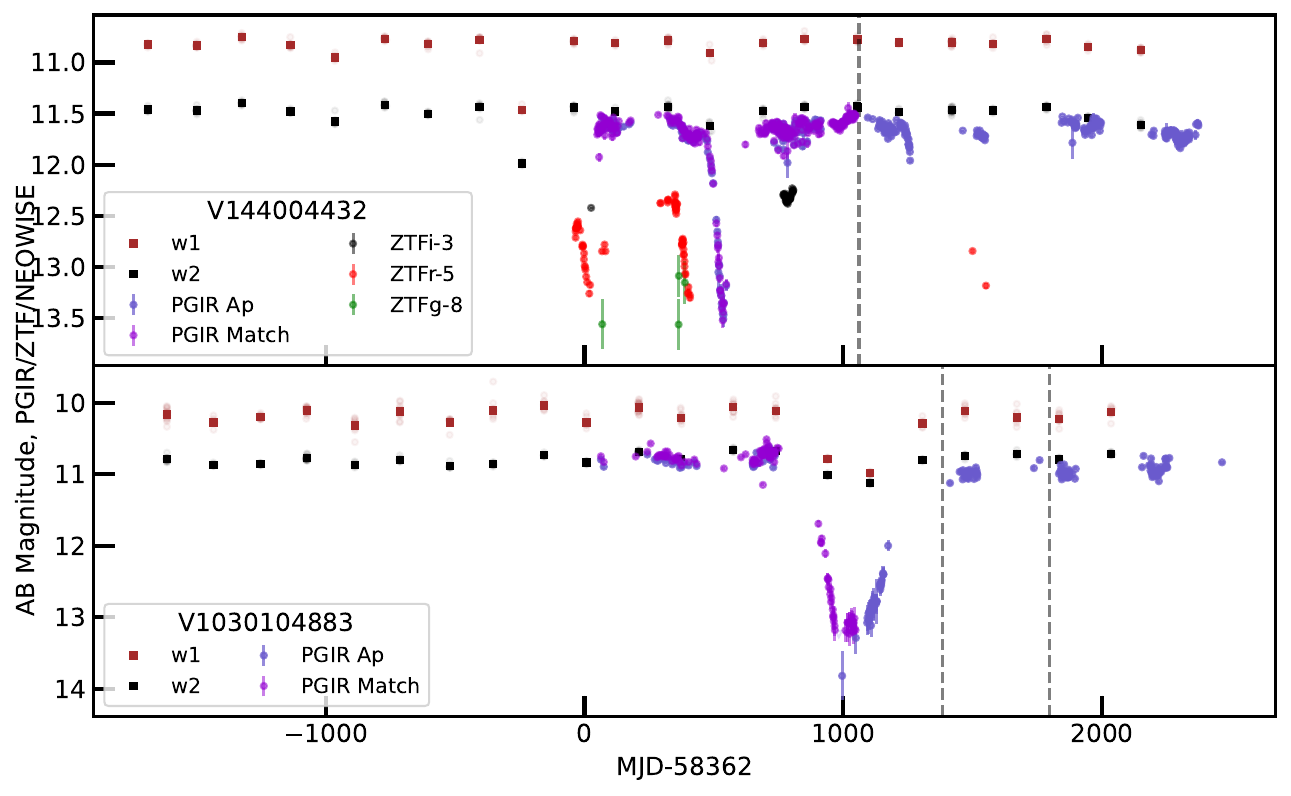}
    \caption{A small subset of variables exhibit erratic dimming phases with oxgyen-rich spectral features at maximum light. The absorption lines are shared by K-type and M-type giants. 2 spectra were obtained of V1030104883 exhibiting minimal spectral evolution in quiescence. No spectra were acquired during periods of minimum brightness. The reference stars HD196610 (M6III) and HD3346 (K6IIIa) are plotted at the top.}
    \label{fig:erratic}
\end{figure*}

\begin{figure}[htbp]
    \centering
    \includegraphics[width=\columnwidth]{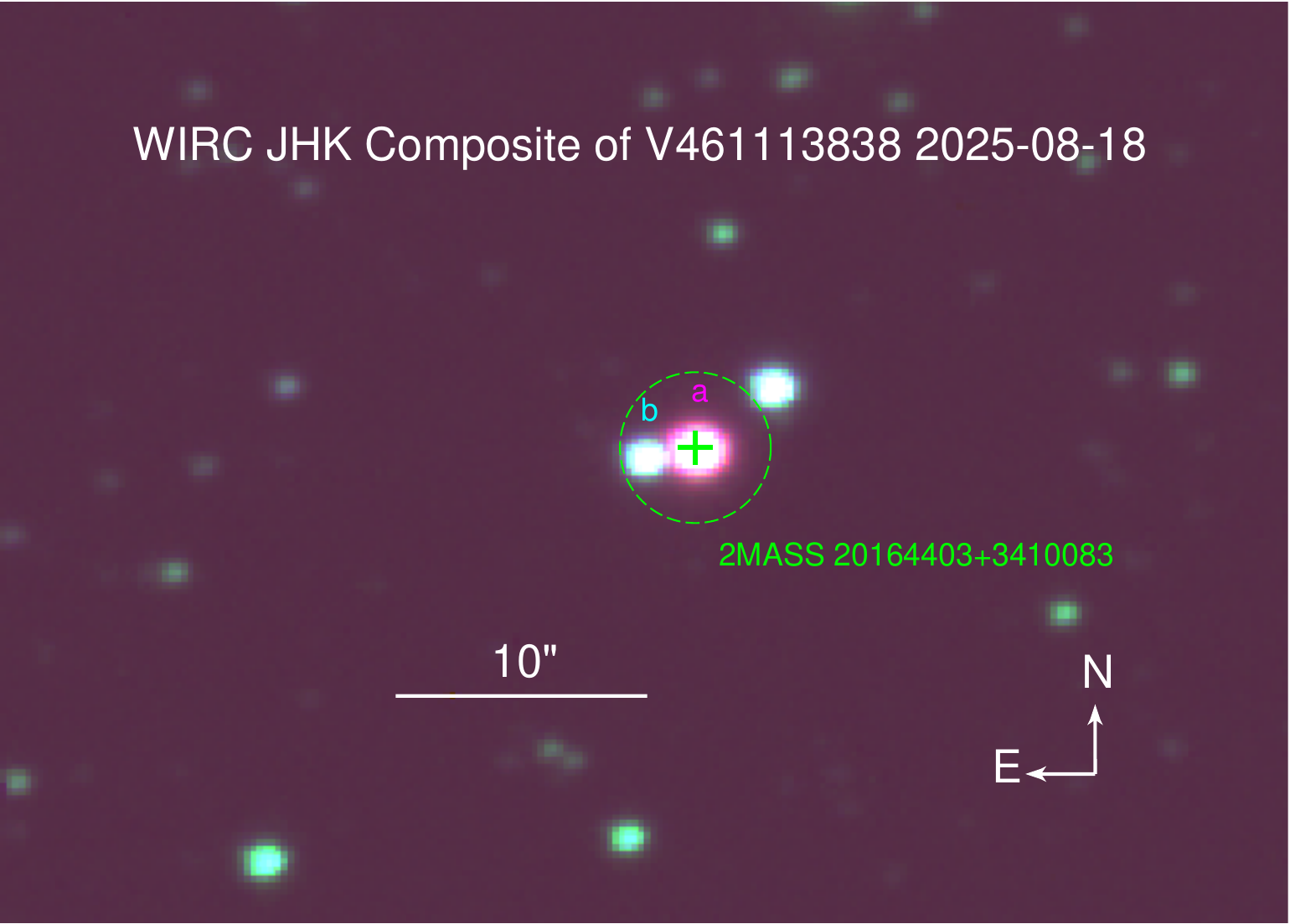}
    \includegraphics[width=\columnwidth]{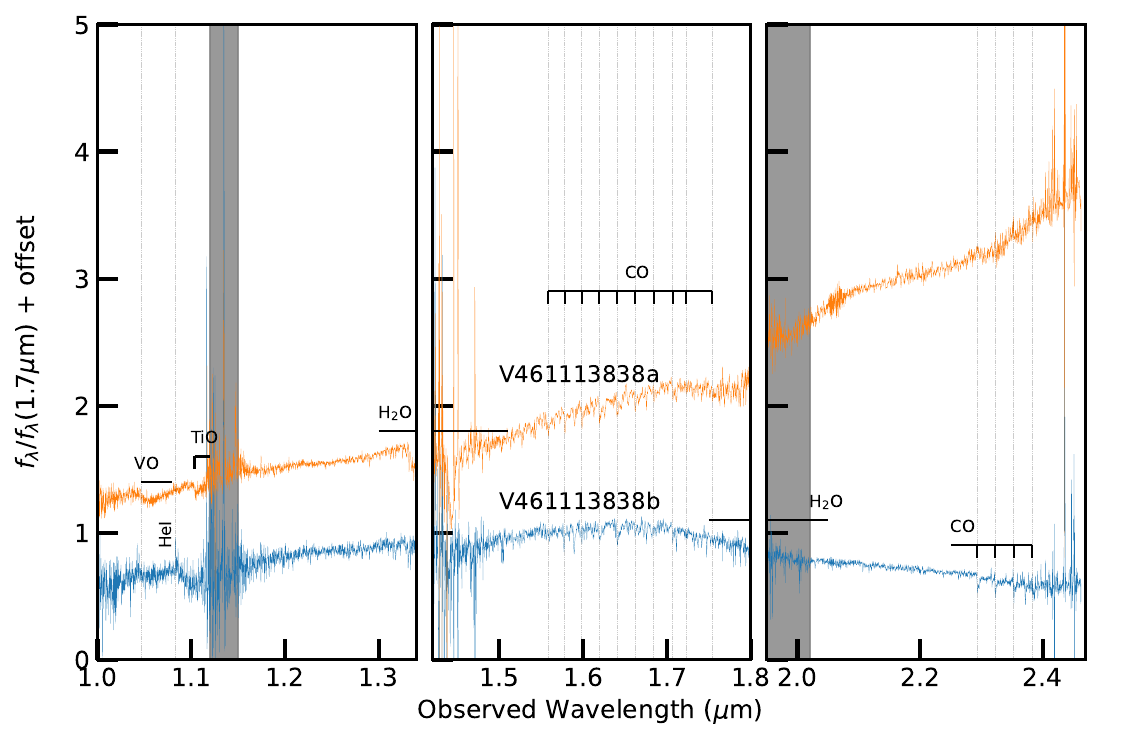}
    \includegraphics[width=\columnwidth]{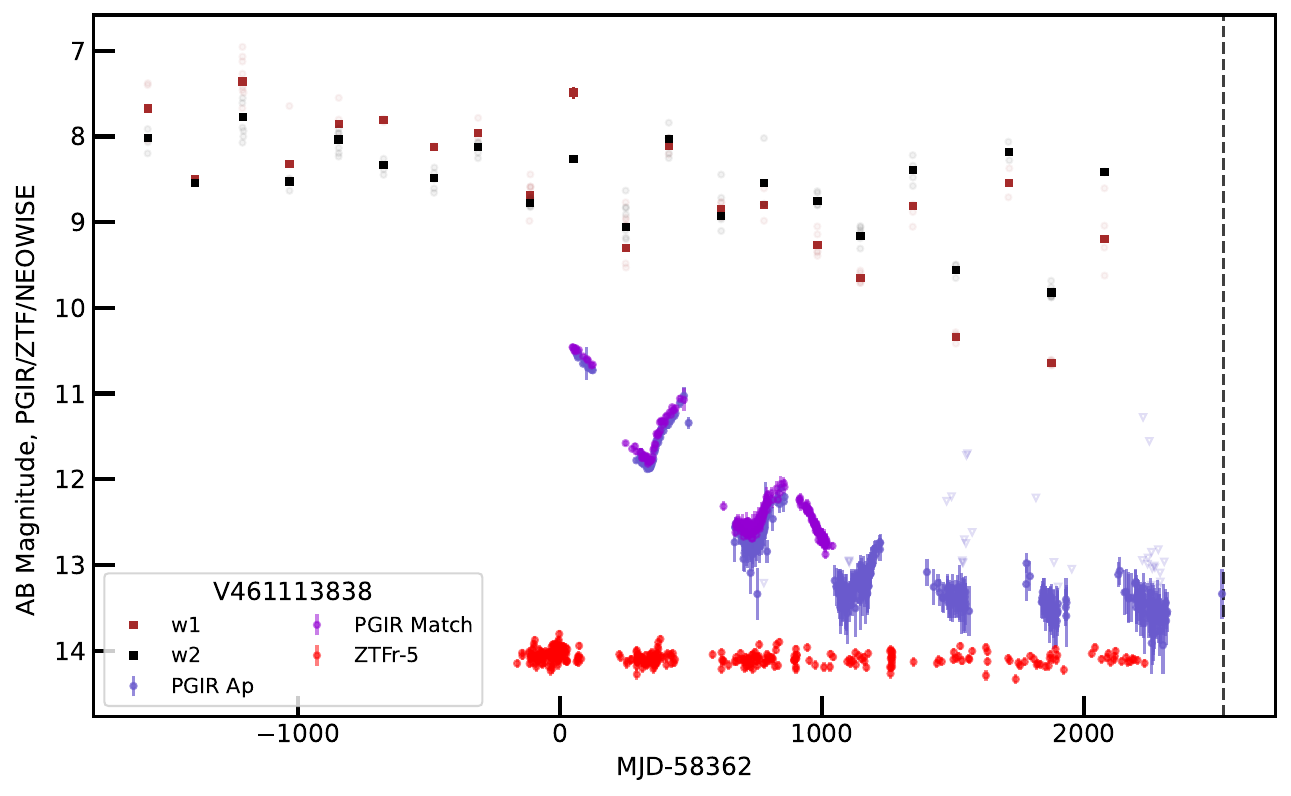}
    \caption{Top: WIRC NIR composite false-color (green $J$, blue $H$, and red $K_s$), image of the field of the PGIR large-amplitude variable V461113838. The coordinates of the 2MASS counterpart 2016440+3410083 are indicated by the cross. An additional source is located within the dashed region centered on the 2MASS source coordinates (3" radius). Middle: Spectra obtained of the two sources with TripleSpec ten days earlier. The primary source (\emph{a}) is marked by a rising $K$-band continuum and the other (\emph{b}) hosts a HeI signature along with CO absorption bands. Both spectra exhibit cool molecular markers common to M/K-type giants as shown in previous classes. Bottom: The lightcurve is atypical compared to other O-rich LPVs, exhibiting reddening of $W1-W2$ at the onset of the decline in the $J$-band lightcurve.}
    \label{fig:erratic_bad}
\end{figure}

\begin{figure*}[htbp]
    \centering
    \includegraphics[width=\textwidth]{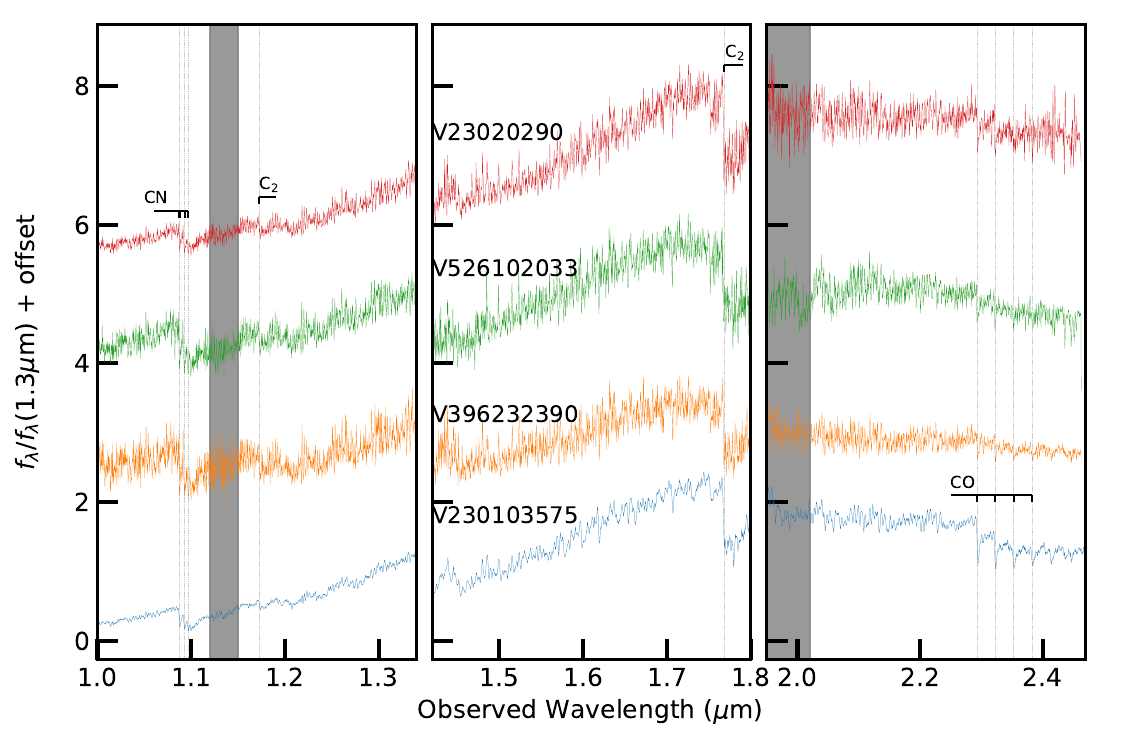}
    \includegraphics[width=\textwidth]{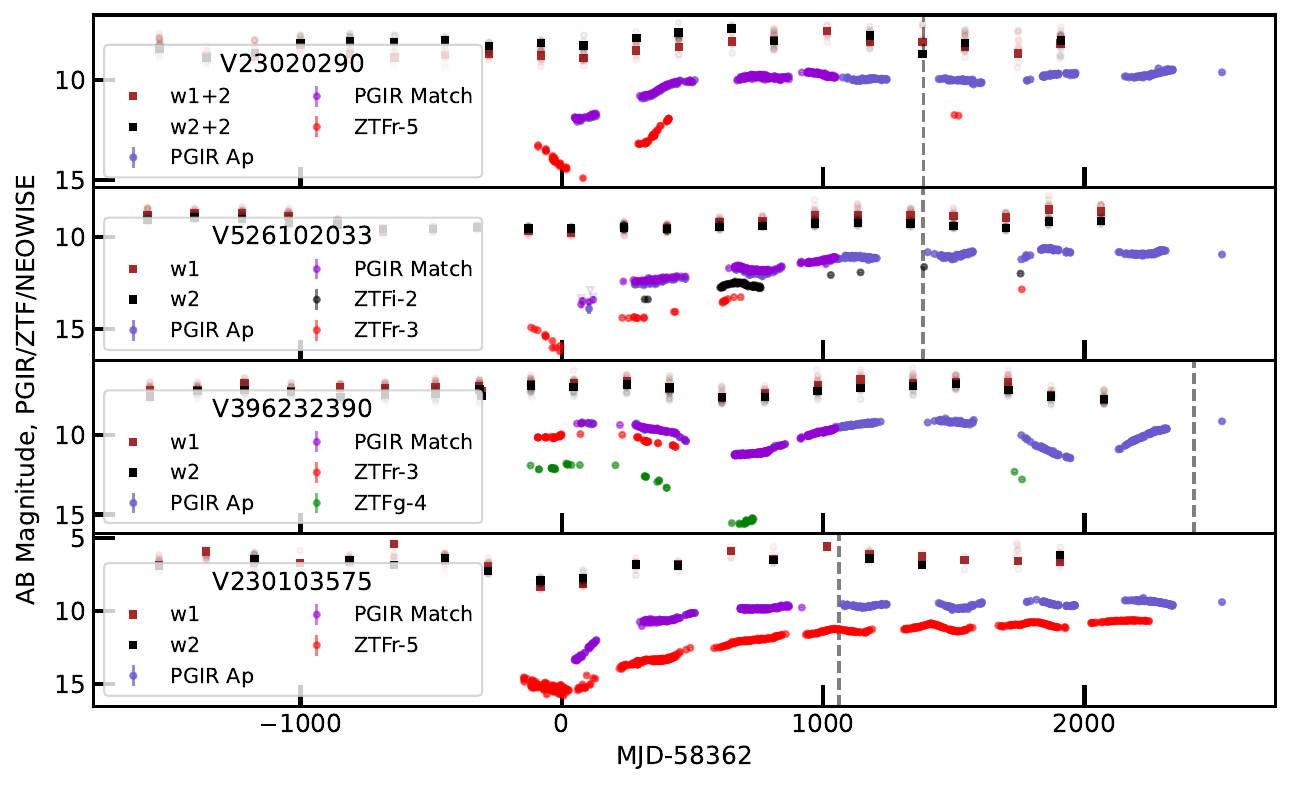}
    \caption{A small subset of erratic variables exhibit typical absorption markers common to the C-rich LPVs in Figure~\ref{fig:hbandflatk0_spec} and \ref{fig:hbandflatk1_spec}, or the RCBs at or near maximum light in Figure~\ref{fig:rcbflatk}, such as CN, C$_2$, and CO, without prominent emission line signatures. The lightcurves demonstrate extremely long-period ($>1000$ days) variability marked by shallow, symmetric dimming events (V396232390), or smoothly rising NIR lightcurves with evidence of prior symmetric declines in the optical (V23020290, V526102033, V230103575).}
    \label{fig:erratic_crich}
\end{figure*}

\begin{figure}[htbp]
    \centering
    \includegraphics[width=\columnwidth]{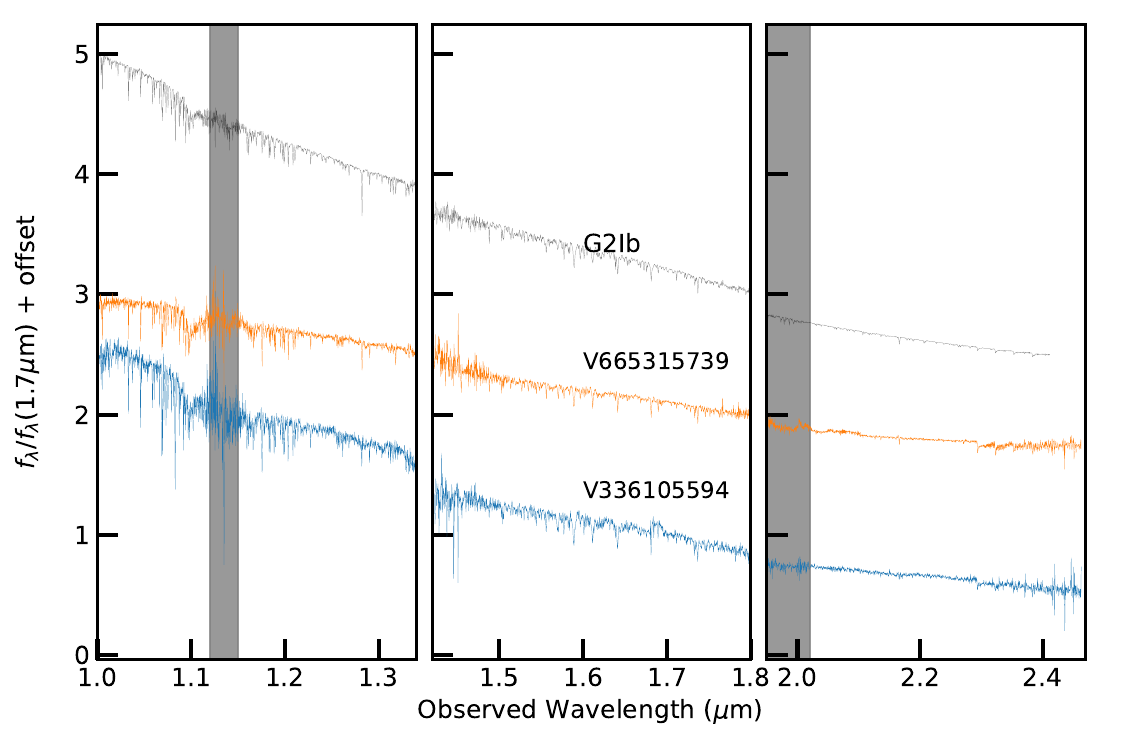}
    \includegraphics[width=\columnwidth]{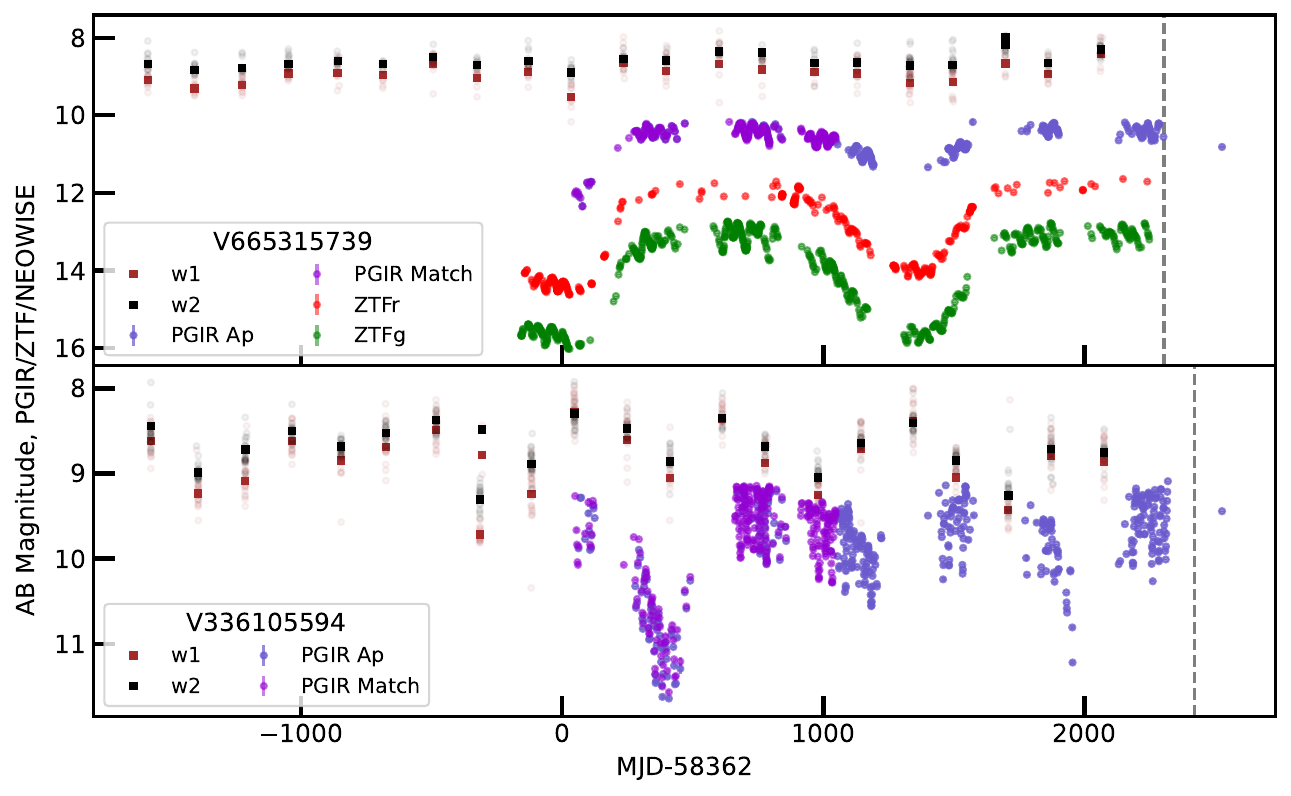}
    \caption{The supergiant RV Tauri stars DF Cyg (V336105594) and RZ Vul (V665315739)}
    \label{fig:rvtauri}
\end{figure}

The remaining stellar variables within the census include known RV Tauri supergiants and ill-defined erratic objects with common O-rich or C-rich spectral properties.

Three sources in the census include stars that compare well with IRTF spectral templates of M- or K-type stars and host erratic variability in their multiband photometry. These stars experienced sudden dimming events, with generally 100 day durations, by up to 4 mags. The erratic events are in contrast to the smoothly varying or semiregular lightcurves of LPVs in the previous subclass. The stellar spectra collectively have deep molecular features, predominantly due to water and CO absorption throughout the H and K bands, with markedly diminished or absent TiO and VO absorption bands. Furthermore, the sources do not host any strong emission lines.

{\bf V1030104883 (2MASS J18300017-0539288)}, presented in Figure~\ref{fig:erratic}, has a deep and broadly asymmetric dimming/rebrightening event (3 mag variation from its quiescent state), roughly 1000 days following PGIR's first light. Two spectra were collected, one in the latter end of the rebrightening phase and the other in the succeeding quiescent state at 11 mag in J, both closely comparing with M-type stars. Outside of slightly weakening absorption features in the $J$-band in the later spectrum, the spectra are not markedly different, suggesting that the features are not strongly phase dependent at these time points. 

{\bf V144004432 (2MASS J23272528+5917136)}, also presented in Figure~\ref{fig:erratic}, experienced two dimming events of $1.5-2$ mags in the $J$-band separated by nearly 1000 days. There is prior history of dimming in the optical via ZTF $r$-band photometry, which precede the dimming in the NIR. Its spectral type compares well with K-type stars with a featureless, linearly rising $J$-band continuum and CO absorption features in the $H$- and $K$-bands.

\textbf{V461113838 (IRAS 20147+3400)} is presented in Figure~\ref{fig:erratic_bad} and hosts a semiperiodic, dimming $J$-band lightcurve, with the $J$-band brightness decreasing by more than 3 magnitudes since the beginning of PGIR observations. Evidence of substantial reddening in the color evolution of the MIR lightcurve is contemporaneous with the NIR decline. An additional source is located within $\sim2$" of the closest 2MASS counterpart.  When spectra were obtained, the slit angle was adjusted such that the slit was perpendicular to the vector connecting the centroids of the source positions to minimize contamination in the spectral extractions. The primary source, closest to the 2MASS counterpart coordinates (designated V461113838a in Figure~\ref{fig:erratic_bad}), hosts cool molecular features in common with M/K-type giants, yet with a rising $K$-band continuum and the absence of CO absorption bandheads. A spectrum was also obtained of the nearby source (designated V461113838b), which exhibits cool molecular features along with an emission line coincident with HeI-10833, similar to the spectra of sources in \S\ref{sec:symbiotic}. Imaging of the field of V461113838 was obtained on UT 2025-08-18, ten days after spectra were obtained, using the Wide-field Infrared Camera \citep[WIRC,][]{Wilson:2003SPIE.4841..451W} on the 200-inch telescope at Palomar Observatory.

A variety of mechanisms can drive erratic variability in cool M-type red giants. For targets that have seemingly one-off or irregular dimming events in the current dataset, the targets may simply be components of eclipsing binary systems, with periods potentially ranging from three to five years or longer. Exceedingly long eclipsing binary systems with M-type red giants have been proposed for these one-off dimming events \citep[e.g.,][]{Smith:2021MNRAS.505.1992S, Tzanidakis:2023ApJ...955...69T} Continued NIR photometric monitoring and further transit modeling would uncover these optically hidden eclipsing binaries in the Galactic plane. Moreover, optically thick dust shells ejected by the M-type host star could also contribute to the irregular dimming events and provide insight into dust production phases for these stars. 

Four sources, shown in Figure~\ref{fig:erratic_crich}, exhibit C-rich spectral features in common with the LPVs detailed in \S\ref{sec:lpv}, yet atypical lightcurves with extremely long-period variability, shallow and symmetric dimming events, and/or gradual large-amplitude behavior. The spectral features and photometric behavior are similar to the RCBs in \S\ref{sec:rcb}. {\bf V230103575 (BC 279)} , {\bf V396232390 ([NC50] 6)}, and {\bf (IRAS 21210+4922)} have been proposed as candidate Galactic DY Per stars \citep{viraj:2024PASP..136h4201K}. It is debated whether DY Per stars constitute a subset of cold RCBs ($T_{\rm eff} \sim 3500$ K), or a distinct class of dust-forming carbon-rich, hydrogen-deficient giants \citep[e.g.,][]{Alksnis:1994BaltA...3..410A,zav:2007A&A...472..247Z,Garcia:2023ApJ...948...15G,Crawford:2025MNRAS.537.2635C}. {\bf V526102033 (IRAS 19437+2812)} is an additional candidate that has not been characterized before.  While sharing similar carbon-rich features as RCBs, including $J$-band CN absorption bandheads, the C$_2$ absorption at 1.77{\um}, and varying degrees of CO absorption in the $K$-band, none of the stars in this subset exhibit He-rich features such as the HeI-10833 line. In normal RCBs, the He line arises from the radiative acceleration of dust dragging He gas along with it \citep{Clayton:2013AJ....146...23C}. The absence of the He line in these sources may be due to lower luminosities of these stars compared to typical RCBs. Alternatively, the signature can be suppressed in the presence of circumstellar dust.

Two supergiant G-type RV Tauri stars also satisfied the census selection criteria. The properties of {\bf V336105594} \citep[{\bf DF Cyg},][]{Harwood:1927BHarO.847....5H} and {\bf V665315739} \citep[{\bf RZ Vul},][]{Tsessevich:1977IBVS.1371....1T} are shown in Figure~\ref{fig:rvtauri}. DF Cyg is a post-RGB binary \citep{Manick:2019A&A...628A..40M}. RZ Vul in particular has been identified of interest due its lightcurve exhibiting simultaneous modes of rapid periodicity and smooth large-amplitude variation.

\section{Projections for future infrared census campaigns}
\label{sec:future}
\begin{figure*}
\centering
\includegraphics[width=\columnwidth]{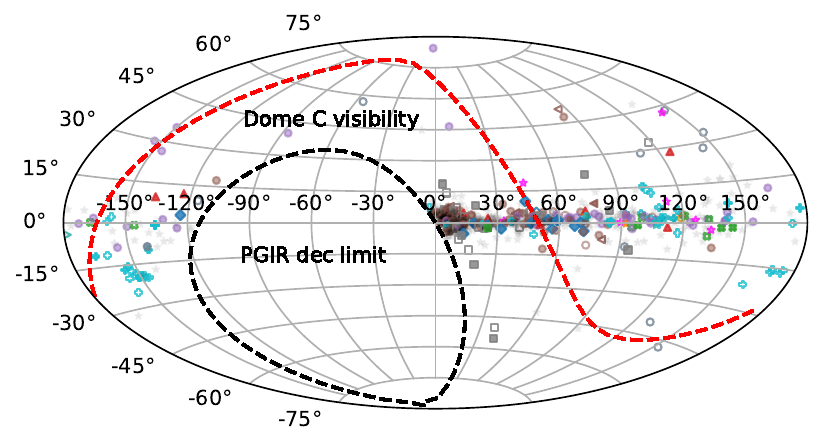}\includegraphics[width=1.2\columnwidth]{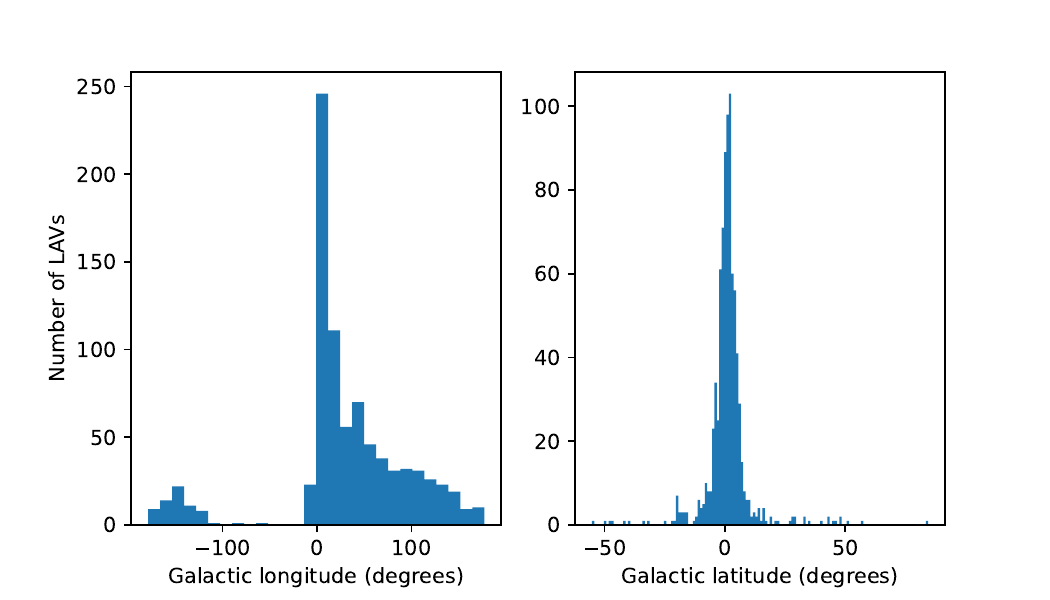}
\caption{Left: Future infrared surveyors in the Southern Hemisphere will extend the census of LAVs in terms of depth and sky coverage. Visibility limits are shown for the planned Cryoscope Pathfinder at Dome C, Antarctica \citep{cryoscope:2025PASP..137f5001K} which is expected to survey Galactic coordinates below the red dashed line to point-source sensitivities of $\sim 19$ mag in the $K_{dark}$ bandpass between $2.2-2.5$ microns ($3-5$ mags deeper than PGIR in $J$-band). The PGIR LAV census sources are colored the same as in Figure~\ref{fig:census}. Right: The histogram of Galactic coordinates, binned according to Knuth's rule. The spatial distribution of LAVs is sharply peaked and clustered in the Galactic center which will be better sampled by a Southern surveyor.}
\label{fig:domec}
\end{figure*}
The PGIR LAV census has aimed to create a repository of long-lived LAVs visible from the Northern sky in the $J$-band with $ptp>2$ magnitudes. Depths were largely limited to $\sim 13$ magnitude for the majority of targets located within the Galactic plane. As evidenced by the color-diagrams in \S\ref{sec:demographics}, MIR colors are optimal for identifying carbon-rich AGBs and YSO LAVs. While certain populations are clustered in the 2MASS $J-H$ and $H-K_s$ (and associated combinations), contaminants within these color-cuts plague spectroscopic follow-up. All-sky MIR coverage via SPHEREx \citep{SPHEREX} will be essential for filling in gaps left with WISE, providing low-resolution spectral classification for the entirety of the PGIR LAV sample.

Future high-cadence, wide field-of-view IR surveyors are setting the stage for deeper and more complete campaigns. The PRime-focus Infrared Microlensing Experiment \citep[PRIME,][]{SumiPRIME:2025arXiv250814474S} located at the South African Astronomical Observatory is a 1.8\,m, 1.3{\sqdeg} active field-of-view telescope with coverage in the $z$, $y$, $J$, and $H$ bandpasses ($0.83-1.8${\um}), that has recently begun operations. The Dynamic REd All-sky Monitoring Survey \citep[DREAMS,][]{SoonDREAMS:2020SPIE11203E..07S} based at the Siding Spring Observatory in Australia is a 0.5\,m, 4{\sqdeg} field-of-view telescope with similar photometric coverage in the $y$, $J$, and $Hs$ bandpasses ($\sim 0.9-1.7${\um}). Together, the two telescopes will be able to deliver comprehensive censuses of stellar variables from the Southern Hemisphere, thereby complementing the PGIR survey from the north. Moving into even longer wavelengths such as the $K$-band will, however, yield the greatest gains in assembling a complete census of the dustiest and coldest stellar variables. 

Cryoscope Pathfinder \citep{cryoscope:2025PASP..137f5001K} is a wide-field infrared cryogenic telescope for Antarctica. The telescope leverages a novel optomechanical design to instantaneously image $\sim$16\,deg$^2$ on-sky in the $K_{dark}$ passband which falls between the last airglow lines at 2.35{\um} and the onset of water absorption at 2.5{\um}. In the following year, it will be installed at Dome C in Antarctica where it will benefit from an extremely low sky brightness \citep{1996PASP..108..721A, 1996PASP..108..718N}, ideal photometric conditions \citep{2010A&A...511A..36C,2018A&A...619A.116C}, and continuous winter nights from the pole. At 2.35{\um}, Cryoscope Pathfinder will be well positioned to monitor the coldest stars with effective temperatures $\sim$ 1200\,K and the dustiest circumstellar environments. In nominal winter survey operations, the Pathfinder will be able to cover the entire Antarctic sky (20,000 deg$^2$) every 4.5 days to a limiting $m_{AB} \sim 16-18$ in the $K_{dark}$ bandpass, depending on confusion noise in the Galactic plane.

Based on the spectra presented in the census, the majority of the variables have integrated fluxes of $\sim10^{-10}$\,erg\,cm$^{-2}$\,s$^{-1}$ in $K_{dark}$, corresponding to $m_{AB} \sim 10$. The Cryoscope Pathfinder will thus be well suited to complement PGIR and the $J$-band campaign presented above, as well as future slightly deeper $yJH_s$ coverage with PRIME and DREAMS, which will fill in the Southern hemispheric gaps in the LAV distribution (Figure~\ref{fig:domec}). Cryoscope Pathfinder will be primed to extend PGIR's Galactic census of LAVs, not only in terms of achieving greater limiting depths, but also in expanding coverage of the Galactic Bulge. 

\begin{figure*}
\centering
\includegraphics[width=0.5\textwidth]{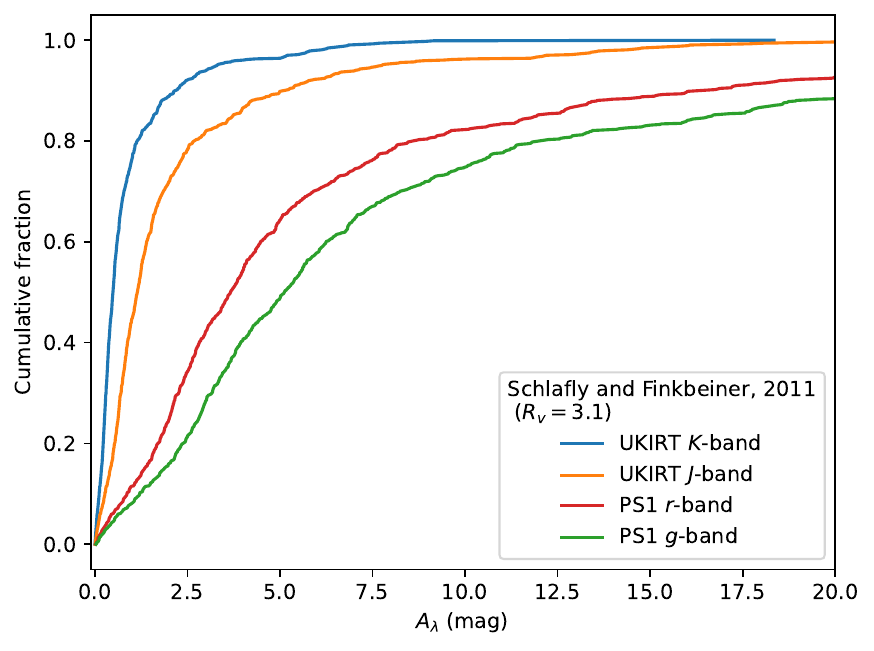}\includegraphics[width=0.5\textwidth]{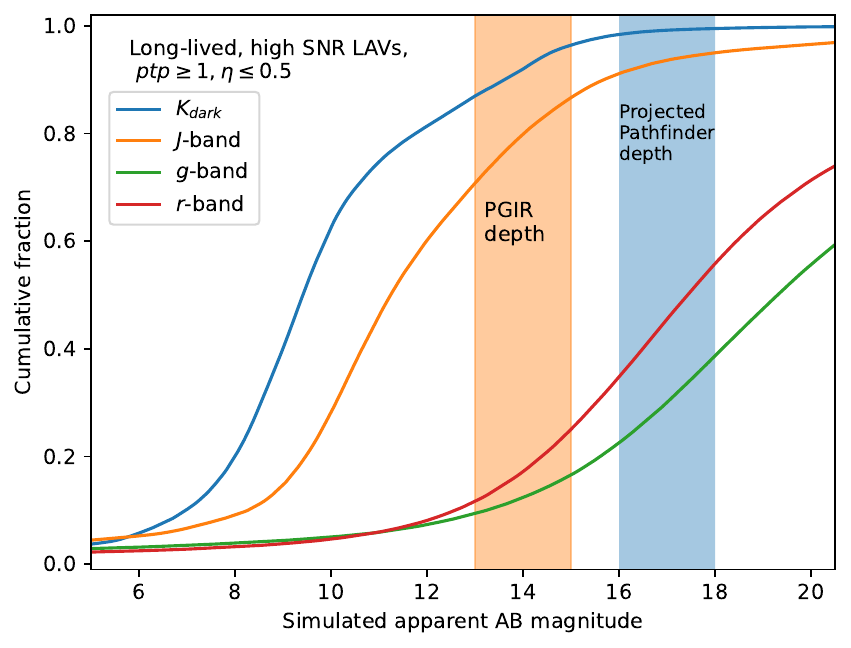}
\includegraphics[width=0.5\textwidth]{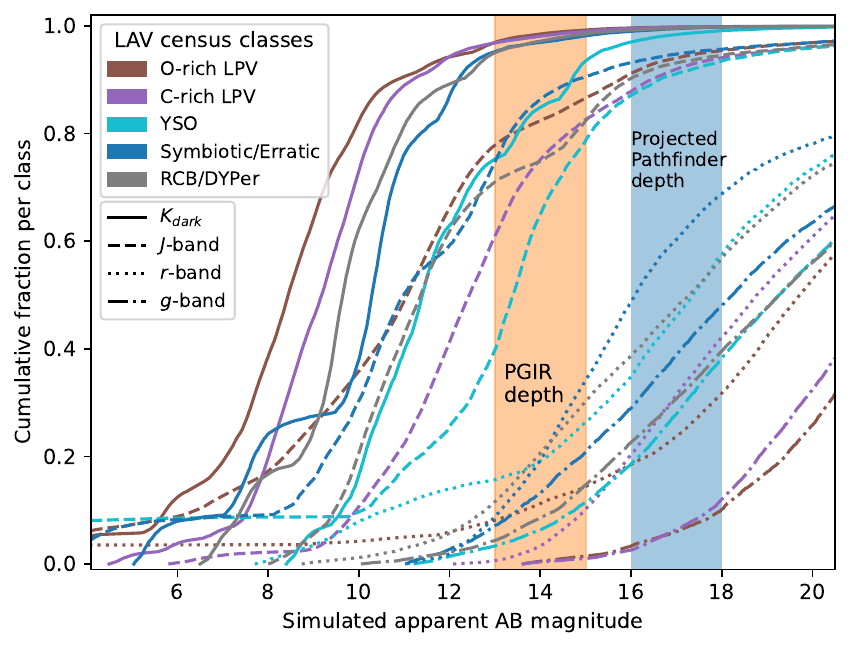}
\caption{{\it Top left}: Cumulative fraction of bandpass-specific extinction values computed up to $A_\lambda= 20$\,mag at the source coordinates for all long-lived, high SNR PGIR LAVs with $ptp\geq1$ and $\eta \leq0.5$. Line-of-sight extinction values are derived from
\citet{Schlafly:2011ApJ...737..103S}, which presume the reddening law from \citet[]{Fitzpatrick:1999PASP..111...63F} with $R_V = 3.1$. 
NIR surveys are optimized for LAV censuses due to the reduction in line-of-sight extinction by several magnitudes as wavelength coverage moves from the optical to $K$-band. {\it Top right}: A simulated cumulative fraction of apparent magnitudes for all long-lived, high SNR LAVs in each band. Nominal sensitivities for PGIR and the $K$-band survey Cryoscope Pathfinder, including variations in depth from confusion noise in the Galactic plane, are indicated by shaded regions. A modestly deeper $K$-band survey would conservatively uncover $20-30\%$ more LAVs. {\it Bottom}: Same as top right, but cumulative fraction per class for the smaller sample of 128 sources in the census. All classes are expected to be fully detectable with a deeper survey in the $K$-band.}
\label{fig:extinction_cdf}
\end{figure*}

In the top left panel of Figure~\ref{fig:extinction_cdf}, we show the cumulative fraction of the bandpass-specific line-of-sight extinction values for the entire PGIR LAV population with $ptp \geq 1$ and $\eta \leq 0.5$, which includes sources outside of the census selection criteria. We employ line-of-sight extinction values from \citet{Schlafly:2011ApJ...737..103S} derived from \citet{Schlegel:1998ApJ...500..525S} $E(B-V)$ maps.

Because of the novelty of Cryoscope Pathfinder's $K_{dark}$ filter, we compute the $K_{dark}$ magnitudes of all stars in the detected sample by convolving the $K_{dark}$ filter response with the observed spectral flux densities. Due to the decreased atmospheric transmission, and thereby decreased SNR, redward of 2.47{\um}, we extrapolate the pseudo-continuum from $2-2.47${\um} in each spectrum to 2.6{\um} when integrating the flux density over the bandpass. Since the sources are variable by definition, the distribution of source magnitudes represents a single snapshot in time of the census population. For sources outside of the census selection which we did not acquire spectra for, we use 2MASS $JHK_s$ and NEOWISE median $W1$ and $W2$ photometry when available to construct SEDs that we interpolate to approximate LAV $K_{dark}$ magnitudes. These interpolated magnitudes are best approximated by 2MASS $K_s$ measurements. Approximating the PGIR LAV spatial distribution (e.g., the sky coordinates for all sources with $ptp \geq 1$ and $\eta \leq 0.5$) as a representative tracer of stellar positions, we construct a uniformly distributed set of 1000 sources in each of the observed census classes at a random location from the complete LAV sample. The apparent magnitudes of the simulated sources depend on the extinction at each location on-sky. Assuming that the sources approximately trace this spatial distribution, and thus extinction distribution, we compute the cumulative distribution fraction of simulated apparent magnitudes for each class of objects in the census (top right panel, Figure~\ref{fig:extinction_cdf}). 
We similarly approximate extinction corrections in $K_{dark}$ as equivalent to the corrections in $K_s$. For more than half of the sources in the full sample, quality ZTF photometry is not available and thus only the cumulative distribution of detected optical magnitudes are shown. The near-infrared is the ideal wavelength range for a complete census of LAVs. Based on the observed population with PGIR, we would have discovered $20-30\%$ more LAVs missed in the current sample via a $K$-band survey with the Cryoscope Pathfinder ($\gtrsim 1000-1200$ LAVs as opposed to 838 in the $J$-band). 

In the bottom panel of Figure~\ref{fig:extinction_cdf}, we plot the cumulative fractions of the possible apparent magnitudes for census objects. We combine certain classes with limited numbers such as the symbiotic and erratic stars as well as RCBs and DYPer stars. All PGIR LAVs under the same criteria, regardless of extinction or confusion noise in the plane, would be recovered with a deeper $K$-band survey. Moreover, from the pole, Cryoscope Pathfinder would be better positioned to sample the Galactic Bulge than a Northern $J$-band survey, which would reveal an even greater trove of variables, those with source magnitudes well in excess of PGIR's shallow depth. Over the same three-year baseline, it is thus expected that Cryoscope Pathfinder will conservatively observe $1050 - 1200$ long-lived, high SNR LAVs with amplitudes exceeding 1 magnitude, $\eta \leq 0.5$, and source magnitudes $\leq 15$ AB mag, extending PGIR's brightness-limited campaign to the Southern Hemisphere.

\section{Conclusion}
With the emergence of wide and fast time-domain surveyors across the electromagnetic spectrum, we are primed for the most accurate censuses of stellar variables in the coming decade, uninhibited by high extinction or limited on-sky areal coverage. Whereas previous surveys have had to contend with such dust extinction or surveys of limited fields-of-view given detector formats, dedicated surveyors are primed to overcome these obstacles. 

In preparation for these projects, we have conducted a shallow, synoptic survey by leveraging the robotic Palomar Gattini-IR telescope as a discovery engine for large-amplitude stellar variables in the $J$-band over a three-year baseline. A complete spectroscopic campaign was mounted to characterize the largest-amplitude and longest-duration variables in PGIR match files (e.g., $ptp \geq2$, $\eta \leq0.2$, $>50$ detections, and $>5$ high-amplitude detections exceeding the minimum brightness by more than 1 mag). Spectra were obtained with medium resolution spectrographs TripleSpec on the 200-inch telescope at the Palomar Observatory and SpeX on the NASA Infrared Telescope Facility. As a result, we have generated an extensive photometric and spectral atlas of near-infrared large-amplitude stellar variables in the $J$-band. 

AGB stars form the bulk of the census, divided into O-rich long period variables, some of which are extremely long period variables with counterpart maser emission, and carbon-rich long-period-variables, some of which exhibit a 1.53{\um} absorption feature indicative of the presence of HCN and C$_2$H$_2$ in their atmospheres. As detailed extensively in previous works, RCBs and DYPer stars are also present in the census. Remaining sources include young stellar objects such as FUOr, EXOr, objects with UX Ori-like variability, and HAeBe variables. One confirmed symbiotic variable was recovered in the census, and several candidate stars exhibit prominent high-excitation emission lines along with cool molecular features. We propose that sustained NIR surveys aim to discover missing symbiotic variables in Galactic census estimates. Some sources with typical M/K-type spectral properties exhibit irregular erratic variability in their timeseries data. 

With SPHEREx, NEO Surveyor, and the Cryoscope Pathfinder, current and future IR surveys aim to further expand areal coverage and depth at longer wavelengths. As a simple exercise into the utility of moving into longer wavelengths, we simulate the distribution of possible apparent magnitudes for sources in the PGIR LAV sample based on their line-of-sight extinction distributions. As a consequence of the higher line-of-sight extinction in the $J$-band relative to longer wavelength surveys, the PGIR LAV survey presented herein is likely missing $20-30\%$ of the largest amplitude variables. Cryoscope Pathfinder's $K$-band coverage will conservatively reveal more than 1000 of the largest amplitude stellar LAVs in the Galaxy, extending PGIR's Northern census and providing a more complete view of the large amplitude variable sky.

\begin{acknowledgments}
{Palomar Gattini-IR (PGIR) is generously funded by Caltech, Australian National University, the Mt Cuba Foundation, the Heising Simons Foundation, the Binational Science Foundation. PGIR is a collaborative project among Caltech, Australian National University, University of New South Wales, Columbia University and the Weizmann Institute of Science. MMK acknowledges generous support from the David and Lucille Packard Foundation. MMK and EO acknowledge the US-Israel Bi-national Science Foundation Grant 2016227. MMK and JLS acknowledge the Heising-Simons foundation for support via a Scialog fellowship of the Research Corporation. MMK and AMM acknowledge the Mt Cuba foundation. J. Soon is supported by an Australian Government Research Training Program (RTP) Scholarship.}

{Spectra taken with IRTF were acquired as part of program 2020A111 (PI: K. De).}

{This research has made use of the NASA/IPAC Infrared Science Archive, which is funded by the National Aeronautics and Space Administration and operated by the California Institute of Technology. This publication makes use of data products from the Two Micron All Sky Survey, which is a joint project of the University of Massachusetts and the Infrared Processing and Analysis Center/California Institute of Technology, funded by the National Aeronautics and Space Administration and the National Science Foundation.}

{This work has made use of data from the European Space Agency (ESA) mission
{\it Gaia} (\url{https://www.cosmos.esa.int/gaia}), processed by the {\it Gaia}
Data Processing and Analysis Consortium (DPAC,
\url{https://www.cosmos.esa.int/web/gaia/dpac/consortium}). Funding for the DPAC has been provided by national institutions, in particular the institutions
participating in the {\it Gaia} Multilateral Agreement.}

{Some results are based on observations obtained with the Samuel Oschin Telescope 48-inch and the 60-inch Telescope at the Palomar Observatory as part of the Zwicky Transient Facility project. ZTF is supported by the National Science Foundation under Grants No. AST-1440341 and AST-2034437 and a collaboration including current partners Caltech, IPAC, the Oskar Klein Center at Stockholm University, the University of Maryland, University of California, Berkeley, the University of Wisconsin at Milwaukee, University of Warwick, Ruhr University, Cornell University, Northwestern University and Drexel University. Operations are
conducted by COO, IPAC, and UW.}
\end{acknowledgments}
\facilities{Hale (TSPEC), IRTF (SpeX), IRSA, Gaia, CTIO:2MASS, FLWO:2MASS, WISE, NEOWISE, PO:1.2m (ZTF)}
\software{\texttt{astropy} \citep{astropy:2013, astropy:2018, astropy:2022}, \texttt{matplotlib} \citep{Hunter:2007},
\texttt{numpy} \citep{harris2020array},
\texttt{pandas} \citep{2022zndo...3509134T},
\texttt{scipy} \citep{2020SciPy-NMeth}}
\bibliography{pgirv}{}
\bibliographystyle{aasjournal}

\appendix
\section{PGIR LAV census source observations}
\label{sec:appendix_observation}
We provide the summary for the PGIR LAV census sources in Appendix Table~\ref{tab:lavselection}. 2MASS IDs and counterpart source coordinates are nominally listed for RA/Dec. PGIR coordinates are provided for sources without 2MASS counterparts.

\begin{longrotatetable}
\begin{deluxetable*}
{cccccc}
\tablecaption{Census LAV observations and classifications}
\label{tab:lavselection}
\tablehead{\colhead{PGIR ID} & \colhead{2MASS ID} & \colhead{RA} & \colhead{Dec} & \colhead{Classification} & \colhead{Instrument/Observation Date} }
\startdata
V1030104883 & 18300017-0539288 & 277.501 & -5.658 & Erratic O-rich & TSPEC 20220614 and 20230804\\
V144004432 & 23272528+5917136 & 351.855 & 59.287 & Erratic O-rich & IRTF 20210727\\
V461113838 & 20164403+3410086 & 304.183 & 34.169 & Erratic O-rich & TSPEC 20250808\\
V230102266 & 21051068+5219189 & 316.295 & 52.322 & C-rich LPV & TSPEC 20250416\\
V843234598 & 06051499+0321542 & 91.312 & 3.365 & C-rich LPV & TSPEC 20250411\\
V80901345 & 18523456+0722082 & 283.144 & 7.369 & C-rich LPV & TSPEC 20250411\\
V230036354 & 21095516+5119269 & 317.480 & 51.324 & C-rich LPV & TSPEC 20241222\\
V14220564 & 22351759+5938129 & 338.823 & 59.637 & C-rich LPV & TSPEC 20241222\\
V108301608 & 01055890+6143489 & 16.495 & 61.730 & C-rich LPV & TSPEC 20220718\\
V337035237 & 20210497+4135466 & 305.271 & 41.596 & C-rich LPV & TSPEC 20241224\\
V703221306 & 08292902+1046241 & 127.371 & 10.773 & C-rich LPV & TSPEC 20220510\\
V1213302850 & 07251747-2026519 & 111.323 & -20.448 & C-rich LPV & TSPEC 20241222\\
V57532555 & 13095363+2325408 & 197.473 & 23.428 & C-rich LPV & TSPEC 20241222\\
V593331693 & 19271183+2430446 & 291.799 & 24.512 & C-rich LPV & TSPEC 20220718\\
V526111970 & 19453535+2931360 & 296.397 & 29.527 & C-rich LPV & TSPEC 20250416\\
V460016668 & 19533441+3219115 & 298.393 & 32.320 & C-rich LPV & TSPEC 20241222\\
V230325277 & 21254899+5259363 & 321.454 & 52.993 & C-rich LPV & TSPEC 20241222\\
V625232734 & 06222612+1709136 & 95.609 & 17.154 & C-rich LPV & TSPEC 20250416\\
V184322720 & 22182109+5737020 & 334.588 & 57.617 & C-rich LPV & TSPEC 20211129\\
V112214225 & 03333040+6048251 & 53.377 & 60.807 & C-rich LPV & TSPEC 20241222\\
V196225744 & 04230758+5006281 & 65.782 & 50.108 & C-rich LPV & TSPEC 20241222\\
V84520786 & 06422307+0121084 & 100.596 & 1.352 & C-rich LPV & TSPEC 20220510\\
V110122046 & 02103980+6251335 & 32.666 & 62.859 & C-rich LPV & TSPEC 20241222\\
V249111511 & 05565061+4834447 & 89.211 & 48.579 & C-rich LPV & TSPEC 20241222\\
V555035043 & 06010664+2224350 & 90.278 & 22.410 & C-rich LPV & TSPEC 20241222\\
V336223002 & 20094971+4031082 & 302.457 & 40.519 & C-rich LPV & TSPEC 20241222\\
V338005971 & 20394755+4047040 & 309.948 & 40.784 & C-rich LPV & TSPEC 20220718\\
V23012483 & 21151001+5202179 & 318.792 & 52.038 & C-rich LPV & TSPEC 20220718\\
V231206202 & 21464552+5003588 & 326.690 & 50.066 & C-rich LPV & TSPEC 20241222\\
V184128304 & 21544167+5752261 & 328.674 & 57.874 & C-rich LPV & TSPEC 20241222\\
V185101619 & 22260866+5715476 & 336.536 & 57.263 & C-rich LPV & TSPEC 20241222\\
V73610693 & 19080090+1401022 & 287.004 & 14.017 & C-rich LPV & TSPEC 20250411\\
V737101497 & 19271638+1402395 & 291.818 & 14.044 & C-rich LPV & TSPEC 20250411\\
V108132682 & 00535194+6255096 & 13.466 & 62.919 & C-rich LPV & TSPEC 20241222\\
V15031115 & 03133582+5850237 & 48.399 & 58.840 & C-rich LPV & TSPEC 20211028\\
V1069206089 & 07240347-1252280 & 111.014 & -12.874 & C-rich LPV & TSPEC 20241222\\
V150312997 & 03091563+5846242 & 47.315 & 58.773 & C-rich LPV & TSPEC 20211129\\
V230103575 & 21040971+5157365 & 316.040 & 51.960 & Erratic C-rich & IRTF 20210727\\
V396232390 & 20020874+3627551 & 300.536 & 36.465 & Erratic C-rich & TSPEC 20250416\\
V526102033 & 19454402+2820054 & 296.433 & 28.335 & Erratic C-rich & TSPEC 20220614\\
V23020290 & 21224796+4935165 & 320.700 & 49.588 & Erratic C-rich & TSPEC 20220614\\
V1028021949 & 17574977-0753149 & 269.457 & -7.887 & RCB & TSPEC 20191023\\
V277102608 & 18375125+4723234 & 279.464 & 47.390 & RCB & TSPEC 20230804\\
V663232970 & 19132516+1737020 & 288.355 & 17.617 & RCB & TSPEC 20220718\\
V1173001132 & 17431753-1824024 & 265.823 & -18.401 & RCB & TSPEC 20250411\\
V1246036587 & 18363125-2059154 & 279.130 & -20.988 & RCB & TSPEC 20250416\\
V341323400 & 22155887+4222464 & 333.995 & 42.380 & RCB & TSPEC 20250416\\
V1170122138 & 16472973-1525229 & 251.874 & -15.423 & RCB & TSPEC 20250416\\
V19331245 & 15020132+8303485 & 225.506 & 83.063 & RCB (Z Umi) & \nodata \\
V1171035455 & 17055280-1634165 & 256.470 & -16.571 & RCB (ToI 1213) & \nodata\\
V1243011042 & 17324259-2151407 & 263.177 & -21.861 & RCB (V532 Oph) & \nodata\\
V1173014283 & 17413888-1615466 & 265.412 & -16.263 & RCB (ToI 182) & \nodata\\
V1104114234 & 18451484-0925360 & 281.312 & -9.427 & RCB (FH Sct) & \nodata\\
V735302247 & 18572648+1349096 & 284.360 & 13.819 & RCB (ToI 1309) & \nodata\\
V1247321719 & 19101182-2029420 & 287.549 & -20.495 & RCB (V1157 Sgr) & \nodata\\
V810026928 & 19124303+0553130 & 288.179 & 5.887 & RCB (ToI 274) & \nodata\\
V118601604 & 22031969-1637352 & 330.832 & -16.626 & RCB (U Aqr) & \nodata\\
V282302182 & \nodata & 317.356 & 48.181 & Nova (V2891 Cyg) & \nodata\\
V59412386 & 19383060+2354384 & 294.628 & 23.911 & Symbiotic (D-type) & TSPEC 20230804\\
V736201539 & 19171298+1144092 & 289.304 & 11.736 & Symbiotic candidate & TSPEC 20220614 and 20250416\\
V460302396 & 20074571+3247317 & 301.940 & 32.792 & Symbiotic candidate & TSPEC 20210916, 20230804, 20230929, and 20250416\\
V594121401 & 19374724+2340578 & 294.447 & 23.683 & Symbiotic candidate & TSPEC 20241222 and 20250416\\
V734306757 & 18344488+1419162 & 278.687 & 14.321 & Symbiotic candidate & TSPEC 20220614 and 20250416\\
V664315364 & 19271431+1950575 & 291.810 & 19.849 & Symbiotic candidate & IRTF 20210725\\
V124611295 & 18350188-1935275 & 278.758 & -19.591 & Symbiotic candidate & TSPEC 20220614\\
V1066333453 & 06305614-0931590 & 97.734 & -9.533 & YSO candidate & TSPEC 20241222\\
V1068325804 & 07092228-1030569 & 107.343 & -10.516 & YSO & TSPEC 20211028 and 20211212\\
V1103313459 & 18362900-0922384 & 279.121 & -9.377 & YSO & TSPEC 20250411 and 20250416\\
V113126105 & 03580766+6244253 & 59.532 & 62.740 & YSO (EXor) & TSPEC 20211129 and 20230804\\
V1068325830 & 07092139-1029344 & 107.339 & -10.493 & YSO (EXor) & TSPEC 20250416\\
V694232281 & 05324305+1221083 & 83.179 & 12.352 & YSO & TSPEC 20250411\\
V916304637 & 05442924-0122167 & 86.122 & -1.371 & YSO & TSPEC 20250411\\
V1066302885 & 06265390-1015349 & 96.725 & -10.260 & YSO (HAeBe) & TSPEC 20250411\\
V422332632 & 05570242+3355342 & 89.260 & 33.926 & YSO (HAeBe) & TSPEC 20241222\\
V397016649 & 20050603+3629135 & 301.275 & 36.487 & YSO (Gaia 18dvy, FUor) & TSPEC 20230804\\
V10213291 & 20455394+6757386 & 311.475 & 67.961 & YSO (PV Cep) & TSPEC 20250416\\
V338316063 & \nodata & 312.860 & 44.090 & YSO (PTF10nvg) & TSPEC 20241222\\
V916133218 & 05362855+0004456 & 84.119 & 0.079 & YSO & TSPEC 20250411\\
V1030216476 & 18384243-0723191 & 279.677 & -7.389 & O-rich LPV & TSPEC 20250411\\
V1174031288 & 18082990-1635363 & 272.125 & -16.593 & O-rich LPV & TSPEC 20220510\\
V1102011162 & 18081600-1201323 & 272.067 & -12.026 & O-rich LPV & TSPEC 20250411\\
V1313324888 & 18172324-2451436 & 274.347 & -24.862 & O-rich LPV & TSPEC 20250411\\
V1312102351 & 17414535-2448267 & 265.439 & -24.807 & O-rich LPV & TSPEC 20250411\\
V1243032760 & 17370135-2121506 & 264.256 & -21.364 & O-rich LPV & TSPEC 20250411\\
V1243001876 & 17320393-2241389 & 263.016 & -22.694 & O-rich LPV & IRTF 20210721 and TSPEC 20230804\\
V59500241 & 19540536+2041263 & 298.522 & 20.691 & O-rich LPV & TSPEC 20241224\\
V735336485 & 19035277+1435321 & 285.970 & 14.592 & O-rich LPV & TSPEC 20250411\\
V1244214862 & 18010916-2130204 & 270.288 & -21.506 & O-rich LPV & TSPEC 20240517\\
V1175008049 & 18185365-1718281 & 274.724 & -17.308 & O-rich LPV & TSPEC 20250411\\
V117422755 & 18184943-1743528 & 274.706 & -17.731 & O-rich LPV & TSPEC 20250411\\
V956221163 & 18464926-0322330 & 281.705 & -3.376 & O-rich LPV & TSPEC 20250416\\
V81010554 & 19115576+0831491 & 287.982 & 8.530 & O-rich LPV & TSPEC 20250411\\
V1175008024 & 18184943-1743528 & 274.706 & -17.731 & O-rich LPV & TSPEC 20220614\\
V1245001441 & 18133219-2228356 & 273.384 & -22.477 & O-rich LPV & TSPEC 20250411\\
V738114031 & 19445539+1521227 & 296.231 & 15.356 & O-rich LPV & TSPEC 20220718\\
V663311501 & 19092709+1912230 & 287.363 & 19.206 & O-rich LPV & TSPEC 20250411\\
V110330904 & 18381141-1052268 & 279.548 & -10.874 & O-rich LPV & TSPEC 20250411\\
V1103124578 & 18305088-1023545 & 277.712 & -10.398 & O-rich LPV & TSPEC 20250411\\
V957135151 & 18533746+0014296 & 283.406 & 0.242 & O-rich LPV & TSPEC 20250411\\
V1103017185 & 18234487-1209207 & 275.937 & -12.156 & O-rich LPV & TSPEC 20250411\\
V883012822 & 18510313+0256568 & 282.763 & 2.949 & O-rich LPV & TSPEC 20250411\\
V664306776 & 19261159+1816178 & 291.548 & 18.272 & O-rich LPV & TSPEC 20241224\\
V1243312856 & 17421256-1845193 & 265.552 & -18.755 & O-rich LPV & TSPEC 20220614\\
V1243211283 & 17430636-2149038 & 265.777 & -21.818 & O-rich LPV & TSPEC 20250411\\
V1243308242 & 17385220-2024330 & 264.718 & -20.409 & O-rich LPV & TSPEC 20250411\\
V736132342 & 19115703+1422168 & 287.988 & 14.371 & O-rich LPV & TSPEC 20250411\\
V131103674 & 17270964-2551563 & 261.790 & -25.866 & O-rich LPV & TSPEC 20250411\\
V1312211315 & 17531884-2656374 & 268.329 & -26.944 & O-rich LPV & TSPEC 20220614\\
V396024711 & 19481595+3522060 & 297.066 & 35.368 & O-rich LPV & TSPEC 20241222\\
V3383041 & 20505862+4248115 & 312.744 & 42.803 & O-rich LPV & TSPEC 20220614\\
V80911496 & 18523277+1023310 & 283.137 & 10.392 & O-rich LPV & TSPEC 20250411\\
V74032703 & 00195127+6559304 & 4.964 & 65.992 & O-rich LPV & TSPEC 20240517\\
V39901735 & 20561522+3643199 & 314.063 & 36.722 & O-rich LPV & TSPEC 20241222\\
V1243111319 & 17321516-1933091 & 263.063 & -19.553 & O-rich LPV & TSPEC 20250411\\
V73902474 & 20120916+1116516 & 303.038 & 11.281 & O-rich LPV & TSPEC 20241224\\
V461215108 & 20280266+3128485 & 307.011 & 31.480 & O-rich LPV & TSPEC 20241222\\
V1218233219 & 09111015-2202116 & 137.792 & -22.037 & O-rich LPV & TSPEC 20241222\\
V883133415 & 18553022+0511406 & 283.876 & 5.195 & O-rich LPV & TSPEC 20250416\\
V736214315 & 19152912+1202399 & 288.871 & 12.044 & O-rich LPV & TSPEC 20250416\\
V526131716 & 19512119+2913013 & 297.838 & 29.217 & O-rich LPV & TSPEC 20250416\\
V1312335822 & 17543653-2407541 & 268.652 & -24.132 & O-rich LPV & TSPEC 20250416\\
V664317076 & 19311725+2019210 & 292.822 & 20.323 & O-rich LPV & TSPEC 20250416\\
V810133542 & 19145391+1032116 & 288.725 & 10.537 & O-rich LPV & TSPEC 20220614\\
V882316882 & 18384242+0541298 & 279.677 & 5.692 & O-rich LPV & TSPEC 20250416\\
V881306400 & 18193355+0354498 & 274.890 & 3.914 & O-rich LPV & TSPEC 20250416\\
V131231726 & 17534741-2343139 & 268.448 & -23.721 & O-rich LPV & TSPEC 20250416\\
V336105594 & 19485394+4302145 & 297.225 & 43.037 & RV Tauri supergiant (DF Cyg) & TSPEC 20250416\\
V665315739 & 19471412+1929173 & 296.809 & 19.488 & RV Tauri supergiant (RZ Vul) & TSPEC 20241222\\
\enddata
\end{deluxetable*}
\end{longrotatetable}

\section{Census source colors}
\label{sec:appendix_color}
In this section, we provide color data for the census sources presented in Figure~\ref{fig:colors} of the main text. As stated in \S\ref{sec:demographics}, only sources with quality photometry in the ALLWISE Source Catalog (Table~\ref{tab:allwise_colors}), 2MASS PSC (Table~\ref{tab:tmass_colors}), and Gaia DR3 (Table~\ref{tab:gaia}) are included. 

\startlongtable
\begin{deluxetable*}{cccccccc}
\label{tab:allwise_colors}
\tablecaption{Census LAVs in ALLWISE}
\tablehead{\colhead{PGIR ID} & 
\colhead{ALLWISE ID} & \colhead{RA} & \colhead{Dec} & \colhead{$W1-W2$} & \colhead{$W3-W4$} & \colhead{$W2-W3$} & \colhead{Classification}}
\startdata
\hline
V1030104883 & J183000.17-053928.8 & 277.501 & -5.658 & $0.13\pm0.03$ & $1.20\pm0.04$ & $1.51\pm0.03$ & Erratic O-rich\\
V1030104883 & J183000.17-053928.8 & 277.501 & -5.658 & $0.13\pm0.03$ & $1.20\pm0.04$ & $1.51\pm0.03$ & Erratic O-rich\\
V80901345 & J185234.56+072208.2 & 283.144 & 7.369 & $0.79\pm0.04$ & $0.16\pm0.04$ & $0.93\pm0.03$ & C-rich LPV\\
V703221306 & J082929.01+104624.2 & 127.371 & 10.773 & $0.25\pm0.03$ & $0.17\pm0.09$ & $0.53\pm0.03$ & C-rich LPV\\
V57532555 & J130953.63+232540.8 & 197.473 & 23.428 & $0.46\pm0.03$ & \nodata & $0.58\pm0.03$ & C-rich LPV\\
V593331693 & J192711.83+243044.7 & 291.799 & 24.512 & $0.76\pm0.03$ & $0.26\pm0.04$ & $0.85\pm0.02$ & C-rich LPV\\
V460016668 & J195334.41+321911.6 & 298.393 & 32.320 & $1.04\pm0.08$ & $0.38\pm0.03$ & $1.06\pm0.05$ & C-rich LPV\\
V184322720 & J221821.10+573702.0 & 334.588 & 57.617 & $0.89\pm0.09$ & $0.45\pm0.03$ & $0.69\pm0.05$ & C-rich LPV\\
V249111511 & J055650.61+483444.0 & 89.211 & 48.579 & $1.82\pm0.06$ & $0.86\pm0.03$ & $1.81\pm0.05$ & C-rich LPV\\
V526102033 & J194544.02+282005.4 & 296.433 & 28.335 & $0.68\pm0.08$ & $0.10\pm0.03$ & $0.77\pm0.05$ & Erratic C-rich\\
V1028021949 & J175749.76-075315.0 & 269.457 & -7.887 & $1.04\pm0.04$ & $0.73\pm0.03$ & $1.51\pm0.03$ & RCB\\
V277102608 & J183751.25+472323.5 & 279.464 & 47.390 & $0.96\pm0.03$ & $0.85\pm0.03$ & $1.76\pm0.03$ & RCB\\
V663232970 & J191325.16+173701.9 & 288.355 & 17.617 & $0.84\pm0.03$ & $0.87\pm0.03$ & $2.60\pm0.02$ & RCB\\
V1170122138 & J164729.74-152522.9 & 251.874 & -15.423 & $1.43\pm0.13$ & $0.67\pm0.02$ & $1.60\pm0.09$ & RCB\\
V734306757 & J183444.88+141916.1 & 278.687 & 14.321 & $0.12\pm0.03$ & $0.78\pm0.06$ & $1.31\pm0.03$ & Symbiotic candidate\\
V1066333453 & J063056.15-093158.9 & 97.734 & -9.533 & $0.72\pm0.03$ & $1.38\pm0.06$ & $2.24\pm0.03$ & YSO candidate\\
V694232281 & J053243.05+122108.4 & 83.179 & 12.352 & $0.77\pm0.03$ & $2.02\pm0.03$ & $2.38\pm0.02$ & YSO\\
V397016649 & J200505.98+362913.0 & 301.275 & 36.487 & $0.86\pm0.03$ & $2.37\pm0.07$ & $2.03\pm0.04$ & YSO (FUor)\\
V916133218 & J053628.55+000445.6 & 84.119 & 0.079 & $0.89\pm0.03$ & $1.75\pm0.03$ & $2.24\pm0.03$ & YSO (ttauri)\\
V1243001876 & J173203.92-224138.8 & 263.016 & -22.694 & $0.63\pm0.03$ & $1.15\pm0.02$ & $2.27\pm0.03$ & O-rich LPV\\
V1243001876 & J173203.92-224138.8 & 263.016 & -22.694 & $0.63\pm0.03$ & $1.15\pm0.02$ & $2.27\pm0.03$ & O-rich LPV\\
V663311501 & J190927.08+191222.9 & 287.363 & 19.206 & $0.68\pm0.07$ & $0.70\pm0.03$ & $0.73\pm0.03$ & O-rich LPV\\
V883012822 & J185103.12+025657.2 & 282.763 & 2.949 & $1.36\pm0.07$ & $1.69\pm0.03$ & $1.43\pm0.04$ & O-rich LPV\\
V131103674 & J172709.64-255156.4 & 261.790 & -25.866 & $0.65\pm0.04$ & $1.10\pm0.03$ & $2.41\pm0.03$ & O-rich LPV\\
V336105594 & J194853.94+430214.5 & 297.225 & 43.037 & $1.01\pm0.08$ & $1.01\pm0.02$ & $1.87\pm0.05$ & RV Tauri supergiant (DF Cyg)\\
V665315739 & J194714.13+192917.2 & 296.809 & 19.488 & $1.17\pm0.10$ & $1.17\pm0.02$ & $2.03\pm0.06$ & RV Tauri supergiant (RZ Vul)\\
\enddata
\tablecomments{Only classified sources with quality photometry as defined in \S\ref{sec:demographics} in all bands are presented.}
\end{deluxetable*}

\startlongtable
\begin{deluxetable*}{ccccccc}
\tablecaption{Census LAVs in 2MASS PSC}
\tablehead{\colhead{PGIR ID} & 
\colhead{2MASS ID} & \colhead{RA} & \colhead{Dec} & \colhead{$H-K_s$} & \colhead{$J-H$} & \colhead{Classification}}
\startdata
\hline
V1030104883 & 18300017-0539288 & 277.501 & -5.658 & $0.64\pm0.05$ & $1.40\pm0.05$ & Erratic O-rich\\
V1030104883 & 18300017-0539288 & 277.501 & -5.658 & $0.64\pm0.05$ & $1.40\pm0.05$ & Erratic O-rich\\
V144004432 & 23272528+5917136 & 351.855 & 59.287 & $0.83\pm0.03$ & $1.77\pm0.03$ & Erratic O-rich\\
V461113838 & 20164403+3410083 & 304.183 & 34.169 & $0.95\pm0.02$ & $1.60\pm0.03$ & Erratic O-rich\\
V230102266 & 21051068+5219189 & 316.295 & 52.322 & $2.17\pm0.04$ & $3.08\pm0.04$ & C-rich LPV\\
V843234598 & 06051499+0321542 & 91.312 & 3.365 & $1.98\pm0.06$ & $2.42\pm0.06$ & C-rich LPV\\
V80901345 & 18523456+0722082 & 283.144 & 7.369 & $1.47\pm0.03$ & $2.15\pm0.03$ & C-rich LPV\\
V230036354 & 21095516+5119269 & 317.480 & 51.324 & $2.07\pm0.04$ & $2.85\pm0.04$ & C-rich LPV\\
V14220564 & 22351759+5938129 & 338.823 & 59.637 & $1.79\pm0.03$ & $2.30\pm0.03$ & C-rich LPV\\
V108301608 & 01055890+6143489 & 16.495 & 61.730 & $1.24\pm0.03$ & $1.60\pm0.03$ & C-rich LPV\\
V337035237 & 20210497+4135466 & 305.271 & 41.596 & $1.77\pm0.02$ & $2.81\pm0.03$ & C-rich LPV\\
V703221306 & 08292902+1046241 & 127.371 & 10.773 & $0.87\pm0.03$ & $1.25\pm0.03$ & C-rich LPV\\
V1213302850 & 07251747-2026519 & 111.323 & -20.448 & $1.70\pm0.04$ & $2.16\pm0.04$ & C-rich LPV\\
V57532555 & 13095363+2325408 & 197.473 & 23.428 & $0.98\pm0.04$ & $1.29\pm0.04$ & C-rich LPV\\
V526111970 & 19453535+2931360 & 296.397 & 29.527 & $1.43\pm0.05$ & $1.86\pm0.09$ & C-rich LPV\\
V460016668 & 19533441+3219115 & 298.393 & 32.320 & $1.46\pm0.03$ & $2.02\pm0.03$ & C-rich LPV\\
V230325277 & 21254899+5259363 & 321.454 & 52.993 & $1.42\pm0.03$ & $2.13\pm0.03$ & C-rich LPV\\
V184322720 & 22182109+5737020 & 334.588 & 57.617 & $1.47\pm0.07$ & $2.14\pm0.07$ & C-rich LPV\\
V112214225 & 03333040+6048251 & 53.377 & 60.807 & $3.04\pm0.04$ & $3.55\pm0.06$ & C-rich LPV\\
V196225744 & 04230758+5006281 & 65.782 & 50.108 & $1.60\pm0.03$ & $2.10\pm0.03$ & C-rich LPV\\
V84520786 & 06422307+0121084 & 100.596 & 1.352 & $1.53\pm0.04$ & $1.98\pm0.03$ & C-rich LPV\\
V110122046 & 02103980+6251335 & 32.666 & 62.859 & $2.01\pm0.03$ & $2.46\pm0.03$ & C-rich LPV\\
V249111511 & 05565061+4834447 & 89.211 & 48.579 & $2.14\pm0.03$ & $2.73\pm0.03$ & C-rich LPV\\
V555035043 & 06010664+2224350 & 90.278 & 22.410 & $2.29\pm0.04$ & $2.70\pm0.04$ & C-rich LPV\\
V336223002 & 20094971+4031082 & 302.457 & 40.519 & $1.81\pm0.03$ & $2.54\pm0.03$ & C-rich LPV\\
V338005971 & 20394755+4047040 & 309.948 & 40.784 & $2.29\pm0.03$ & $3.52\pm0.03$ & C-rich LPV\\
V23012483 & 21151001+5202179 & 318.792 & 52.038 & $2.12\pm0.03$ & $2.98\pm0.03$ & C-rich LPV\\
V231206202 & 21464552+5003588 & 326.690 & 50.066 & $2.38\pm0.04$ & $3.04\pm0.04$ & C-rich LPV\\
V184128304 & 21544167+5752261 & 328.674 & 57.874 & $2.13\pm0.05$ & $2.83\pm0.04$ & C-rich LPV\\
V185101619 & 22260866+5715476 & 336.536 & 57.263 & $1.52\pm0.03$ & $2.09\pm0.04$ & C-rich LPV\\
V73610693 & 19080090+1401022 & 287.004 & 14.017 & $2.12\pm0.02$ & $2.61\pm0.03$ & C-rich LPV\\
V737101497 & 19271638+1402395 & 291.818 & 14.044 & $1.46\pm0.03$ & $2.31\pm0.03$ & C-rich LPV\\
V108132682 & 00535194+6255096 & 13.466 & 62.919 & $2.17\pm0.04$ & $2.65\pm0.04$ & C-rich LPV\\
V15031115 & 03133582+5850237 & 48.399 & 58.840 & $1.77\pm0.06$ & $2.14\pm0.06$ & C-rich LPV\\
V1069206089 & 07240347-1252280 & 111.014 & -12.874 & $1.86\pm0.05$ & $2.29\pm0.05$ & C-rich LPV\\
V150312997 & 03091563+5846242 & 47.315 & 58.773 & $1.77\pm0.03$ & $2.34\pm0.03$ & C-rich LPV\\
V230103575 & 21040971+5157365 & 316.040 & 51.960 & $1.08\pm0.03$ & $1.93\pm0.03$ & Erratic C-rich\\
V396232390 & 20020874+3627551 & 300.536 & 36.465 & $0.97\pm0.05$ & $1.59\pm0.04$ & Erratic C-rich\\
V526102033 & 19454402+2820054 & 296.433 & 28.335 & $1.18\pm0.04$ & $1.72\pm0.04$ & Erratic C-rich\\
V23020290 & 21224796+4935165 & 320.700 & 49.588 & $1.15\pm0.03$ & $1.95\pm0.04$ & Erratic C-rich\\
V1028021949 & 17574977-0753149 & 269.457 & -7.887 & $0.72\pm0.04$ & $0.85\pm0.04$ & RCB\\
V277102608 & 18375125+4723234 & 279.464 & 47.390 & $0.66\pm0.02$ & $0.50\pm0.03$ & RCB\\
V663232970 & 19132516+1737020 & 288.355 & 17.617 & $0.44\pm0.03$ & $0.70\pm0.03$ & RCB\\
V1173001132 & 17431753-1824024 & 265.823 & -18.401 & $1.03\pm0.06$ & $0.98\pm0.06$ & RCB\\
V1246036587 & 18363125-2059154 & 279.130 & -20.988 & $1.50\pm0.03$ & $1.50\pm0.03$ & RCB\\
V341323400 & 22155887+4222464 & 333.995 & 42.380 & $2.41\pm0.04$ & $3.01\pm0.04$ & RCB\\
V1170122138 & 16472973-1525229 & 251.874 & -15.423 & $1.21\pm0.03$ & $1.14\pm0.03$ & RCB\\
V59412386 & 19383060+2354384 & 294.628 & 23.911 & $1.55\pm0.03$ & $2.25\pm0.03$ & Symbiotic (D-type)\\
V736201539 & 19171298+1144092 & 289.304 & 11.736 & $1.49\pm0.04$ & $2.49\pm0.05$ & Symbiotic candidate\\
V460302396 & 20074571+3247317 & 301.940 & 32.792 & $0.73\pm0.04$ & $1.47\pm0.03$ & Symbiotic candidate\\
V594121401 & 19374724+2340578 & 294.447 & 23.683 & $1.08\pm0.03$ & $1.39\pm0.03$ & Symbiotic candidate\\
V594121401 & 19374724+2340578 & 294.447 & 23.683 & $1.08\pm0.03$ & $1.39\pm0.03$ & Symbiotic candidate\\
V734306757 & 18344488+1419162 & 278.687 & 14.321 & $0.66\pm0.02$ & $1.32\pm0.03$ & Symbiotic candidate\\
V664315364 & 19271431+1950575 & 291.810 & 19.849 & $0.65\pm0.06$ & $1.27\pm0.06$ & Symbiotic candidate\\
V124611295 & 18350188-1935275 & 278.758 & -19.591 & $0.57\pm0.03$ & $1.10\pm0.04$ & Symbiotic candidate\\
V1066333453 & 06305614-0931590 & 97.734 & -9.533 & $1.02\pm0.03$ & $1.05\pm0.03$ & YSO candidate\\
V1068325804 & 07092228-1030569 & 107.343 & -10.516 & $1.08\pm0.04$ & $1.49\pm0.04$ & YSO\\
V113126105 & 03580766+6244253 & 59.532 & 62.740 & $1.57\pm0.04$ & $1.71\pm0.04$ & YSO (EXor)\\
V1068325830 & 07092139-1029344 & 107.339 & -10.493 & $1.12\pm0.03$ & $1.41\pm0.04$ & YSO (EXor)\\
V694232281 & 05324305+1221083 & 83.179 & 12.352 & $0.73\pm0.03$ & $1.02\pm0.03$ & YSO\\
V916304637 & 05442924-0122167 & 86.122 & -1.371 & $0.62\pm0.03$ & $0.92\pm0.03$ & YSO\\
V1066302885 & 06265390-1015349 & 96.725 & -10.260 & $1.20\pm0.05$ & $1.18\pm0.04$ & YSO (HAeBe) \\
V422332632 & 05570242+3355342 & 89.260 & 33.926 & $1.23\pm0.04$ & $1.33\pm0.04$ & YSO (HAeBe)\\
V10213291 & 20455394+6757386 & 311.475 & 67.961 & $2.21\pm0.03$ & $2.96\pm0.04$ & YSO (PV Cep)\\
V916133218 & 05362855+0004456 & 84.119 & 0.079 & $0.39\pm0.03$ & $0.32\pm0.03$ & YSO\\
V1030216476 & 18384243-0723191 & 279.677 & -7.389 & $1.66\pm0.03$ & $2.91\pm0.03$ & O-rich LPV\\
V1174031288 & 18082990-1635363 & 272.125 & -16.593 & $1.30\pm0.05$ & $2.06\pm0.05$ & O-rich LPV\\
V1102011162 & 18081600-1201323 & 272.067 & -12.026 & $1.33\pm0.04$ & $2.16\pm0.04$ & O-rich LPV\\
V1313324888 & 18172324-2451436 & 274.347 & -24.862 & $1.66\pm0.03$ & $2.48\pm0.04$ & O-rich LPV\\
V1243032760 & 17370135-2121506 & 264.256 & -21.364 & $0.80\pm0.05$ & $1.31\pm0.05$ & O-rich LPV\\
V1243001876 & 17320393-2241389 & 263.016 & -22.694 & $0.90\pm0.06$ & $1.36\pm0.06$ & O-rich LPV\\
V59500241 & 19540536+2041263 & 298.522 & 20.691 & $0.86\pm0.04$ & $1.44\pm0.04$ & O-rich LPV\\
V1244214862 & 18010916-2130204 & 270.288 & -21.506 & $1.95\pm0.05$ & $2.49\pm0.05$ & O-rich LPV\\
V1175008049 & 18185365-1718281 & 274.724 & -17.308 & $1.41\pm0.04$ & $2.20\pm0.04$ & O-rich LPV\\
V117422755 & 18184943-1743528 & 274.706 & -17.731 & $1.00\pm0.05$ & $1.56\pm0.05$ & O-rich LPV\\
V81010554 & 19115576+0831491 & 287.982 & 8.530 & $1.58\pm0.05$ & $2.45\pm0.05$ & O-rich LPV\\
V1175008024 & 18184943-1743528 & 274.706 & -17.731 & $1.00\pm0.05$ & $1.56\pm0.05$ & O-rich LPV\\
V1245001441 & 18133219-2228356 & 273.384 & -22.477 & $0.69\pm0.06$ & $1.05\pm0.06$ & O-rich LPV\\
V738114031 & 19445539+1521227 & 296.231 & 15.356 & $0.88\pm0.03$ & $1.15\pm0.03$ & O-rich LPV\\
V663311501 & 19092709+1912230 & 287.363 & 19.206 & $0.73\pm0.03$ & $1.24\pm0.03$ & O-rich LPV\\
V110330904 & 18381141-1052268 & 279.548 & -10.874 & $1.72\pm0.03$ & $2.19\pm0.03$ & O-rich LPV\\
V957135151 & 18533746+0014296 & 283.406 & 0.242 & $1.39\pm0.05$ & $2.49\pm0.05$ & O-rich LPV\\
V1103017185 & 18234487-1209207 & 275.937 & -12.156 & $1.37\pm0.04$ & $2.29\pm0.04$ & O-rich LPV\\
V883012822 & 18510313+0256568 & 282.763 & 2.949 & $1.66\pm0.04$ & $2.76\pm0.03$ & O-rich LPV\\
V664306776 & 19261159+1816178 & 291.548 & 18.272 & $1.82\pm0.02$ & $2.38\pm0.02$ & O-rich LPV\\
V1243312856 & 17421256-1845193 & 265.552 & -18.755 & $1.50\pm0.05$ & $2.14\pm0.04$ & O-rich LPV\\
V1243211283 & 17430636-2149038 & 265.777 & -21.818 & $1.19\pm0.05$ & $1.77\pm0.05$ & O-rich LPV\\
V1243308242 & 17385220-2024330 & 264.718 & -20.409 & $1.19\pm0.06$ & $1.83\pm0.06$ & O-rich LPV\\
V736132342 & 19115703+1422168 & 287.988 & 14.371 & $1.00\pm0.02$ & $1.51\pm0.03$ & O-rich LPV\\
V131103674 & 17270964-2551563 & 261.790 & -25.866 & $0.96\pm0.04$ & $1.47\pm0.03$ & O-rich LPV\\
V396024711 & 19481595+3522060 & 297.066 & 35.368 & $2.74\pm0.03$ & $3.75\pm0.04$ & O-rich LPV\\
V3383041 & 20505862+4248115 & 312.744 & 42.803 & $2.15\pm0.03$ & $3.29\pm0.03$ & O-rich LPV\\
V80911496 & 18523277+1023310 & 283.137 & 10.392 & $1.99\pm0.03$ & $2.97\pm0.03$ & O-rich LPV\\
V74032703 & 00195127+6559304 & 4.964 & 65.992 & $1.75\pm0.06$ & $2.35\pm0.06$ & O-rich LPV\\
V39901735 & 20561522+3643199 & 314.063 & 36.722 & $2.06\pm0.03$ & $2.66\pm0.04$ & O-rich LPV\\
V1243111319 & 17321516-1933091 & 263.063 & -19.553 & $1.30\pm0.06$ & $2.08\pm0.06$ & O-rich LPV\\
V73902474 & 20120916+1116516 & 303.038 & 11.281 & $1.79\pm0.06$ & $2.34\pm0.06$ & O-rich LPV\\
V461215108 & 20280266+3128485 & 307.011 & 31.480 & $1.79\pm0.03$ & $2.56\pm0.03$ & O-rich LPV\\
V1218233219 & 09111015-2202116 & 137.792 & -22.037 & $1.72\pm0.04$ & $2.25\pm0.04$ & O-rich LPV\\
V883133415 & 18553022+0511406 & 283.876 & 5.195 & $2.30\pm0.04$ & $3.72\pm0.03$ & O-rich LPV\\
V736214315 & 19152912+1202399 & 288.871 & 12.044 & $1.92\pm0.04$ & $2.91\pm0.04$ & O-rich LPV\\
V1312335822 & 17543653-2407541 & 268.652 & -24.132 & $1.86\pm0.04$ & $2.93\pm0.04$ & O-rich LPV\\
V664317076 & 19311725+2019210 & 292.822 & 20.323 & $1.89\pm0.03$ & $2.74\pm0.03$ & O-rich LPV\\
V810133542 & 19145391+1032116 & 288.725 & 10.537 & $1.43\pm0.05$ & $2.46\pm0.05$ & O-rich LPV\\
V882316882 & 18384242+0541298 & 279.677 & 5.692 & $0.99\pm0.04$ & $1.55\pm0.04$ & O-rich LPV\\
V131231726 & 17534741-2343139 & 268.448 & -23.721 & $0.94\pm0.04$ & $1.66\pm0.04$ & O-rich LPV\\
V336105594 & 19485394+4302145 & 297.225 & 43.037 & $1.16\pm0.02$ & $0.94\pm0.03$ & RV Tauri supergiant (DF Cyg)\\
V665315739 & 19471412+1929173 & 296.809 & 19.488 & $0.80\pm0.04$ & $0.76\pm0.04$ & RV Tauri supergiant (RZ Vul)\\
\hline
\enddata
\label{tab:tmass_colors}
\tablecomments{Only classified sources with quality photometry as defined in \S\ref{sec:demographics} in all bands are presented.}
\end{deluxetable*}

\startlongtable
\begin{deluxetable*}{cccccccc}
\tablecaption{Census LAVs in Gaia DR3}
\tablehead{\colhead{PGIR ID} & 
\colhead{Gaia DR3 ID} & \colhead{RA} & \colhead{Dec} & \colhead{$(BP-RP) - E(BP-RP)$} & \colhead{$M_G$} & \colhead{rpgeo (kpc)} & \colhead{Classification}}
\startdata
\hline
V1030104883 & 4256311701322804736 & 277.501 & -5.658 & $2.14$ & $-4.16$ & $5.70$ & Erratic O-rich\\
V14220564 & 2200652495570094080 & 338.823 & 59.637 & $1.40$ & $-3.94$ & $11.80$ & C-rich LPV\\
V108301608 & 522549365040711424 & 16.495 & 61.730 & $1.13$ & $-2.82$ & $6.52$ & C-rich LPV\\
V1213302850 & 2930166640852891904 & 111.323 & -20.448 & $0.99$ & $-1.11$ & $4.67$ & C-rich LPV\\
V460016668 & 2033865480850545536 & 298.393 & 32.320 & $1.26$ & $-2.98$ & $9.56$ & C-rich LPV\\
V230325277 & 2172813995026952064 & 321.454 & 52.993 & $1.98$ & $-4.38$ & $9.26$ & C-rich LPV\\
V184322720 & 2198572597533906816 & 334.588 & 57.617 & $1.79$ & $-4.43$ & $8.03$ & C-rich LPV\\
V196225744 & 270485224851806720 & 65.782 & 50.108 & $1.91$ & $-2.68$ & $2.55$ & C-rich LPV\\
V110122046 & 514646827070604416 & 32.666 & 62.859 & $1.51$ & $-1.35$ & $4.50$ & C-rich LPV\\
V73610693 & 4314416321039627904 & 287.004 & 14.017 & $1.42$ & $-4.26$ & $13.34$ & C-rich LPV\\
V108132682 & 523657947638181120 & 13.466 & 62.919 & $1.60$ & $-5.26$ & $12.89$ & C-rich LPV\\
V150312997 & 460488290195134976 & 47.315 & 58.773 & $1.26$ & $-3.58$ & $3.66$ & C-rich LPV\\
V230103575 & 2169557928773377024 & 316.040 & 51.960 & $2.04$ & $-4.03$ & $3.11$ & Erratic C-rich\\
V460302396 & 2055104643862870400 & 301.940 & 32.792 & $3.21$ & $-2.58$ & $4.58$ & Symbiotic candidate\\
V664315364 & 4515694706300129536 & 291.810 & 19.849 & $0.80$ & $-1.80$ & $3.50$ & Symbiotic candidate\\
V1068325830 & 3046391406515862912 & 107.339 & -10.493 & $1.86$ & $5.49$ & $2.17$ & YSO (EXor)\\
V916304637 & 3218175740278049664 & 86.122 & -1.371 & $0.55$ & $3.37$ & $0.40$ & YSO\\
V916133218 & 3220829441656781824 & 84.119 & 0.079 & $0.67$ & $3.69$ & $0.40$ & YSO\\
V1243308242 & 4118436379211298048 & 264.718 & -20.409 & $2.71$ & $-3.95$ & $5.50$ & O-rich LPV\\
\enddata
\label{tab:gaia}
\tablecomments{Only classified sources with defined distances and extinction-corrected and de-reddened photometry are presented. $M_G$ is corrected by $A_G$ and the photogeometric distances from \citet{2021AJ....161..147B} are considered.}
\end{deluxetable*}

For the Gaia dereddened, extinction-corrected color-absolute magnitude diagram, we also provide the ADQL query used to generate the random sample in the reference CaMD, adapted from \citet{Andrae:2018A&A...616A...8A} and \citet{Gaia:2019A&A...623A.110G}: 

\texttt{
SELECT bp\_rp\_index AS bp\_rp, g\_abs\_index AS g\_abs, n \\
FROM (\\
SELECT (bp\_rp-ebpminrp\_gspphot) AS bp\_rp\_index, (phot\_g\_mean\_mag - ag\_gspphot + 5 * LOG10(parallax) - 10) AS g\_abs\_index, COUNT(*) AS n \\
FROM gaiadr3.gaia\_source \\
    WHERE visibility\_periods\_used > 5\\
    AND astrometric\_excess\_noise < 0.5\\
    AND parallax > 0\\
    AND parallax\_over\_error > 5\\
    AND phot\_bp\_mean\_flux\_over\_error > 20\\
    AND phot\_rp\_mean\_flux\_over\_error > 20\\
    AND phot\_g\_mean\_flux\_over\_error > 50\\
    AND phot\_bp\_rp\_excess\_factor < 1.2*(1.2+0.03*power(phot\_bp\_mean\_mag-phot\_rp\_mean\_mag,2))\\
	GROUP BY bp\_rp\_index, g\_abs\_index\\
    )\\
	AS subquery} 

\end{document}